\documentclass[manuscript,screen]{acmart}
\AtBeginDocument{%
  }

\setcopyright{acmlicensed}
\copyrightyear{2018}
\acmYear{2018}
\acmDOI{XXXXXXX.XXXXXXX}
\acmISBN{978-1-4503-XXXX-X/2018/06}

\usepackage{amsmath}
\usepackage{subcaption}
\usepackage{enumitem}
\usepackage{float}
\usepackage{pdflscape} 
\usepackage{adjustbox}
\usepackage{booktabs}
\usepackage{array}
\usepackage{orcidlink}
{\hfuzz=40pt \hbadness=10000\relax}
\begin{document}

\title{Architectural Design and Performance Analysis of FPGA based AI Accelerators: A Comprehensive Review}

\author{Soumita Chatterjee}
\orcid{0009-0006-4559-5775}
%\email{webmaster@marysville-ohio.com}
\affiliation{%
  \institution{Indian Institute of Engineering Science and Technology, Shibpur}
  \city{Howrah}
  \country{India}}
\email{csoumita45@gmail.com}

% \affiliation{%
% \author{Habibur Rahaman}
% \orcid{0009-0000-2841-5572}
% \affiliation{%
%  \institution{University of Florida}
%  \city{Gainesville}
%  \country{United States of America}}
% \email{rahaman.habibur@ufl.edu}
\author{Sudip Ghosh}
\orcid{0000-0001-5750-2501}
\affiliation{%
 \institution{Indian Institute of Engineering Science and Technology, Shibpur}
 \city{Howrah}
 \country{India}}
\email{sudip.ghosh@vlsi.iiests.ac.in}

\author{Tamal Ghosh}
\orcid{0000-0003-3032-0163}
\affiliation{%
 \institution{Indian Institute of Engineering Science and Technology, Shibpur}
 \city{Howrah}
 \country{India}}
\email{tamalghosh.vlsi@faculty.iiests.ac.in}

\author{Hafizur Rahaman}
\orcid{0000-0001-9012-5437}
\affiliation{%
  \institution{Indian Institute of Engineering Science and Technology, Shibpur}
  \city{Howrah}
  \country{India}}
\email{hafizur@vlsi.iiests.ac.in}

\renewcommand{\shortauthors}{ Chatterjee et al.}

\begin{abstract}
Deep learning (DL) has emerged as a rapidly developing advanced technology, enabling the performance of complex tasks involving image recognition, natural language processing, and autonomous decision-making with high levels of accuracy. However, as these technologies evolve and strive to meet the growing demands of real-life applications, the complexity of DL models continues to increase. These models require  processing of massive volumes of data, demanding substantial computational power and memory bandwidth. This gives rise to the critical need for hardware accelerators that can deliver both high performance and energy efficiency. Accelerator types include ASIC based solutions, GPU accelerators, and FPGA based implementations. The limitations of ASIC and GPU accelerators have led to FPGAs becoming one of the prominent solutions, offering distinct advantages for DL workloads. FPGAs provide a flexible and reconfigurable platform, allowing model specific customization while maintaining high efficiency. This article explores various hardware level optimizations for DL. These optimizations include techniques such as loop pipelining, parallelism, quantization, and various memory hierarchy enhancements. In addition, it provides an overview of state-of-the-art FPGA-based neural network accelerators. Through the study and analysis of these accelerators, several challenges have been identified, paving the way for future optimizations and innovations in the design of FPGA-based hardware accelerators.

\end{abstract}




\maketitle

\section{Introduction}
\label{sec1}
Deep learning (DL) models are advanced machine learning systems composed of multiple layers of interconnected nodes that progressively transform input data into higher level representations. These models are designed to automatically extract complex patterns and features from large volumes of data without manual intervention. This capability makes DL highly effective for solving complex problems in domains like computer vision, natural language processing, speech recognition and autonomous systems. Using vast datasets and substantial computational power, DL models achieve high performance across a wide range of applications while remaining adaptable to innovations across diverse domains such as agriculture, healthcare, Automation, robotics, etc.

In recent years, deep learning has experienced significant growth in prominence and scalability across various domains. Its applications span image classification \cite{10.1007/978-981-97-1335-6_31, FAN2024e38104}, image segmentation \cite{ZHANG2025129395}, speech recognition \cite{8632885, 10924161}, text processing \cite{8978121, SINTHUJA2024789}, object detection \cite{10098596}, disease prediction \cite{Doshi2025}, medical diagnosis \cite{9303459}, fault prediction \cite{5948412} and detection \cite{BATOOL2022107886}, climate modeling \cite{YANG2024120797}, and fraud detection \cite{9755930}. As DL models scale, the need to process and manage large volumes of data grows accordingly. Spectacular hardware accelerators that offload computation from the Central Processing Unit (CPU) address performance bottlenecks by enabling more efficient and faster execution. The increasing complexity of neural networks has fuelled the demand for hardware based Artificial Intelligence (AI) accelerators, which provide high throughput, low latency, and concurrent multiprocessing capabilities. These characteristics allow them to process large datasets with greater speed than traditional computing platforms.

\begin{figure*}[tb]
\centering
\includegraphics[width=\linewidth, height= 0.46\linewidth]{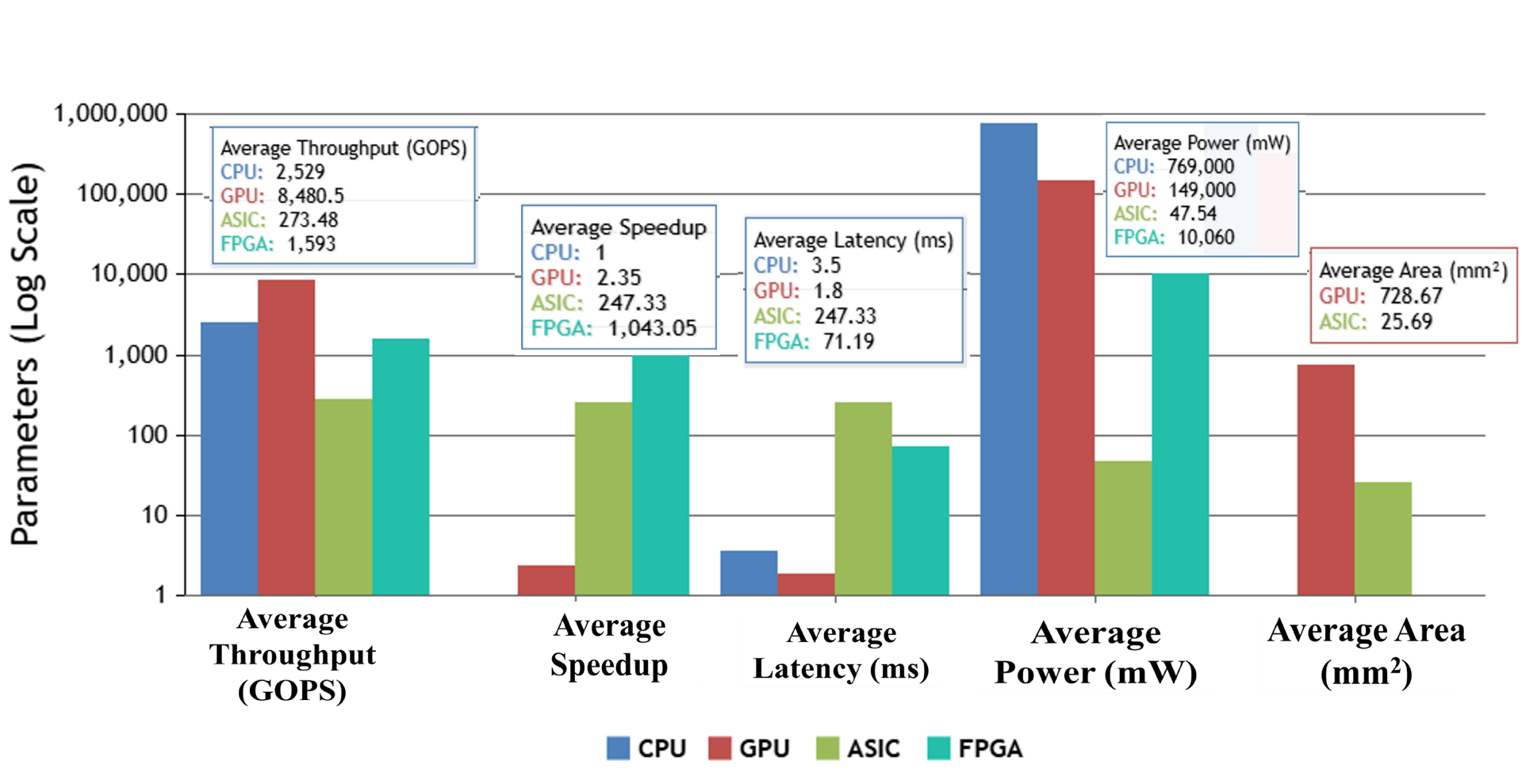}
\caption{Performance Comparison Metrics Across CPU, GPU, ASIC and FPGA.}
\Description{Motivation chart.}
\label{chart_metrics}
\end{figure*}

\begin{figure*}[!ht]
\centering
\includegraphics[width=0.8\linewidth]{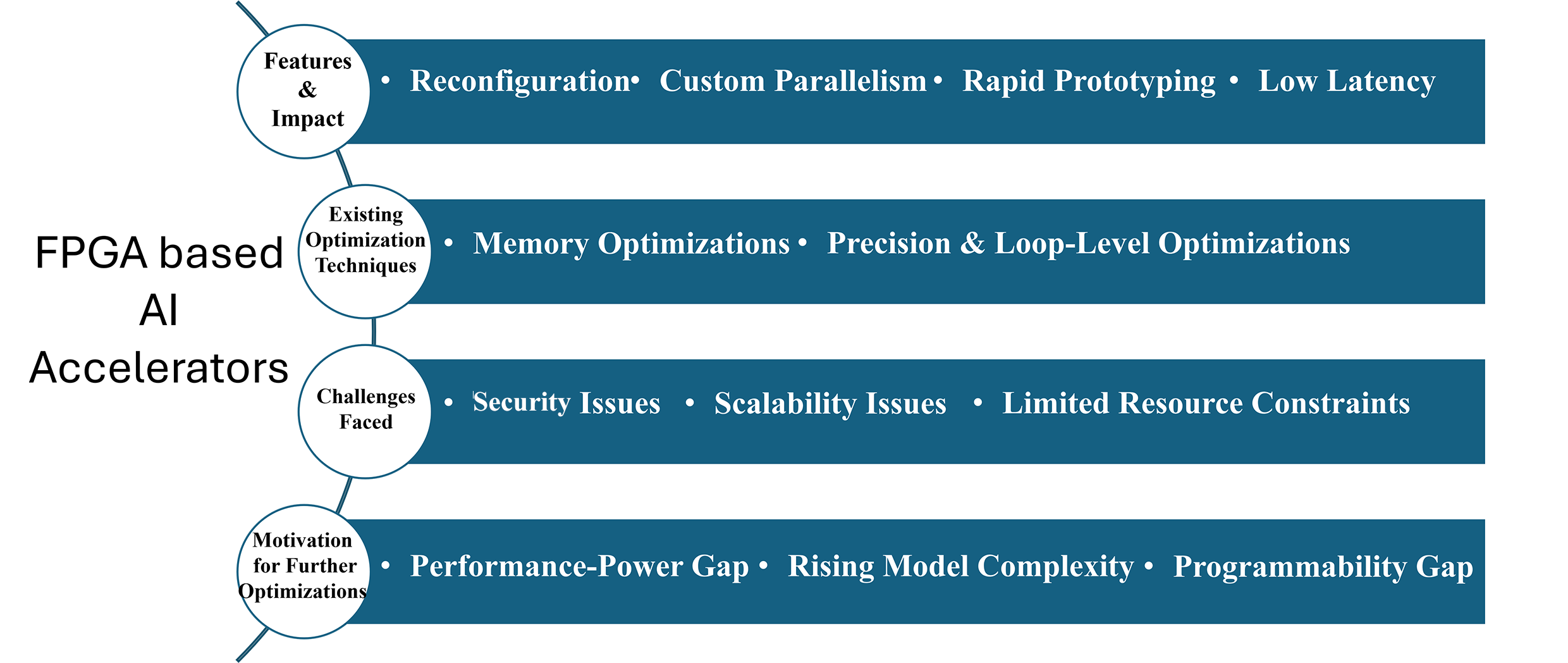}
\caption{Key features, existing optimization techniques, limitations \& need for further optimization of FPGA-based hardware accelerators.}
\Description{Motivation chart.}
\label{motiv}
\end{figure*}

Among the various accelerators proposed over the years, ranging from Graphics Processing Units (GPUs) to Application-Specific Integrated Circuits (ASICs), each offers distinct advantages and limitations. GPUs provide substantial parallel processing capability through standardized instruction sets, while ASICs offer high computational efficiency but lack reconfigurability and require long development cycles. As DL models continue to expand on a larger scale, GPUs and ASICs alone are no longer sufficient. Field Programmable Gate Arrays (FPGAs), known for their low latency, parallelism, and reconfigurable architecture, have emerged as promising candidates for accelerating large scale neural networks. However, the growing diversity of FPGA-based accelerators presents challenges in identifying appropriate optimization techniques and methodologies. Recent survey efforts \cite{8594633} highlight these trends and limitations. A brief overview of the architecture, operation, merits and demerits of the different types of AI hardware accelerators if given in the Table \ref{comparison_accelerators}. 
\begin{table*}[!ht]
\centering
\caption{Comparison of GPU, ASIC, and FPGA-based AI Accelerators}
\small
\label{comparison_accelerators}

\newcolumntype{P}[1]{>{\centering\arraybackslash}p{#1}}

\begin{tabular}{
P{1.5cm}
P{2.35cm}
P{2.35cm}
P{3.45cm}
P{3.45cm}
}
\toprule
\textbf{Accelerator Type} &
\textbf{Architecture} &
\textbf{Operation} &
\textbf{Merits} &
\textbf{Demerits} \\
\midrule
\textbf{GPU} \cite{10372220}, \cite{6100451},
\cite{7828382}, \cite{10323722} &
Massively parallel SIMD/SIMT architecture with thousands of cores; optimized for throughput-based computing. &
Executes large numbers of parallel threads; ideal for matrix multiplications and convolution operations. &
\vspace{-1em}
\begin{itemize}[leftmargin=*, labelsep=0.4em]
    \item Mature and well-supported software ecosystem (CUDA, ROCm)
\item High computational throughput
\item Flexible for both training and inference
\item Widely available and cost-effective
\end{itemize} &
\vspace{-1em}
\begin{itemize}[leftmargin=*, labelsep=0.4em]
\item High power consumption
\item Lower energy efficiency compared to ASICs
\item Limited memory bandwidth for certain workloads
\item General-purpose, not fully optimized for specific networks
\end{itemize}
\\
\hline
\textbf{ASIC} (NPUs, TPUs) \cite{10.1145/2541940.2541967}, 
\cite{7011421},
\cite{10.1145/2694344.2694358},
\cite{WANG2025109887},
\cite{8682197} &
Custom dataflow architectures optimized for neural network operations; fixed-function hardware tailored for AI kernels. &
Executes highly optimized tensor and matrix operations using specialized systolic arrays or dataflow engines. &
\vspace{-1em}
\begin{itemize}[leftmargin=*, labelsep=0.4em]
\item Highest performance and energy efficiency
\item Predictable latency
\item Optimized for specific DNN workloads
\item Suitable for large-scale datacenter deployments
\end{itemize} &
\vspace{-1em}
\begin{itemize}[leftmargin=*, labelsep=0.4em]
\item Very high development cost and long design cycle
\item Inflexible—cannot be reconfigured after fabrication
\item Limited adaptability to evolving AI models
\end{itemize}  \\
\hline
\textbf{FPGA} \cite{9252000},
\cite{8050816},
\cite{ABDELBAKY2025155845},
\cite{10867080}  &
Reconfigurable logic fabric with LUTs, DSP blocks, BRAM, and interconnects; adaptable hardware architecture. &
Implements custom data paths and operators through reconfiguration; supports pipelined and parallel execution. &
\vspace{-1em}
\begin{itemize}[leftmargin=*, labelsep=0.4em]
\item Highly flexible and reprogrammable
\item Low latency and energy-efficient for edge workloads
\item Customizable precision (e.g., INT8, binary networks)
\item Suitable for prototyping custom accelerators
\end{itemize} &
\vspace{-1em}
\begin{itemize}[leftmargin=*, labelsep=0.4em]  
\item Lower peak performance than GPUs/ASICs
\item Programming requires hardware expertise (HDL, HLS)
\item Long compilation and bitstream generation times
\item Limited on-chip memory
\end{itemize} \\
\hline
\end{tabular}
\end{table*}
The diagram shown in Fig. \ref{chart_metrics} provides a view of the various performance metrics in the acceleration of the neural network, including average throughput, speedup, latency, power efficiency, and area. FPGA based AI accelerators offer a balanced middle ground between the flexibility of GPUs and the efficiency of ASICs. Although GPUs provide high throughput and benefit from mature programming ecosystems~\cite{markidis2018nvidia,shi2018graph}, they often suffer from high power consumption and limited architectural specialization for new or rapidly evolving AI models. ASIC accelerators, although capable of delivering exceptional performance and energy efficiency through custom dataflow designs~\cite{jouppi2017datacenter, chen2016eyeriss}, are expensive to develop, unflexible after fabrication, and slow to adapt to emerging neural architectures. In contrast, FPGAs combine reconfigurable logic with energy-efficient computation, enabling designers to tailor hardware precisely to model requirements and update implementations as algorithms evolve~\cite{ zhang2015optimizing,venieris2018toolflows}. This reconfigurability, along with their support for low-latency execution and customizable precision, makes FPGAs particularly valuable for edge and domain-specific AI applications where adaptability, efficiency, and rapid prototyping are essential. \\
In this article, we present a comprehensive overview of accelerator advancements along with the optimization methodologies at both the network and hardware levels. We also discuss the challenges and future research directions in accelerator optimization. This study analyzes various approaches from multiple perspectives, as illustrated in Fig.~\ref{motiv}. 

Section~\ref{sec2} reviews the structures of different types of existing hardware accelerators and the methodologies employed within them. Section~\ref{sec3} surveys network specific FPGA accelerators including Convolutional Neural Networks (CNNs), Spiking Neural Networks (SNNs), Recurrent Neural Networks (RNNs), and Graph Neural Networks (GNNs) and discusses their design considerations, acceleration techniques, and associated challenges. Section~\ref{sec4} examines hardware level optimization strategies designed to minimize bottlenecks. A performance analysis of the state-of-the art model specific FPGA accelerators is provided by Section~\ref{sec5} which gives an estimate of achieved throughput and resource utilization. Although FPGAs demonstrate strong potential as AI accelerators, they face limitations such as restricted hardware resources. These challenges are discussed in Section~\ref{sec6}. Section~\ref{sec7} outlines potential directions for further optimization, and Section~\ref{sec8} concludes the article.

% \begin{table*}[ht]
% \centering
% \caption{Performance Comparison Metrics Across CPU, GPU, ASIC and FPGA}
% \label{comparison_clean}
% \small

% \begin{tabular}{c c c c c}
% \toprule
% \textbf{Performance Metrics} & 
% \textbf{CPU} & 
% \textbf{GPU} & 
% \textbf{ASIC} & 
% \textbf{FPGA}  
%  \\
% \midrule
% Average Throughput (GOPS) & 2529 & 8480.5 & 273.48 & 1593 \\
% Average Speedup & 1 & 2.35 & 247.33 & 1043.05 \\
% Average Latency (ms) & 55.84 & 937.97 & 49 & 71.19 \\
% Average Power (mW) & 7,69,000 & 1,49,000 & 47.54 & 10,060 \\
% Average Area (mm2) & N/A & 728.67 & 25.69 & N/A\\
% \bottomrule
% \end{tabular}
% \end{table*}

\section{Background}
\label{sec2}
Hardware accelerators have emerged as a preferred solution for speeding up deep learning workloads, offering significantly higher computational capability and efficiency than traditional CPUs. Among these, GPUs are typically the first choice due to their massive parallelism and support from mature, standardized software libraries. ASIC-based accelerators form the next category, providing custom designed architectures tailored to specific network requirements; these primarily include Neural Processing Units (NPUs) and Tensor Processing Units (TPUs). Finally, FPGA based accelerators are gaining prominence for their reconfigurable architectures, which allow flexible hardware customization. The following subsections provide a detailed architectural review of different types of hardware accelerators.

\subsection{Graphics Processing Units (GPU)}
\label{subsec2.1}
\begin{figure*}[!ht]
\centering
 \includegraphics[width=0.7\linewidth]{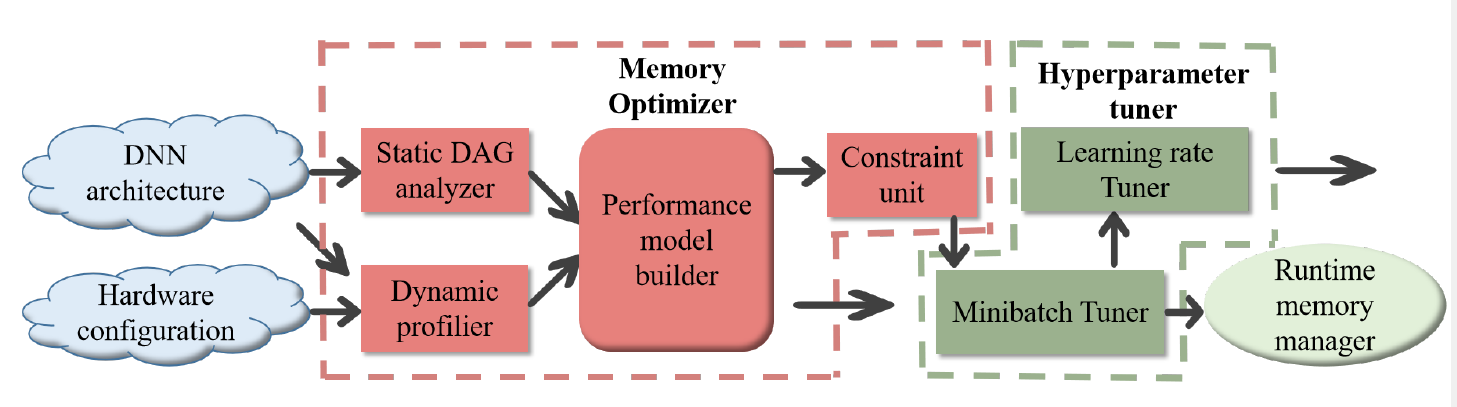}
  \caption{Architecture of AccUDNN showing the process flow between the memory optimizer and hyperparameter tuner modules. 
  \cite{8988598}}
  \Description{diagram for hardware accelerator}
  \label{AccUDNN}
\end{figure*}

GeForce 256 accredited as the "the world's first 'GPU' was described as "a single-chip processor with integrated transform, lighting, triangle setup/clipping, and rendering engines that is capable of processing a minimum of 10 million polygons per second" \cite{nvidia1999}. GPUs have evolved and now have become one of the foremost AI accelerator due to its ability to speed up the training and inference processes. DG-RePlAce, a new age GPU accelerator based on OpenROAD infrastructure (Ajayi et al. \cite{10.1145/3316781.3326334}) is proposed by Kahng et al. \cite{10620224}. The methodology starts with the conversion of a structural netlist to a clustered one with the help of open source platform Hier-RTLMP (Kahng et al. \cite{10372220}). This is followed by inclusion of datapath to find instances for each cluster and its placement according to proximity rules. A parallel analytical placement framework built on OpenROAD system is employed to reduce memory overhead and enhance computational speed.\\ 
 Xie et al. \cite{6100451} has exploited the vast network of streaming multiprocessors in a GPU to execute parallel thread computations for a SNN network. The design has a reported speedup of 31x compared to its CPU counterpart. Compute Unified Device Architecture (CUDA), an extension of C language uses serial and parallel coding to swiftly switch from CPU to GPU operations. Computation intensive processes such as training of the accelerator and matrix multiplications are carried out in the GPU. Due to parallel processing capabilities the processes are accelerated to a certain extent which also makes future prediction possible. Li. et al. \cite{7828382} proposes design of a core feedforward module. in which module takes input from the parameter training module and adds a new feature to the feedforward network accordingly. This model also reduces the processing time of each image with the help of a sliding window sample module. Xie et al. \cite{10323722} proposes a GPU accelerator for Graph Convolution Network (GCN) for optimization and provision of multi-level memory efficiency and reduction of workload imbalance. It consists of a load aware balancing scheme that evenly distributes work across warps which is further localized with the help of an execution model. Block-warp mapping combines several warps into one to handle entire workload. This ensures memory continuity and optimizes memory bandwidth. \\
 Juang et al. \cite{6887358} proposed parallelization of the learning phase of Interval Type-2 Neural Fuzzy Systems  approach to speedup the process.  With the help of the CUDA model, dataset parallelism is ensured with the distribution of datasets over the shared memory. This feature ensures high throughput while maintaining a low latency. Each thread of a single core is responsible for obtaining the inference on different data samples. This approach is used for derivation of system outputs which trumps the traditional iterative method and obtains a 30x training speed improvement. Fei et al. \cite{9826082} have proposed an alternative method to accelerate the training phase in the form of the processing-in-memory-GPU architecture. The network consists of a multiple 3D stack structure made of streaming multiprocessors. Each stack stores a set of weights for each layer which then operates concurrently thus providing model parallelism. For parallelization of data transmission, gradient transmission is used to reduce the communication overhead. A tradeoff between these two optimizations is made according to the model requirements. Guo et al. \cite{8988598} proposes "AccUDNN" (given by Fig. \ref{AccUDNN}), a GPU accelerator to reduce the  memory resource scarcity issue. The design employs a memory optimizer module which uses a dynamic technique to find the optimal tradeoff between trainability and enhancement of training efficiency. This is decided by the optimal swapping of the feature maps onto the host memory which ranges from the extreme of transferring all data to the host to zero. Once trainabilty with the limited memory resource is attained the hyperparameter tuner proceeds onto ensure optimum training efficiency. Though the model makes the training process highly optimized in terms of memory usage but it heavily relies on the Peripheral Component Interconnect Express (PCIe) Bandwidth which can become a bottleneck due to its transfer speed.

% \subsubsection{Disadvantages and Challenges}
% The huge processing capabilities come at a cost of high power consumption which leads to requirement of additional cooling systems. Though highly parallel, but these are incapable of handling complex sequential and control functionalities. This leads to significant hampering of the communication bus (PCIe) between the CPU and the acceleration unit. This poses as a significant bottleneck and increases the computation time significantly for small datasets. It also suffers from low latency inspite having an impressive throughput. 

\subsection{ASIC based Accelerator}
\label{subsec2.2}

\begin{figure*}[!ht]
\centering
 \includegraphics[width=0.7\linewidth]{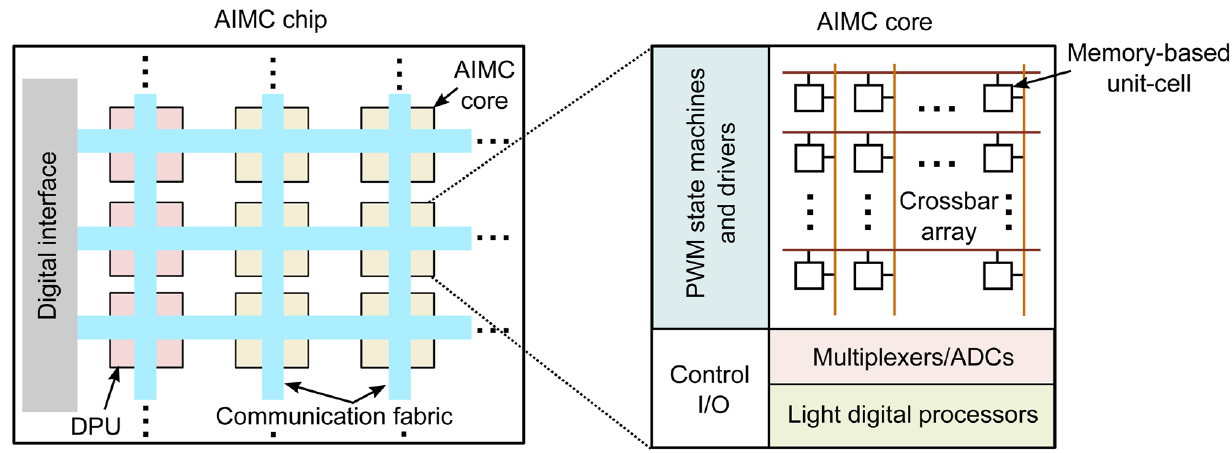}
  \caption{Architecture of an analog in-memory computing AIMC core with integrated crossbar arrays \& memory-based unit cells. \cite{10.1063/5.0168089}}
  \Description{AIMC Core Architectural Diagram}
  \label{asic}
\end{figure*}

Chen et al. \cite{10.1145/2541940.2541967} proposed an 65 nm technology ASIC-based accelerator for very large-scale convolutional and deep learning models. Employment of staggering pipeline methodology streamlines the convolutional, maxpooling and the sigmoid layers. The authors exploited the locality properties of the layers to design dedicated buffer storage for handling of the input, intermediate and output results. To further optimize memory access, the accelerator (DianNao) employs local transposition of loops and circular rotation of buffers for temporal reuse. After exploration of various fixed point representations, the authors observed that a fixed 16 bit representation is optimal and provides comparable results with an affordable tradeoff between chip area and power. On analyzing the performance of the accelerator, it is seen that it achieves a speedup of (117.87x) while consuming a minimal power of 485 mW. However at the time they had not considered the energy and power overhead and was also faced with a limitation of memory bandwidth usage. \\
DaDianNao, proposed by Chen along with a collaborator, Luo \cite{7011421} a 28 nm technology multi-chip system was an improvised version of the previously stated accelerator. The architecture employs a tiled based design to counter the internal bandwidth issues where the large number of neurons are distributed over multiple Neural Functional Units (NFUs). It also uses biased nodal footprint which not only prioritizes storage than computations but also are placed closely to reduce data access. Though being able to achieve a massive speedup of (450.65x) over a GPU, it is limited by the demerits of interconnects which limits scalability of the deep learning network.\\
Over the next years several improvisations has been implemented over the originally proposed accelerator generating a family of ASIC based accelerators. PuDianNao and Shidiannao proposed by Liu et al. \cite{10.1145/2694344.2694358} and Du et al. \cite{7284058} are a part of this family. The former uses single port Static RAM (SRAM) based buffers connect to the same Direct Memory Access (DMA) which significantly reduces footprint and power consumption. This accelerator is designed in such a way so that it can employ several ML techniques with the help of dedicated novel units. Though lesser than \cite{7011421} this achieves a speedup of (1.20x) over a GPU. Unlike its predecessors, the latter stores all the weights in SRAM itself thus eliminating the need of Dynamic RAM (DRAM) accesses and unnecessary data movements within the layers itself. This makes it (4688.13x) times more energy efficient than a GPU but is not able to achieve a massive speedup.
\\These designs aim to optimize memory accesses but are not adaptable to different types of network. Parallelism is introduced to efficiently use computational resources and make scalable architecture. \cite{7920855} proposed FlexFlow estimates degree of parallelism for each convolution layer and accordingly selects the data flow mechanism to solve the mismatch issues associated with the workload. This makes the design adaptable to a wider range of neural models. UniPRE, a hybrid accelerator given by Zhang et al. \cite{11048618} is one such example which eliminates redundant computations resulting from maxpooling operations. It achieves this by using a channel order queue to perform computations only at maximum channels as predicted from earlier ones. It also reduces energy layer wise achieving an efficiency of 19.32 TSOPS/W. \\
An attempt to design a low power accelerator was taken by Chen et al. \cite{11043330}. The design includes a time reconfigurable neuron architecture which parallelizes multi-time operations thus reducing power significantly. Gallo et al. \cite{10.1063/5.0168089} cites the usage of a specialized ASIC architecture- International Business Machines (IBM) analog accelerator. This is designed for Analog In-Memory Computing (AIMC). In this type of computing, the AIMC core (given by Fig. \ref{asic}) consisting of resistive crossbar arrays directly perform matrix–vector multiplications. The problems of conductance drift and device variability are mitigated by the noise model. The cores are combined with Digital Signal Processor (DSP) units to perform auxiliary Deep Neural (DN) computations which accelerate the computation process. The cores have weight update schemes like in-memory Stochastic Gradient Descent (SGD) for overlapping pulses over rows and columns of the matrix, Tiki-Taka for smoothing of noisy gradients, to increase robustness that often suffers due to analog updates. To streamline the memory computations, tiling and partitioning strategies are introduced. A hybrid digital-analog mode is provided for performing time sensitive operations digitally with the help of affine transformations. To further improve the performance, algorithm-hardware co-optimization is carried out to match the low quantized pre-trained models with the hardware resources. In this way analog efficiency meets with digital precision. 

\subsection{Neural Processing Units (NPUs)}
\label{subsec2.2.1}

A NPU (also known as AI chip or AI accelerator) is a special type of deep learning accelerator designed specially to speedup the neural network computations. The architecture is modeled in such a way to emulate the processing of a human brain \cite{ibm2024npu}. Though computationally not very precise, these have specialized models which can accelerate operations like dot product and matrix multiplication. Unlike GPUs the instruction set is concise and is able to accomplish several tasks just with few instructions at same computing power.\\ 
An in-memory NPU given by Jeon et al. \cite{9469808} addresses the problem of heavy memory computation. In order to speed up the computation capability of a neural model, the design includes a Single Instruction Multiple Data (SIMD) parallel vector processing unit and systolic array structure. The systolic array is composed of a Multiplier Accumulator Unit (MAC) which performs concurrent multiplications thus accelerating the convolution operation. Further optimization is achieved by using library functions which makes it adaptable to any neural network. According to the requirements of the neural network, the BLAS library optimizes operations like dot vector, matrix matrix multiplication, matrix vector multiplication element wise addition and so on. The Network-on-chip (NoC) architecture is introduced by Ouyang et al. \cite{OUYANG2025106684} which comprises of a set of NPUs which handle the computation part whereas the communication functions are carried out by the router. The router is responsible for multicast communication where it sends the relevant weights as packets to the NPUs. The NPUs then handle the processing of the operations. \\
Wang et al. \cite{WANG2025109887} proposes a Kernel Convolution Network for efficient handling of tasks like segmentation and detection. Some channels of the network are allocated for storing the feature maps while the remaining capture fine spatial detail. This makes the design adaptable to large size kernels. Cheng et al. \cite{7738524} presents a more powerful deep CNN accelerator, Eyeriss which is capable of processing large vision based tasks. The spatial architecture contains a 12 × 14 array of 168 Processing Elements (PEs) which is coupled with a unique Row Stationary (RS) dataflow. This reconfigures the PE array to map the computation of various CNN shapes to maximize data reuse locally within a four-level memory hierarchy. The hierarchy is divided into spads, inter-PE communication, Global Buffer, and DRAM. To further enhance performance, it also integrates a NoC for single-cycle data delivery which exploits zero-value statistics and use data gating to save up to 75\% of DRAM bandwidth.\\
The NPUs offer a very low latency and also quite efficient in acceleration of standard deep and convolutional neural network. Being application specific, it has a very low flexibility and suffers from scalability issues. These are unfit for applications where precision is of utmost importance as accuracy may worsen if not properly quantized.

\subsection{Tensor Processing Units (TPUs)}
\label{subsec2.2.2}

\begin{figure*}[!ht]
\centering
 \includegraphics[width=0.5\linewidth]{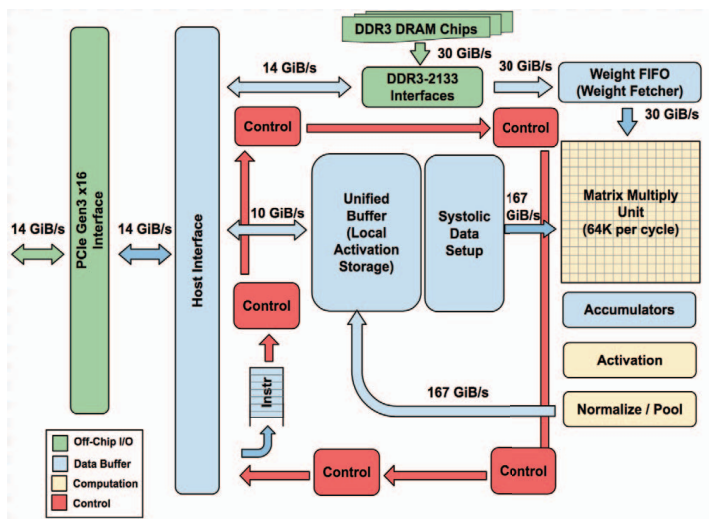}
  \caption{Architecture of TPU v1 showing dataflow between the systolic array, unified buffer \& high-bandwidth memory interface.\cite{8358031}}
  \Description{TPU Architectural Diagram}
  \label{tpu_1}
\end{figure*}

Tensor Processing Unit a custom ASIC accelerator has been developed by Google which are used in datacenter to speedup the inference phase of the models. Jouppi et al. \cite{8192463} explained the architecture (given by Figure \ref{tpu_1}) of the TPU in details. It processes every operation in a predictable manner and omits general purpose tasks making its architecture simpler. This was only used in datacentres and was not made accessible for public use. Joppi et al. \cite{8358031} laid out the implementation details of the first generation TPU. It consists of a single core having a systolic MAC unit which not only optimizes data flow by reusing operands. The architecture is equipped with a small set of operations which retains its simple nature. Like its predecessor, this design also offers high throughput and bandwidth memory for large deep neural networks. Mummoju et al. \cite{10020427} provided a thorough history of TPU. Both TPU versions 2 and 3 have four chips equipped with two cores consisting of a scalar unit, a vector processing unit and a couple matrix multiply units. In addition to this, they proposed two methodologies of diving the datasets required for a k-means algorithm. For both the methods, the batch datasets and the CPU computed centroids are passed onto the TPU for assignment and further computations. Notably, the runtime is significantly lower on TPU as compared to a CPU.

Yuan et al. \cite{8682197}  proposes a dimensionality reduction technique in the form of a tensor-ring decomposition method. It combines two decomposition methods- Tensor Ring Alternative Least Squares (TRALS) and Tensor Ring Singular Value Decomposition (TRSVD) to make the process of tensor decomposition fast and reliable. Soto-Chirinos et al. \cite{10326073} explore three TPU based architectures to solve different types of scalability problems. The first model - TPU Virtual Machine (VM) is a non-scalable architecture that while having high performance is suited only for single-node experiments. Despite being a slower model, the second one is a scalable version suited for moderate scalability. Also, the production time was a bit increased which is corrected by the last model. The last architecture scales horizontally across various nodes while being resource efficient. \\
Other optimizations include time-triggered memory access approach given by Ezekiel \cite{10456805} that fetches data before request for loading data is made. This also consists of an offline phase where the time-triggered schedule is generated to address memory time accesses. For design of modern accelerators, authors often sought the help of automatic Register-Transfer Level (RTL) generation. Using affine-based transformation, Lin et al. \cite{10946782}, LEGO is proposed which synthesizes adjustable memory sizes and mixes different type of dataflow methodologies. The back end of the methodology transforms hardware into a lower level graph for remaining optimizations.\\ 
Latest advancements include replacement of traditional electronic components with light based processing units (Tang et al. \cite{10255124}). It consists of a single modulator having an optical core which performs all parallel tensor convolutions thus making it suitable for large data handling applications. The convolution operations are simultaneously encoded into different channels which are further multiplied with the weights. This is done with the help of light interference, which then passes onto the circuits for application of activation. As a result of this, the architecture consists of a single computing hardware which minimizes data replication and clock synchronization issues. These photonic TPUs are low power systems with high throughput and maximum efficiency (Schwartz et al. \cite{10543636}).\\
A high- speed TPU accelerator given by Fardoost et al. \cite{9975540} combines wavelength-division multiplexing with mode-division multiplexing for multidimensional encoding of matrices of different sizes. This makes the scalability multiplicative and also solves the problem of orthogonal dimensions. An application of this methodology has been given by Tang et al. \cite{10369087} that takes a 64 channel CT scan as input. The reported average extraction time taken by the convolutions and average pooling are ($1.17 x 10^{-5}$) and ($3.8 x 10^{-6}$) respectively. To make the architecture more energy efficient, Li et al. \cite{10809912} cited a convolution compression method employing CP decomposition (Lebedev et al. \cite{lebedev2015speedingupconvolutionalneuralnetworks}).The methodology employs kernel decomposition and fine-tuning to decompose layer-by-layer convolutions. Adam optimizer is used for fine-tuning to adjust the reductions in accuracy. However, these systems suffer from an issue of a memory bottleneck- Von Neumann bottleneck  (Schwartz et al. \cite{10614519}). This rises when the capacity of memory nodes are insufficient for the high computation power. It becomes a compounding problem for DRAM main memory when there are frequent updates from the output. This is due to transmission delay and issues in memory access. The authors also state probable solutions such as cache memory, in-memory computation and advanced packaging of the memory and the processor.\\
% \subsubsection{Disadvantages}
%  However, these accelerators also face certain challenges. ASICs have long design cycles with limited flexibility. Their functionality is hardwired and cannot be reprogrammed once designed. So if there arises a case where the network details needs to be changed or adapted to a larger network, these designs can no longer be used. Furthermore, any design errors require costly re-spins which come with hardware risks. While being highly efficient these are limited to domains suited for traditional prototyping which requires low adaptability and lesser updates.

 \subsection{FPGA based Accelerator}
\label{subsec2.3}
\begin{figure*}[!ht]
\centering
 \includegraphics[width=0.8\linewidth]{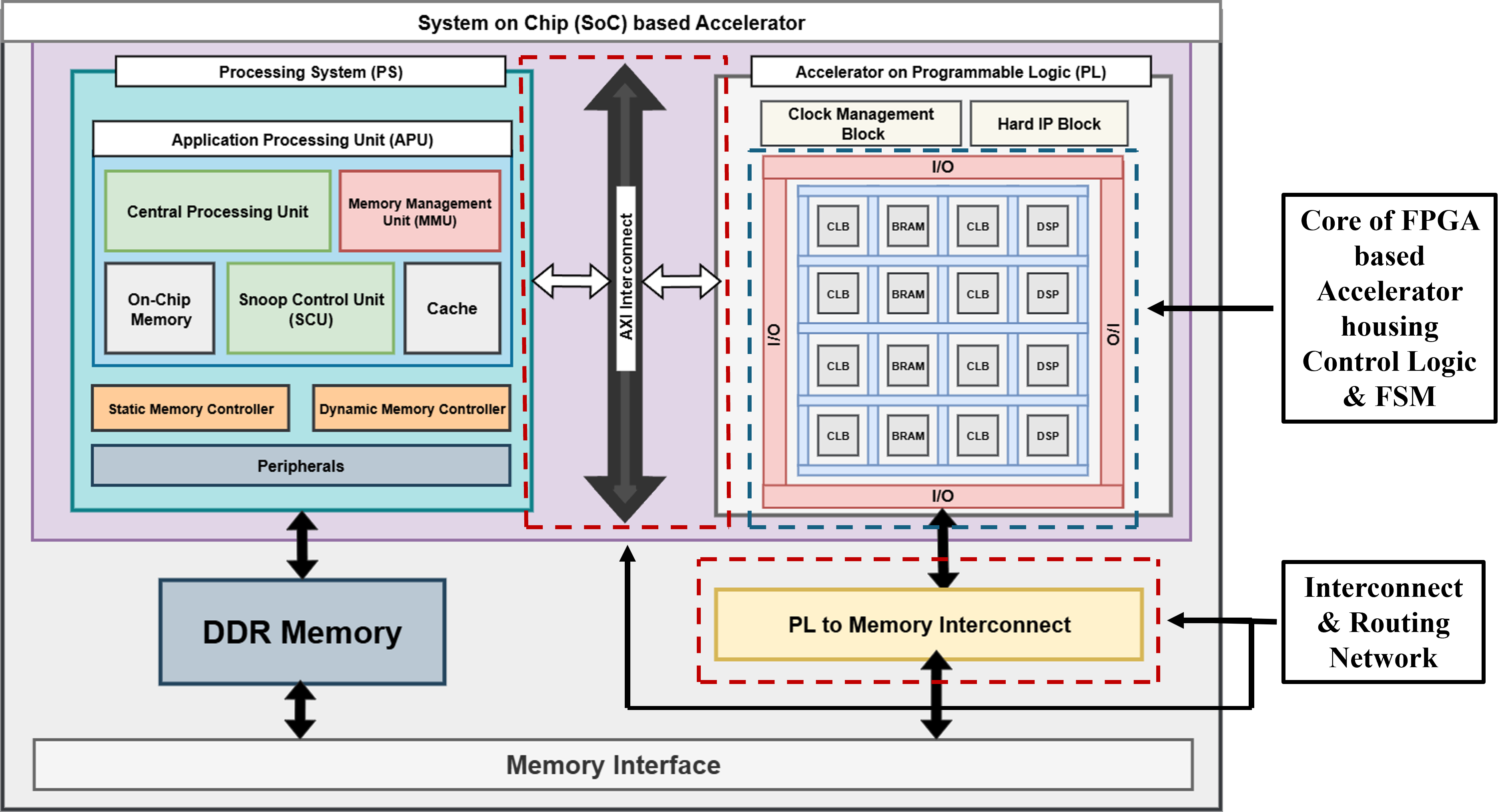}
  \caption{Complete Architecture of SoC based Accelerator depicting the sub-modules- PS, PL \& and shared DDR memory.}
  \Description{DrawIO Diagram for FPGA Architecture}
  \label{bd}
\end{figure*}

GPUs as hardware accelerators achieve high throughput and ASICs provide high energy efficiency with predictable latency for fixed workloads. As discussed in Table \ref{comparison_accelerators}, these have certain demerits and from the statistics given in the Fig. \ref{chart_metrics}, FPGA-based accelerators occupy a unique position by facilitating reconfigurable yet application-specific hardware execution. The next sections discuss the architecture of a FPGA with the structural details and optimizations made at different levels.
\paragraph{Architecture of FPGA and Basic Structure of FPGA as Hardware Accelerators}-
An FPGA is a semiconductor device which lets the user configure according to their application needs at a low cost. Earlier versions of FPGAs lacked heterogeneous structure in which it relies on an external CPU and off-chip DRAM. These FPGA units were restricted in both logic density as well as clock speed. With the advancement in technology, FPGAs have evolved into a System on chip (SoC) based structure in which the memory and the processing unit are integrated. In the current structure, SoC is primarily divided into two parts- Processing System (PS) which incorporates the CPU on chip and the second is the Programmable Logic (PL). The PL part has the reconfigurability architecture which hosts the core of the accelerator. Fig. \ref{bd} illustrates the design of the SoC-based accelerator, highlighting all of its components. \\
The accelerator relies on the parallel processing capabilities and reconfigurability features. The SoC structure integrates the software programmability with hardware acceleration features.  Inside the PS, there is an Application Processing Unit (APU) consisting of CPU cores, on chip memory, cache memory and memory management module. The APU houses CPU consisting typically of ARM cores or hard processors in Intel SoCs for handling of sequential operations and control of accelerators. Snoop Control Unit (SCU) maintains cache coherency across multiple CPU cores. The memory store the frequently stored data thus reducing the number of DRAM accesses. Two memory controllers are provided which manage the communication between PS and Double-Data Rate (DDR) memory. This is also equipped with various peripherals like Universal Asynchronous Receiver/Transmitter (UART), General Purpose Input/Output (GPIO) and so on. The PL forms the base level architecture of the accelerator. Configurable Logic Blocks (CLBs) provide for the implementation of the parallelization of the computation heavy tasks. DSP slices provide for the fixed arithmetic operations of the network. Block Random Access Memory (BRAMs) provide for the buffering of the memory data and often used for storage of intermediate data. Both the units are connected with the help of a  Advanced eXtensible Interface (AXI) interconnect which provides high bandwidth communication and also provides exchanges between the DDR memory. The basic architecture of the accelerator remains the same. Although different models require different routing networks, memory access, and functionality, the accelerator is designed accordingly.

Over the years, researchers have exploited the various features of FPGA to propose a wide range of FPGA based hardware accelerators. Growing complexity of the networks calls for workload management, network specific and hardware level optimizations. The need of real-time data processing has now become inevitable. For this, efficient usage of hardware resources with an acceptable degree of accuracy is required.  Section \ref{sec3} elaborates on the variety of deep learning accelerator models with their structural details and acceleration techniques. Apart from this, optimizations are made at numerous levels for acceleration of the tasks. A review of some of the general hardware level optimization techniques are discussed in Section \ref{sec4}.

\section{Model-Specific Design Approaches for FPGA based AI Accelerators}
\label{sec3}
\begin{figure*}[!ht]
\centering
\includegraphics[width=0.88\textwidth]{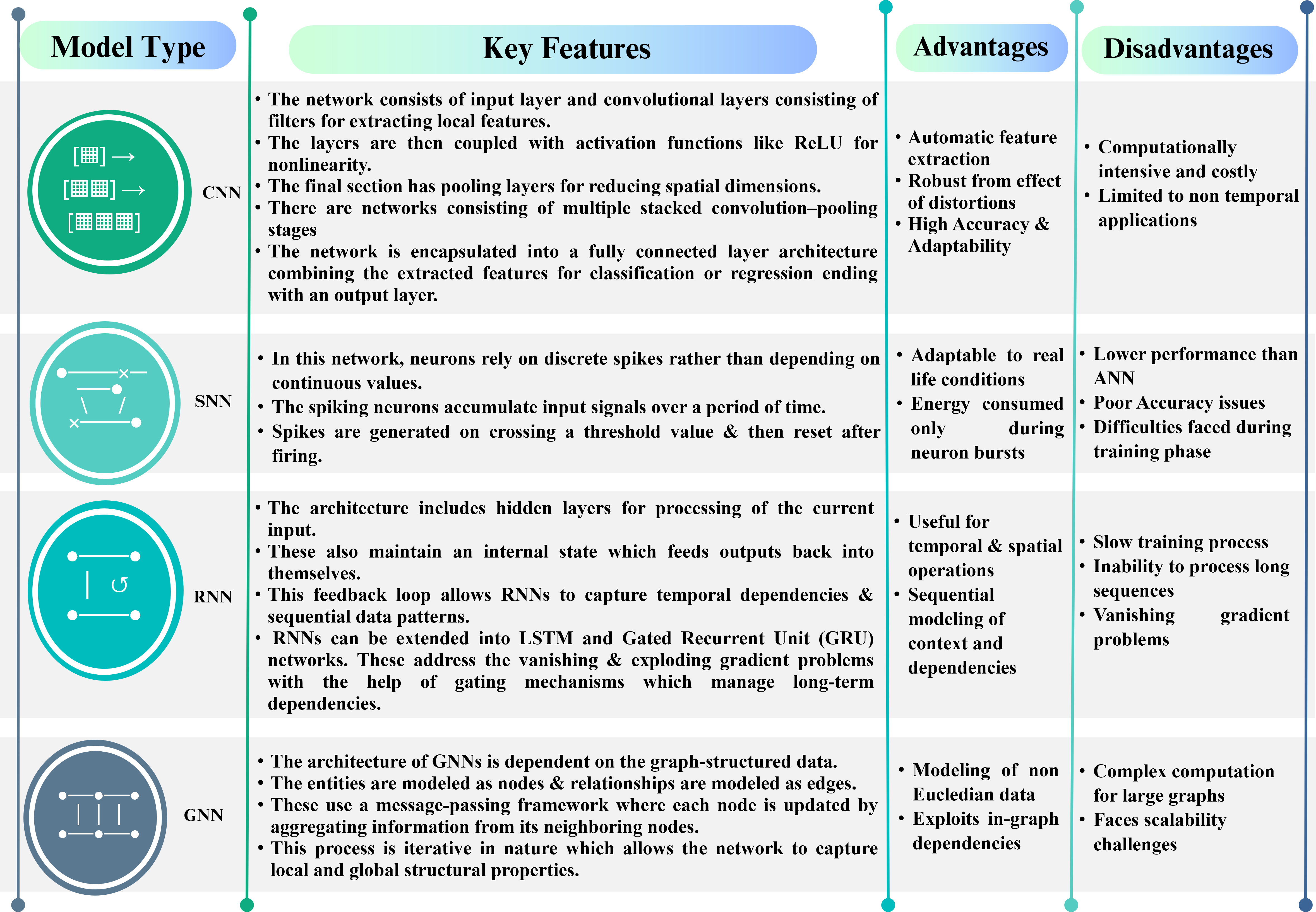}

\caption{Overview of CNN, SNN, RNN, and GNN with  principles, key advantages \& disadvantages.}
\Description{Architectural Details of Various Neural Networks}
\label{arc}
\end{figure*}

AI accelerators often adopt model-specific design strategies to efficiently map neural network computations. These approaches change according to the architecture, dataflow, and memory hierarchy to the characteristics of specific models—such as CNNs, SNNs, RNNs and GNNs. Fig. \ref{arc} provides the architectural details of each model along with its key advantages and disadvantages.

\subsection{Convolutional Neural Network (CNN)}
\label{subsec3.1}
\begin{figure}[tb]
    \centering
    % First image
    \begin{subfigure}[b]{0.48\textwidth}
        \centering
        \includegraphics[width=\textwidth, height =0.26\textheight]{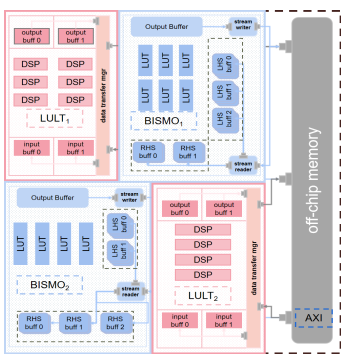}
        \caption{Architecture of FPGA Hybrid Accelerator with combined LUT \& DSP units for streamlined computation.  \cite{10927809}}
        \Description{Acc_1_diagram}
        \label{10_a}
    \end{subfigure}
    \hfill
    % Second image
    \begin{subfigure}[b]{0.46\textwidth}
        \centering
        \includegraphics[width=\textwidth, height =0.26\textheight]{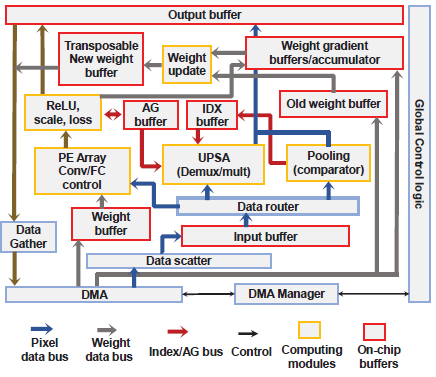}
        \caption{Framework showing unified dataflow control with buffers, \& compute modules of Automatic Compiled Accelerator. \cite{8892195} }
        \Description{Acc_2_diagram}
        \label{10_b}
    \end{subfigure}
    \caption{Block Diagram for CNN Architectures}
    \label{fig:two_images}
\end{figure}

One of the early designs of a FPGA based accelerator, Convolutional Network Processor (CNP) was given by Farabet et al. \cite{5272559}. The main computation engine utilizes the parallel nature of the convolution layers to provide efficient hardware usage. A dedicated hardware unit is included to perform 8 concurrent data accesses by interconnecting with the off-chip memory. The dataflow is maintained with the help of first in first out buffers between the computation unit and the off-chip memory. The computation unit also performs MAC operations concurrently synchronized with each clock cycle. Streaming vector PEs (Farabet et al. \cite{5537908}), 2D clusters (Sankaradas et al. \cite{5200010}, Graf et al. \cite{NIPS2008_31fefc0e}), and other modules are examples of different architectures.\\
Xie et al. \cite{8742284} uses a roofline model to fix the bandwidth and performance limits. To enforce the defined bandwidth limit, the architecture leverages on-chip memory for storage of feature maps and weights. The memory is configurable such that a partitioned Unified Random Access Memory (URAM) stores the feature maps and simultaneously rotates once computed whereas the BRAM stores the respective biases. Like its predecessors, this architecture also uses a 2D systolic array for performing parallel computations. Each systolic array has a PE which computes with the help of pipelined registers. These registers has flexible modes for computing convolutions, vector multiplication and distance calculation which further maximize the DSP efficiency. With a 90\% utilization of hardware resources, the accelerators achieves a performance peak of 9.39 Trillions of Operations Per Second (TOPS). An older version of the roofline model was used by Zhang et al.  \cite{10.1145/2684746.2689060}.\\
For upholding the performance requirements, the accelerator used loop pipelining and loop tiling for maximizing parallel MAC operations. However, as an improvement the authors proposed the usage of a cross-layer unrolling factor so that it can fit a variety of CNN architectures. Zhang et al. \cite{10927809} deliberately introduces memory stalls to address bandwidth limits. Fig. \ref{10_a} shows the architecture which combines bit-serial LUTs with loop unrolling and tiling DSPs to process operands bit by bit with massive parallel processing capabilities.  Also, a hybrid Design Space Exploration (DSE) is proposed for providing layer-wise quantization and scheduling to increase the above stated concurrency. \\
In recent years, some authors have cited optimizations at a RTL level addressing various clock issues directly related to the accelerator performance. Apart from the standard optimization techniques, Kim et al. \cite{10148988} applies low-power techniques on the generated baseline RTL code. Clock-Gating (CG) is used for elimination of redundant clock cycles. To ensure this a multiplexer is designed with a Local Explicit Clock Enable (LECE) which is further connected with a series of flip-flops to only update the required outputs. The enable signal is further controlled by a Enhanced Clock Gating (ECG) based on the input/output signals. Venkataramanaiah et al. \cite{8892195} proposes a custom RTL module shown in Fig. \ref{10_b} which contains Verilog modules for CNN training phase. \\
Based on the library an automatic RTL compiler generates general FPGA-based accelerator designs. Apart from having a 2D systolic array unit, a MAC load balancing unit is present for different kernel sizes. Unlike traditional buffers, this accelerator consists of transposable weight buffers made up of a circular matrix format which allows for both normal and transpose data accesses. These optimizations make a high throughput and highly flexible hardware accelerator. Using the same methodology \cite{8558097} cites for the inference phase of the CNN accelerators with an emphasis on the scalability. The second difference is that it uses a parametric systolic array fitted with a layer-wise compilation method to adapt for different types of computations involved in the inference phase. The compiler also plays a role in deciding the quantization factor and resource allocation strategy for different CNN methodologies. Along with these optimizations, the accelerator uses the traditional memory level computational methods to achieve a peak throughput of 1046 GOPS. Varadarajulu et al. \cite{9465504} presents an improvised RTL library- SentiNet RTL library consisting of RTL modules for convolution, pooling, activation and fully connected layer computations. \\
All the optimization methods not only increase throughput but also make the power efficient to some extent. But the power efficiency comes as an added advantage. G et al. \cite{10927809} proposes a series of methodologies that can offer a trade-off between power consumption and performance. One such is dynamic precision handling which calculates the bit width according to the needs of the operations. A conventional systolic array can be coupled with multi-mode operation to support various operations without needing specialized hardware units. As discussed clock and power not only reduces power switching but also reduces the usage of inactive components to eliminate leakage power. These methods result in near about 11\% in power reduction. So, apart from pruning and quantization, Maksoud et al. \cite{9610053} proposes an extreme model for compression and pruning by quantizing the architecture to 4-bit power-of-two values. This makes the model to fit entirely in the BRAM itself thus eliminating the need of power-hungry DRAMs. The model is designed in such a way that it does not use any DSP. The main operation of the DSPs-MAC is replaced by bit shift operations banking on the fact that the weights are a power of two. This reduces the dynamic power consumed by the accelerator. The custom memory architecture comprising of 256 BRAM banks and multi-level buffers provides data reuse and minimal data transfers which further increases the power efficiency. 

% \begin{figure}[h]
%     \centering
%     % First image
%     \begin{subfigure}[b]{0.48\textwidth}
%         \centering
%         \includegraphics[width=1.10\textwidth, height =0.34\textheight]{Co.png}
%         \caption{Architecture of FPGA Hybrid Accelerator \cite{10927809}}
%         \label{CNN_1}
%     \end{subfigure}
%     \hfill
%     % Second image
%     \begin{subfigure}[b]{0.46\textwidth}
%         \centering
%         \includegraphics[width=\textwidth, height =0.34\textheight]{Automatic.png}
%         \caption{Architecture of Automatic Compiled Accelerator \cite{8892195} }
%         \label{CNN_2}
%     \end{subfigure}
%     \caption{Block Diagram for CNN Architectures}
%     \label{fig:two_images}
% \end{figure}

\subsection{Spiking Neural Network (SNN)}
\label{subsec3.2}
\begin{figure}[ht]
    \centering
    % First image
    \begin{subfigure}[b]{0.50\textwidth}
        \centering
        \includegraphics[width=\textwidth, height =0.60\textwidth]{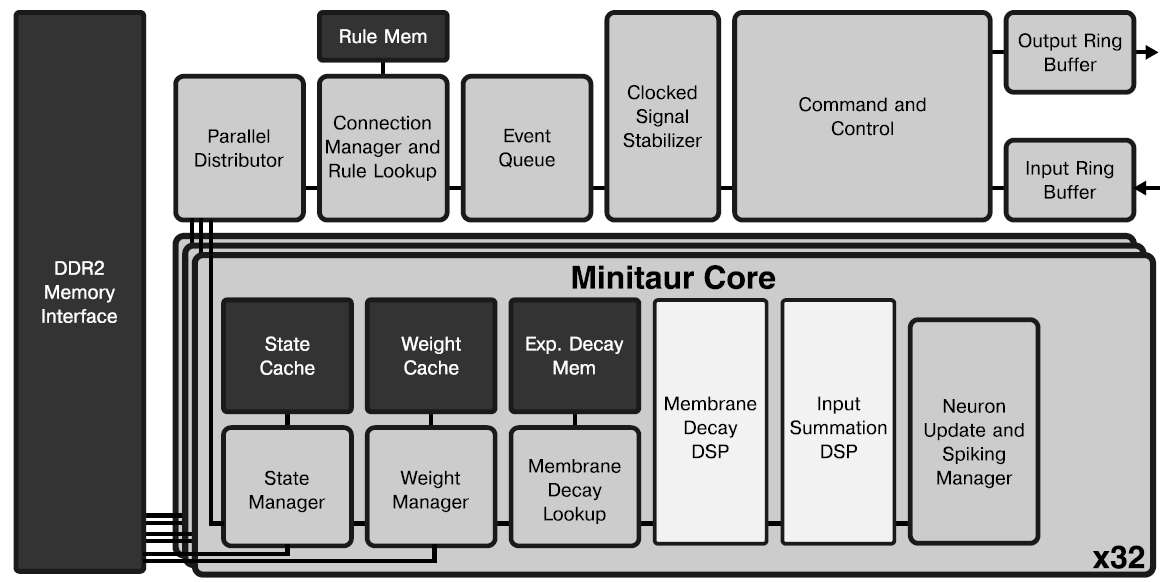}
        \caption{Architecture of a scalable processor Minitaur with dedicated units for synaptic computation.  \cite{6701396}}
        \Description{SNN_1_diagram}
        \label{11_a}
    \end{subfigure}
    \hfill
    % Second image
    \begin{subfigure}[b]{0.48\textwidth}
        \centering
        \includegraphics[width=\textwidth, height =0.24\textheight]{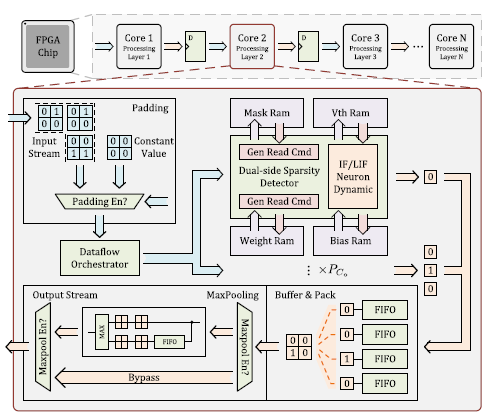}
        \caption{Framework of FireFlyS with optimized modules for convolution, pooling, \& data management.\cite{10754657} }
        \label{11_b}
    \end{subfigure}
    \caption{Block Diagram for SNN Accelerator Architectures}
    \Description{SNN_2_diagram}
    \label{fig:SNN_acc}
\end{figure}

\cite{10143752} proposes FireFly-an SNN accelerator which enhances memory efficiency by utilizing optimized on-chip RAM. The 48-bit ALU is activated in SIMD mode which performs four parallel 12-bit additions. High computational density is achieved by eight cascaded DSPs with each handling a 2×4 synaptic crossbar computation. A specialized line-buffer is used for performing 3x3 Statistical Convolutional Neural Network (SCNN) convolutions which incorporates First In, First Out (FIFO) partial reuse and multi-stage stream buffers to maximize weight reuse and thus reduce latency. Memory efficiency is enhanced with the usage of a Finite State Machine (FSM) controlled partial-sum accumulation and voltage storage membrane. However, membrane voltage value are stored for long periods resulting in burdening of the memory. \\
Yin et al. \cite{YIN2024106377} proposes an accelerator designed for implementation of both inference and training phases of backpropagation algorithm. The input sparsity is exploited through spikes thus avoiding weight passing and accumulation passes in absence of spikes. Backpropagation to inactive outputs during the training phase is eliminated by using cases where spikes are generated by threshold crossing neurons. These techniques allow only most useful computations to take place per systolic array clock cycle and also reduce unnecessary memory accesses leading to increase in throughput. Li et al. \cite{10754657} employs a Bitmap-based sparse decoding logic (shown in Fig.\ref{11_b}) to locate non zero weights and input spikes with assignment of masks. This introduces a performance-limiting bottleneck. Neil et al. \cite{6701396} focuses on the event driven nature of irregular sparse spikes. They propose a deep pipelined structure for simultaneous processing of the spikes as given by Fig. \ref{11_a}. This is done with the help of a series of parallel PEs. The system also uses a multi-bank memory system comprised of partitioned synapse storage to reduce memory stalls during the process of simultaneous fetching of same weights. 

Zhang et al. \cite{ZHANG2025106616} aims to reduce data movement with the help of temporal parallelism. For n timestep operation spikes from all the steps are distributed over n PEs which reuse the same weights and perform parallel computations simultaneously. Parallelism is also provided within each PE by performing parallel channel computations. They also use a streaming dataflow mechanism to incorporate pipelining in passing the input feature maps and the weight data from buffers through linebuffers to spike generation pooling and residual unit. Each stage has concurrent operations in every clock cycle which prevents stalling and contributes to a increase in throughput values. This methodology still relies on dependent Membrane Potentials (MPs) which require extra logic to distribute intermediate MPs. Karakchi \cite{10586065} introduces a novel approach to make the storage type user configurable. Each neuron of the network is encoded into one of the PEs of the 2D array network. The current membrane level of the neuron is encoded as a storage element (i.e., memory). Parameterized Verilog coding configures the memory type from the available URAM, BRAM and registers. The authors observed that the URAM based accelerator (SINK) is low-power as well as highly suitable for scaled up-operations. This observation is limited only to the ZCU104 UltraScale device. Although this methodology gives the user a choice but is limited only to static memory fixing. proposes a model in which the compilation process allocates the task of the processing models according to the required configurations. Like its predecessor models, parallelism is ensured within the module which performs parallel computations for every feature map. In addition to this, intra module parallelism is also provided by duplicating the kernels and hardware reuse for every layer. To increase computational efficiency and reduce memory burden, traditional loop hierarchy is modified by making the input channels as inner loops so that it results in partial sums which are not only very precise but also storage friendly. 
\subsection{Recurrent Neural
Network (RNN)}
\label{subsec3.3}
\begin{figure}[ht]
    \centering
    % First image
  \begin{subfigure}[b]{0.49\textwidth}
        \centering
        \includegraphics[width=1.02\textwidth, height=0.45\textwidth]{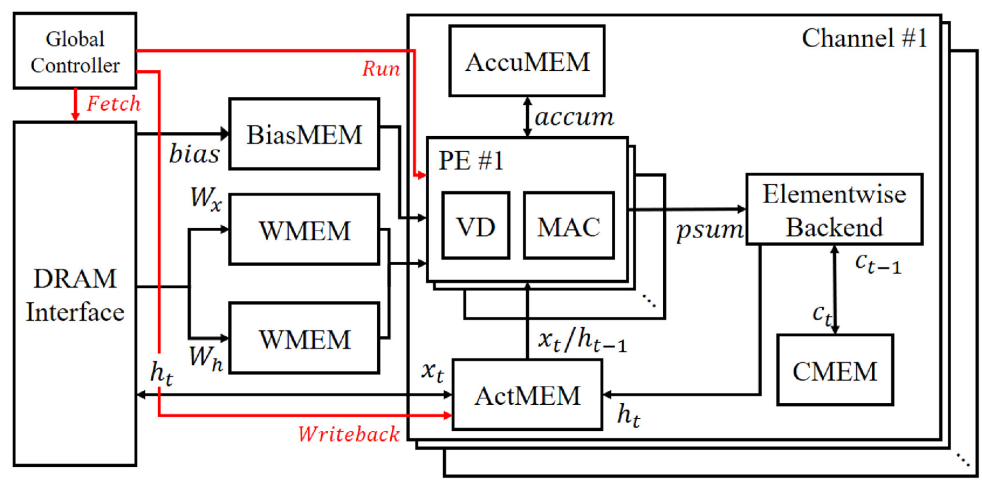}
        \caption{Architecture of V-LSTM Accelerator demonstrating the memory hierarchy which enables efficient LSTM computation.
        \cite{10041123}  }
         \Description{RNN_1_diagram}
        \label{12_a}
    \end{subfigure}
    \hfill
    % Second image
    \begin{subfigure}[b]{0.50\textwidth}
        \centering
        \includegraphics[width=\textwidth, height =0.45\textwidth]{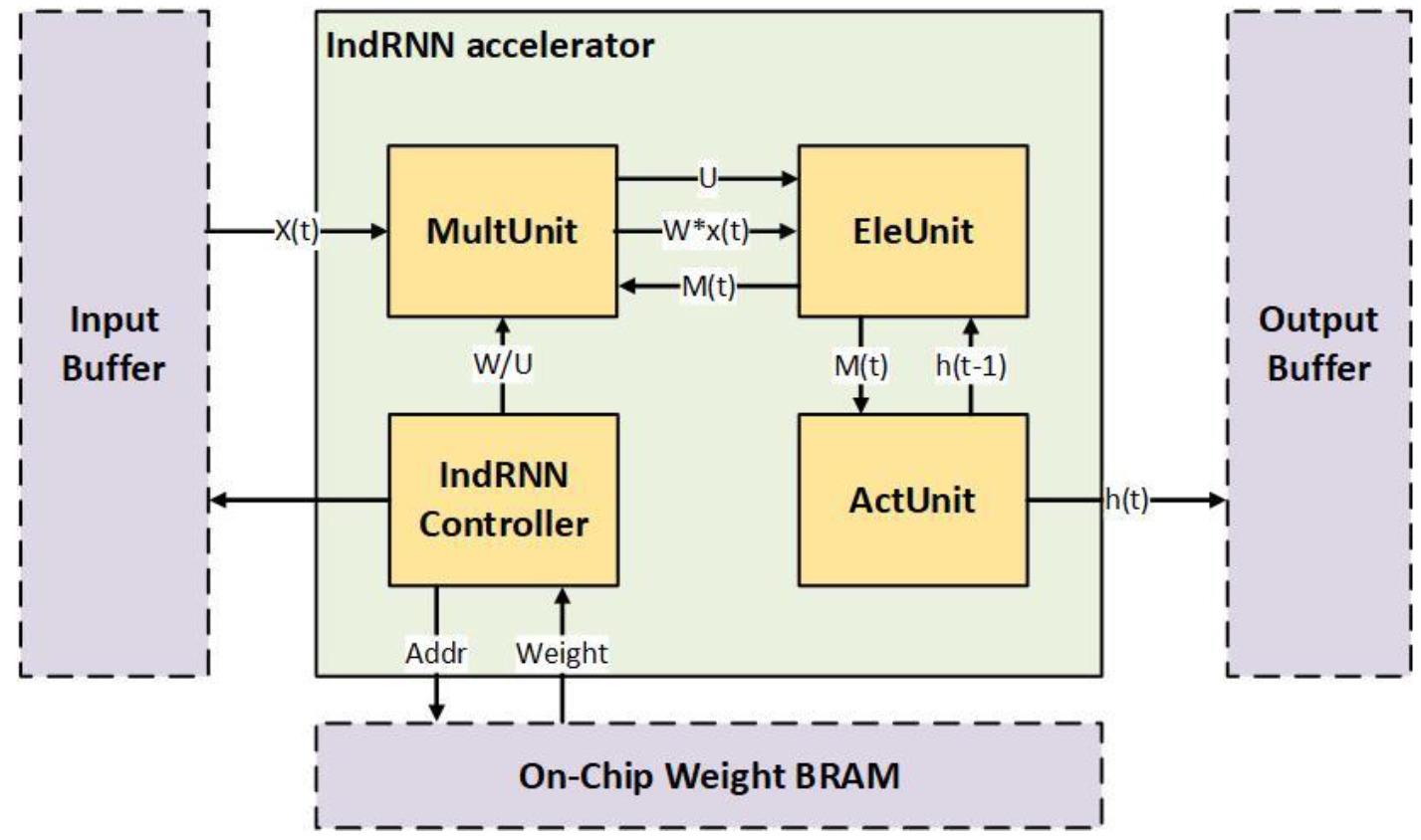}
        \caption{Depiction of architecture of IndRNN accelerator with parallel computational units and on-chip BRAM. \cite{8780644} }
         \Description{RNN_2_diagram}
        \label{12_b}
    \end{subfigure}
    \caption{Block Diagram for RNN Accelerator Architectures}
    \label{fig:RNN_acc}
\end{figure}

Architecture of three hardware accelerators- DeepStream, DeepStore and DeepRNN for RNN applications are presented by Chang et al. \cite{8050816}. DeepStream continuously streams weights and vectors from off-chip memory to the MAC units thus providing maximum utilization. But this memory suffers from high memory bandwidth requirement which limits its scalability. This problem is alleviated to some extent by DeepStore which stores all the weights in on-chip memory. It then preloads them onto the DMA and replicates 128 PEs for achieving a massively high-level of parallel computation. Although, it uses lower memory bandwidth but still suffers from scalability issues. This is due to the dependence of the number of MAC units and internal memory on the height of the weights. DeepRNN approaches the problem in a hybrid manner by both storage and streaming of weights. The MAC units are replaced with a scalable grid of SIMD-style MACs. To further optimize the communication and data access, double buffered memories are used. These techniques eliminate both memory bandwidth and scalability issues.\\
Pacini et al. \cite{11027895} proposes FPG-AI RNN accelerator for faster implementation of LSTM and GRU models. The RNN model is first compressed using a quantization process followed by pruning and weight reduction which keeps recurrent weights on an on-chip BRAM. Gate sharing reduces the number of repeated computations by sharing the intermediate values across the platform. The heart of the accelerator is a layer wise computation engine which consists of a systolic array of PEs for computation of matrix vector operations. It then feeds the optimized network onto an automatic mapping network for mapping different configurations of a RNN network. \\
Very - Long Short-Term Memory (V-LSTM), a LSTM accelerator cited by Kim et al. \cite{10041123} uses an improved Fixed Nonzero-ratio Viterbi based Pruning methodology to efficiently optimize the memory. Fig. \ref{12_a} shows the architecture segregated into front-end, PEs, and a back-end. The role of the front-end is to fetch the activations and compressed sparse weights (Viterbi indices + nonzero values) from DRAM. In order to reduce bandwidth the inputs/hidden states are stored in a  on-chip SRAM. This is followed by sparse weights decompression and computation of MAC operations in the PEs. The back-end handles element-wise operations like activations and additions. In addition to this, double-buffering scheme and sub-block computation are used to further reduce latency and SRAM usage. This pruning methodology uses a nonzero ratio per row or column of the weight matrix to create a regular sparsity pattern. Using this, the problem of low utilization of PEs is fixed by keeping the number of nonzero weights constant. The architecture is structurally divided into three parts- the front end, PE and the back end. It includes a double buffered scheme which streams weights continuously when the front end fetches it. This reduces memory stalls. The PE arrays are distributed in such a way that every part of the array structure provides for one type of parallelism. These arrays compute MAC operations in parallel with the reconstruction of non-zero weights which reduce latency. This methodology also uses ping pong buffers for efficient memory usage. The accelerator is further streamlined with its backend performing partial sum-reductions and reusing it in the next steps. The accelerator achieves an impressive latency of 46.30 $\mu$s.\\
Jang et al. \cite{JIANG2022104417} uses a sparsity pruning methodology- Shared Index Bank-Balanced Sparsity (SIBBS) co-processor for accelerating LSTM applications. SIBBS groups row memory banks into bank-clusters for sharing index masks and also balanced distribution across banks. A set of parallel Sparse-Matrix×Vector processing engine (SpMxV PEs) provide inter row parallelism. Each SpMxV PE is further split into bank-PEs which individually holds private vectors, fetches the weights from memory matrix and subsequently feeds an adder tree. The processes are pipelined in a manner that Similarity Check Unit (SMCU) parallely performs similarity checks with the current inputs. The Central Control Unit further increase the throughput by preloading compressed weights and vectors into the vector unit. These provide low latency single sample inference. \\
In order to reduce memory access with optimum computation efficiency, Gao et al. \cite{8780644} proposes implementation of an Independently Recurrent Neural Network (IndRNN) on an accelerator. As shown in Fig. \ref{12_b}, the hardware architecture comprises of a Controller Unit (IndCtrlUnit), Multiplication Unit (MultUnit), Element-wise Product Unit (EleUnit) and Activation Unit (ActUnit). IndCtrlUnit manages various control signals and regulates the data flow between modules. MAC operations are performed by the multiplier using data from the previous FIFO, on-chip weight BRAM, and the product unit. Using the MultUnit output, EleUnit  computes the Hadamard product required in IndRNN operations. Lastly, the ActUnit applies the Rectified Linear Unit (ReLU) activation function to the computed results for the current timestep, for computing the final activated output. The computation steps are significantly reduced by replacement of hidden layer matrix multiplication with vector multiplication. In addition to this, weights and repeating vectors are stored on a dual mode BRAM to reduce the memory accesses. The array of PEs consist of 128 MAC units. In addition to this, EleUnit and ActUnit operate in a pipelined manner thus significantly increasing the throughput. These techniques make the accelerator energy efficient as well as faster than its predecessor architectures. Guo et al. \cite{GUO2025128871} exploits fine-grained parallelism in an LSTM cell and coarse-grained parallelism during training. During forward propagation, dense vector-matrix multiplications are merged together alongwith loop initiation intervals during HLS that promotes parallel execution and decreases latency. 
To minimize the data movement inside the LSTM cell, local buffering and array partitioning is provided. The errors in backward propagation are computed simultaneously with parallel generation of weight gradients. A coarse grained pipeline architecture for training is provided for complementing the fine grained parallelism.

\subsection{Graph Neural Network (GNN)}
\label{subsec3.4}

\begin{figure}[tb]
    \centering

    % First image (top)
    \begin{subfigure}[t]{\linewidth}
        \centering
        \includegraphics[width=0.7\linewidth,keepaspectratio]{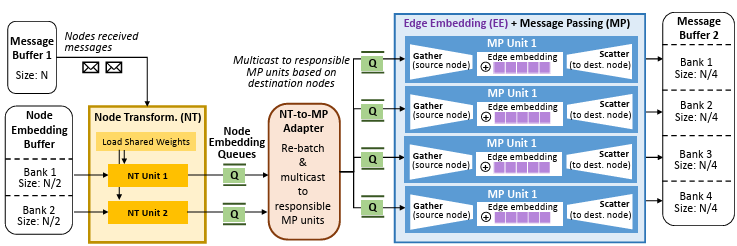}
        \caption{FlowGNN architecture showing parallel node transformations and message passing techniques \cite{10071015}.}
        \Description{GNN_1_diagram}
        \label{13_a}
    \end{subfigure}

    \vspace{1.2ex} % controlled vertical spacing

    % Second image (bottom)
    \begin{subfigure}[t]{\linewidth}
        \centering
        \includegraphics[width=0.7\linewidth,height=0.3\linewidth,keepaspectratio]{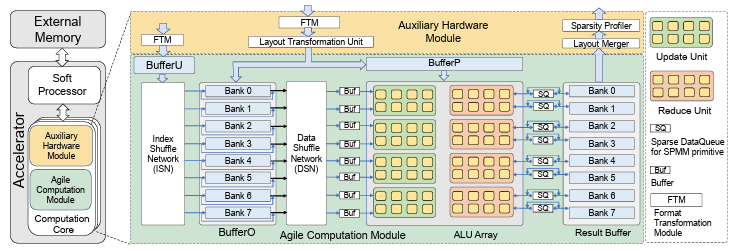}
        \caption{Structure of Dynasparse with modular computation and adaptive dataflow modules \cite{10177394}.}
        \Description{GNN_2_diagram}
        \label{13_b}
    \end{subfigure}

    \caption{Block diagram for GNN accelerator architectures.}
    \label{fig:two_images_4}
\end{figure}

 Zhou et al. \cite{9586181} presents BlockGNN which enhances computational speed by performing compression of the GNN architecture. Block-Circulant -a compression technique where convolutions in frequency domain using Fast Fourier Transform (FFT) and Inverse Fourier Transform (IFFT) on partitioned blocks. The heart of the Block-Circulant- CirCore consists of a vector processing unit which acts as a specialized datapath for performing parallel FFT transformations and complex multiplications. The channels are arranged parallely in a systolic array structure which compute the calculations and transform them back to the spatial domain. The global buffer stores the intermediate results, input vectors and compressed weights thus lowering data accesses.
Sarkar et al. \cite{10071015} applies multi-queue dataflow (given by Fig. \ref{13_a}) to solve the problem faced during parallelizing message passing across several edges. Each message passing unit is provided with its dedicated partitioned memory bank where node embeddings are sent as soon as they are produced. This alleviates memory stalls and thus increases the efficiency.  Following which, two types of edge based scattering operations take place. One is node transformation followed by scattering and the second is gathering then transformation. The optimized dataflow structure promotes parallelism for a generalized GNN structure.
\\
 Geng et al. \cite{9252000} proposes hardware auto tuning methodologies- dynamic distribution smoothing, remote switching, and row remapping to solve the irregularity and imbalance in graph-node structure. Dynamic distribution smoothing methods divides the higher degree nodes into smaller ones while merging the smaller nodes into one. This balances the workload and promotes parallelism of the tasks. These techniques not only improves the throughput and hardware utilization but also accelerate the Graph Convolutional Network (GCN) inference phase.\\
 Liang et al. \cite{9256539} presents an automatic GNN acceleration network which determines the performance bottleneck and accordingly selects the optimum computation, memory and graph manipulation template from the predefined hardware templates. The computation template comprising of an array of systolic elements accelerate both dense and sparse operations which are further optimized with the help of different memory modes. The memory template has partitioned on-chip memory and DRAM specifically allocated for storage of intermediate results and node features. The graph manipulation technique reorders edges and remaps rows/columns to regularize sparsity patterns. Dynasparse, a GNN accelerator proposed by Zhang et al. \cite{10177394}  uses a compiler where each kernel  provides partitioning and generate specific execution scheme. Fig. \ref{13_b} shows the three dynamic switched proposed schemes which are General Matrix-Matrix Multiplication (GEMM), Sparse-Dense Matrix Multiplication (SpDMM), and  Sparse Matrix-Matrix Multiplication (SPMM). When both the matrices are dense, there are no optimizations and the systolic arrays perform the convolution as usual. The second one is used for partially sparse activations. In this a sparse–dense intermediate is multiplied with a dense matrix which allows dense processing for the other operand while maintaining sparse activations. SPMM is chosen when one is dense and the other one is sparse. For sparse where there are zero computations, no hardware resources are allocated thus reducing memory bandwidth. FP-GNN is proposed by Tian et al. \cite{TIAN2022294} which aims to eliminate sparsity problems and solve memory bottleneck issues. It combines node and feature level parallelism to accelerate the workloads. Parallel processing of multiple nodes take place across PEs to balance irregular graph structures which are further subjected to dynamic load balancing. A pipelined dataflow with memory banking not only promotes efficient data movement but also provides conflict free memory access.

\section{Hardware-Level Optimization Strategies}
\label{sec4}

\begin{figure*}[!ht]
\centering
 \includegraphics[height= 5.2 cm, width=\linewidth]
{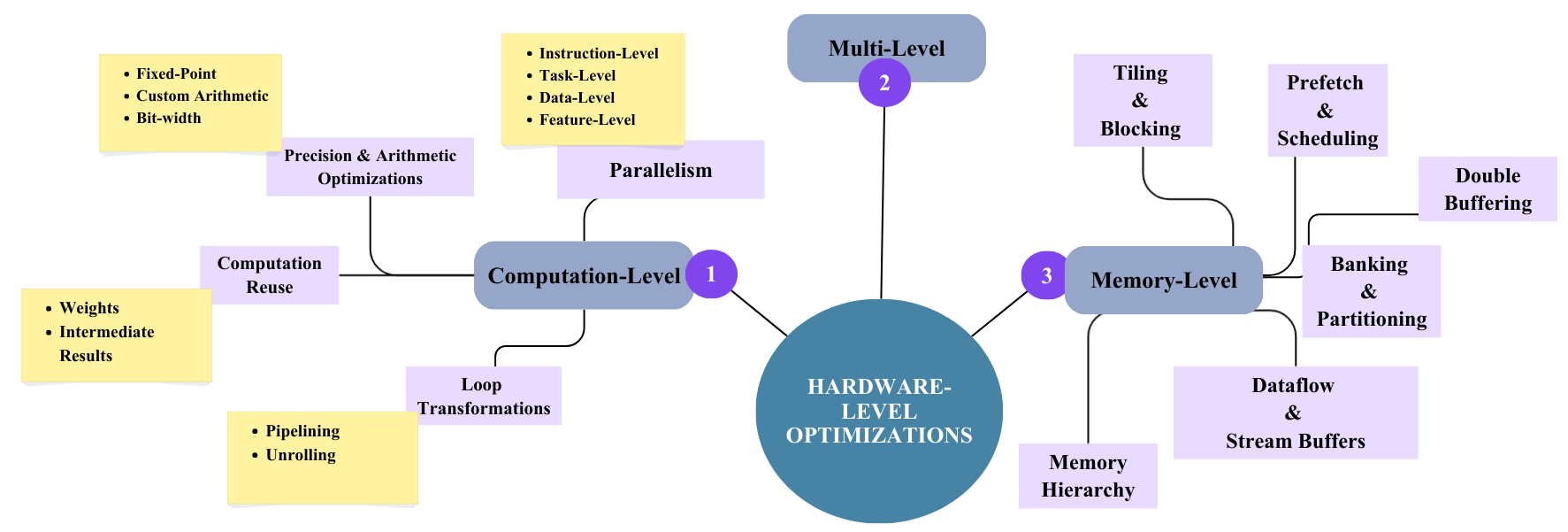}
  \caption{Summary of the different types of Hardware-Level Optimizations- computation, memory and multi-level optimizations.}
  \Description{Chart for Optimizations}
  \label{hard}
\end{figure*}

This section addresses recent hardware level optimizations for FPGA-based accelerators. AI applications are also computationally intensive, needing improvements for both efficient execution and speeding up the computational process. Furthermore, resource-constrained hardware makes it even more critical. As a result, adaptations must be performed at multiple levels to speed up computation, ensure efficient execution, and optimize hardware consumption.
\\Fig. \ref{hard} summarizes the discussed hardware level optimization techniques for FPGA based accelerators. These optimizations improve the performance and efficiency of accelerators by maintaining a balance between computation and memory usage. At the computation level, precision and arithmetic optimizations such as fixed-point representation, custom arithmetic units and reduced bit-width reduce hardware cost and latency. In addition to this there are loop transformation methodologies like pipelining and unrolling which improve the execution throughput. Computation reuse techniques minimize redundant operations by reusing weights and intermediate results. The last type of computation level technique is employment of different forms of parallelism at various levels. These include instruction-level, data-level, task-level, and feature-level parallelism which enable simultaneous processing to boost speed. At the multi-level, tiling and blocking divide large computations into smaller segments to improve data locality and make on-chip memory deployment easier. At the memory level, banking and partitioning allow concurrent memory accesses. The other subtypes include double buffering which overlaps data transfer with computation, memory hierarchy which organizes storage efficiently across levels, prefetching and scheduling to reduce stalls, and dataflow with stream buffers for enabling continuous data movement. These collectively reduce latency with maximization of throughput values. Subsections \ref{subsec4.1}, \ref{subsec4.2} and \ref{subsec4.3} explore the implementation of these optimization techniques in some of the state-of-the accelerators.

\subsection{Computation Level Optimizations}
\label{subsec4.1}

Deep learning techniques have high arithmetic density which spans to billion of operations. Optimizations helps these operations to meet the throughput and latency requirements. These also aid efficient usage of FPGA resources and eliminate stalled or ideal operations. These optimizations also enhance the performance of the FPGA accelerators.\\
CAESAR, a CNN given by Kim et al. \cite{10212679} removes redundant computations and thus significantly reduces the computational overhead. Overlapping convolution windows and weight repetition result in many redundant multiplications cause many redundant computations, making the convolution operation bulky. To address this issue, weight reordering is applied in which the weights are restructured in such a way that the identical weights can be easily identified.  The accelerator eliminates the redundant computations by scattering over several accumulators and keeping the record in a mapping table and sparse activation group.\\
Liu et al. \cite{LIU2024106197} proposes a series of computation optimization strategies such as loop unrolling and pipelining, data reuse, precision scaling and quantization. Tiling first decomposes a large computational tasks into smaller blocks. It is then followed by blocking which uses loop transformations to be applied on the entire block. This facilitates data reuse and reduces off-chip memory operations. Precision scaling and quantization further reduces the computational load by replacing floating point operations to fixed-point and reduced-bit operations. Alongwith tiling, Rai et al. \cite{10900711} proposes a dynamic reconfigurable framework- FPUGen which has customized mantissa and exponent widths which provides for a variety of representations such as- Single Precision-32 (SP-32), Tensor-float-32 (TF-32), Half Precision-16 (HP-16), and Brain-FP format-16 (BF-16). It is a five-stage pipelined architecture with a shared adder/subtractor  and bit-partitioned multiplier units to support this multi-point representation.The dynamic precision configuration comes at a cost of low resource efficiency and increased design complexity. \cite{10822997} proposes a six-stage pipeline architecture for execution of simultaneous operations. It also prevents memory stalls thus reducing idle operations. Additionally these computations are spread across multiple PEs for exploiting data level parallelism.  Pipelining comes with certain disadvantages as well. These include data control and structural hazards. \\
The training process is quite resource and memory intensive, so researchers have often opted for offline training solutions. Yu et al. \cite{9120209} proposes a 3D CNN accelerator which provides a sparsity aware methodology to solve memory bandwidth issues encountered during training phase. It leverages sparsity to opt out from unnecessary MAC operations. A monolothic 3D design employs parallelism by tightly binding computation units with the memory layers thus reducing data movement which enhances the efficiency. Like the predecessor architecture this also employs quantization to further enhance the computational capabilities. Zhang et al. \cite{10702206} uses an improved 3D pulse array structure maximizes parallel Multiply-accumulate (MAC) operations across spatial, input and output channel dimensions. The outputs are computed in such a way that the partial results from different PEs are are computed parallely. The input channels are distributed in such a way so that the multiple channels are processed simultaneously. These techniques not only reduce the pipeline latency but also improve throughput and speed up the inference time.\\
Model wise optimizations also lead to improvement in computational efficiency. One such example is given by Kang et al. \cite{9360939} where the fully connected layers of LeNet-5 (Han et al. \cite{7753296}) are removed thus lowering the computational demand. The architecture also have convolution windows with shift registers and parallel multiply add module for simultaneous computations.

\subsection{Memory Level Optimizations}
\label{subsec4.2}

\begin{figure*}[!ht]
\centering
 \includegraphics[width=0.8\linewidth]{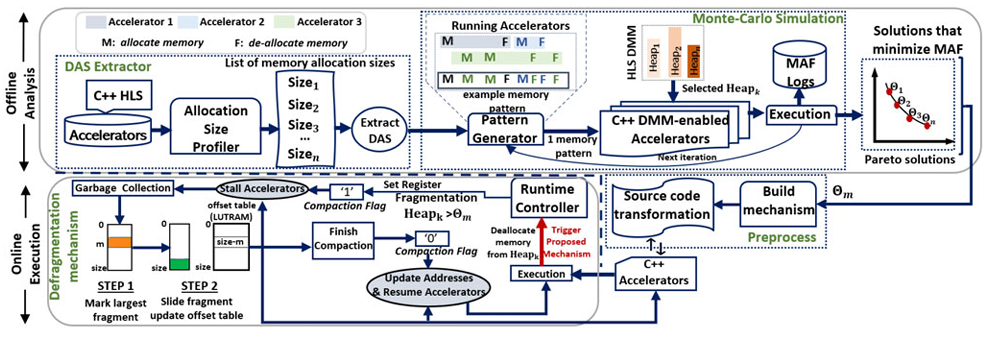}
  \caption{Design flow for a  HLS-based dynamic memory optimized framework with offline pattern analysis \& Monte-Carlo-guided heap optimization. \cite{9826871}}
  \Description{Diagram for memory optimization technique}
  \label{memory}
\end{figure*}

In FPGA based accelerators, the type of memory and its corresponding optimizations are fixed according to the requirements of the implemented model. Memory is classified as two types- on-chip and off-chip memory. Both memory types support memory requests for data transfer and storage of intermediate values. Improper handling of memory access can result in bottleneck and have an effect on the overall system performance. Additionally, because FPGAs are memory constrained devices, optimizations are performed to maximize memory bandwidth and streamline memory access requests.
 \\
Khan et al. \cite{10595963} studies the effects of four types of memory accesses- Multiple Input Single Output (MISO), Single Input Multiple Output (SIMO), Multiple Input Multiple Output (MIMO) and All Input Multiple Output (AIMO). The modes are configured based on the number of inputs and generated outputs. MISO uses multiple inputs and filters to generate a single output tile, whereas MIMO generates several output features. In the second case, a single input is convolved with numerous filters to produce multiple outputs. The final one involves the creation of partial outcomes while computing all outputs in a single phase. But first, tiling is done to accommodate large sized feature maps and weights into on-chip buffers. This is followed by reuse and double buffering to reduce memory access requests. It was observed that out of all the memory access techniques, AIMO offered best tradeoff between memory efficiency and performance for acceleration of CNN operations.\\
Cheng et al. \cite{6239808} presents a framework for providing memory level parallelism configured during High Level Synthesis (HLS). A profile guided tool first analyzes the runtime memory traces and accordingly partitions the memory accesses into independent groups. During HLS, according to the configurations, loop pipelining, instruction rescheduling, and operation reordering is applied. A vulnerability window measures the maximum extent of reordering between dependent memory accesses. Violation is raised if the partitioned memory's access falls outside this window. Following which the execution from a correct start-point. This makes it adaptable to sustain higher bandwidth and achieve upto 52\% performance improvement.  Diamantopoulos et al. \cite{9586181} proposes implementation of a  Long Short-Term Memory (LSTM) accelerator by reshaping on-chip memory during HLS. Like the previous one, partitioning and reshaping of the BRAMs are provided with the help of directives like \verb|#pragma HLS PARTITION| and \verb|#pragma HLS RESHAPE|. This is done in such a way that uniform bit-widths are stored in BRAM and higher precision bit-widths are stored in  Look-Up Table Random Access Memory (LUTRAM). This memory hierarchy and storing the data in a partitioned manner not only increases bandwidth but also provide for maximum parallelism. Kokkinis et al. \cite{9826871} also uses dynamic configurations for memory optimization.  Fig. \ref{memory} provides the details about the operation of the proposed framework. It operates in two phases: offline analysis (top part) and online execution/defragmentation (bottom part). In the offline phase, the DAS extractor profiles accelerator memory allocations and generates allocation patterns. This is followed by evaluation through Monte-Carlo simulation to identify Pareto-optimal configurations that minimize memory allocation failures (MAF). In the online phase, a runtime controller monitors the fragmentation ratio for each heap. When it exceeds the predefined threshold (Q), the garbage collection mechanism is triggered. The lower-left block shows the Mark-Compact process where live objects are marked, fragmented regions are identified, and memory blocks are compacted toward lower heap addresses. During compaction, the offset table updates the relocated addresses, while a 1-bit compaction flag stalls to prevent invalid memory access. After completion of compaction, addresses are updated and accelerator execution resumes which ensures efficient dynamic memory utilization.

Near memory computing (NMC) is the process of moving computation blocks closer to the on-chip memory in order to reduce high cost memory transfers. Tao et al. \cite{9951617}, the NMC core is presented which consists of a shift-and-add multiplier block, adder tree and a quantization block. The weights are loaded in the BRAMs and inputs are passed through the shift registers thus reducing the memory access distance. The pooling layers are also accelerated with the help of the comparators near the memory. This significantly increases the memory usage. Dhar et al. \cite{8839401} also includes in-storage computing which uses partial calculations to further optimize the memory bandwidth.

\subsection{Multi-level Optimizations}
\label{subsec4.3}
\begin{figure*}[t]   % Use top placement, not h!
\centering
\includegraphics[width=\linewidth]{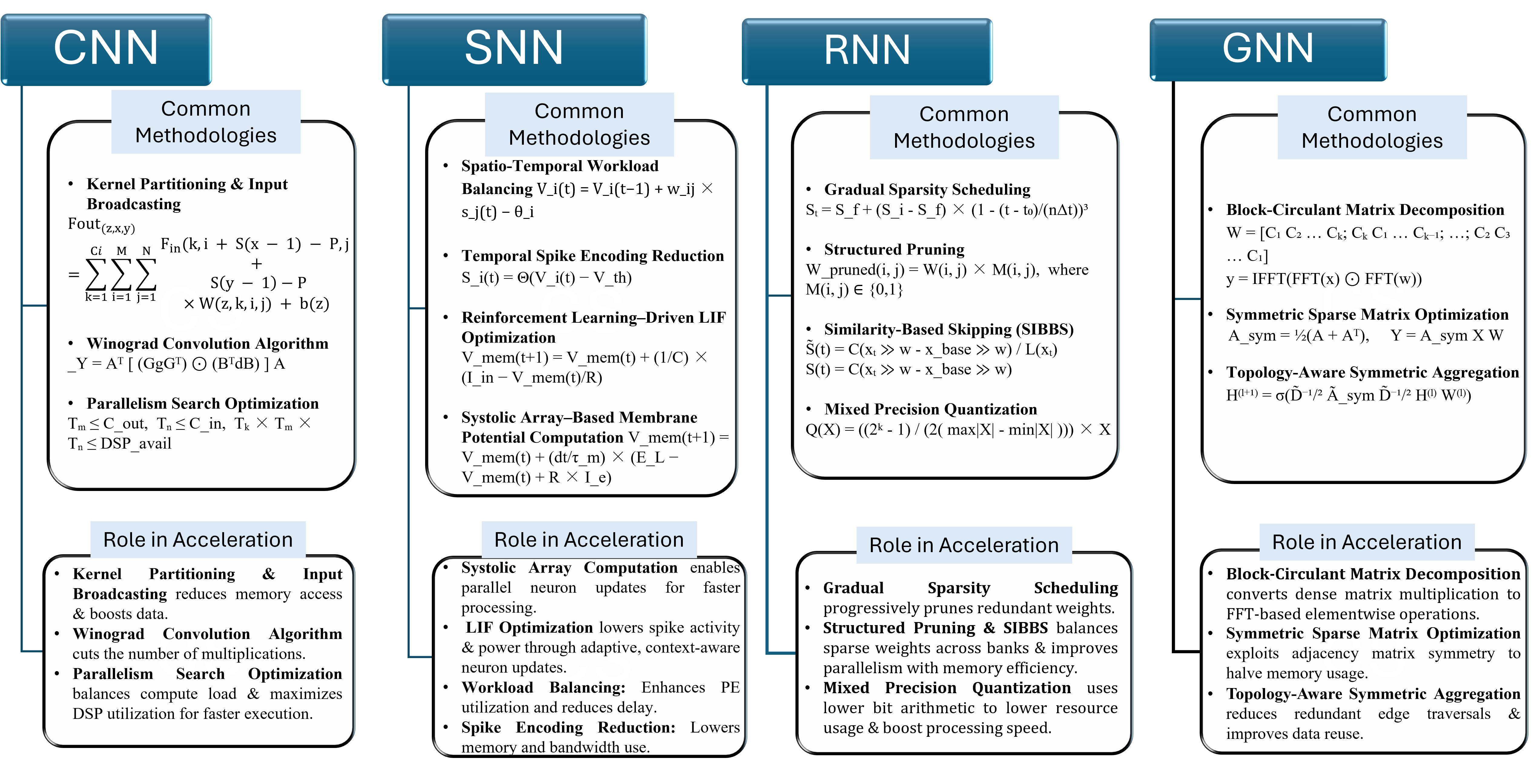}  % exactly fits text width
\caption{Summary of optimization methodologies and roles of acceleration of various neural models}
\Description{A conceptual diagram summarizing FPGA optimization methodologies.}
\label{opt}
\end{figure*}

Single-level optimizations often fail to address all issues.  It may also cause problems that must be addressed by optimizations at other levels. Thus, multi-level optimization improves performance on all fronts.\\
FixyFPGA cited by Meng et al. \cite{9556422} optimizes resource utilization as well as computational overhead.  For optimization of resource utilization, fusion of batch normalization and weights is done. Also the structure is fully pipelined and is equipped with buffers and shift registers. The weights are arranged in a manner such that the important weights are identified and accordingly pruning can be done to reduce the computational load. A 4 bit quantization method is implemented as a part of this approach for reduction of computational load as well as memory usage. Baranwal et al. \cite{9211770} proposes ReLAccS- a multi-level optimization strategy to accelerate the reinforcement learning on FPGA. At computational level loop pipelining and unrolling is done to implement parallelism of multiple operations. Also for precision optimization, fixed point is used. On the other hand, memory is partitioned for simultaneous access to multiple banks. The memory computations and accesses are further accelerated with the help of buffering and prefetching. Also in order to reduce idle units dynamic allocation of hardware resources is provided. Wang et al. \cite{10871700} proposes optimizations for robotic intelligence applications. They propose a hardware estimation model with an empirical calibration with nonlinear fitting ($R^2 \approx 0.98$) to find out the number of BRAM and Lookup Tables (LUTs). AutoML further guided for selection of hardware based configurations to enhance the performance. Further optimizations are made by designing a pipelined and clock cycle balanced architecture. Unlike its predecessor, Zhong et al. \cite{7927161} proposes design space exploration is done to find the balance between memory banks and other hardware resources. On-chip memory tiling, buffering and partitioning with weight mapping reduces external memory accesses and addresses bandwidth issues. Like all the accelerator configurations, loop unrolling, pipelining and quantization is done to enhance the arithmetic efficiency and reduce the computational load to an extent. Sun et al. \cite{8977882} proposes a hybrid CNN-RNN accelerator which comprises of a streaming architecture and mapping strategy. The design of a unified convolution engine is presented which maps matrix-vector onto same hardware resources thus promoting reuse and maximization of resource utilization. Furthermore, scheduling  strategies are provided: layer-by-layer, subgraph-by-subgraph and subnetwork-by-subnetwork. The first one is used for maximum flexibility operations whereas the second one decreases intermediate memory accesses. The accelerator achieves an impressive throughput of 646 Giga Operations Per Second (GOPS).\\
Fig. \ref{opt} summarizes the major numerical hardware optimization methodologies across CNN, SNN, RNN, and GNN accelerators. It also highlights the imapact on acceleration. In CNNs, convolution $F_{\text{out}} = \sum F_{\text{in}} \cdot W + b$ is parallelized via kernel partitioning and input broadcasting. Another important methodology is the Winograd transform $Y = A^T[(GgG^T) \odot (B^T d B)]A$  which reduces multiplications (e.g., a $3\times3$ convolution from 9 to 4 multipliers) There are also tiling factors which satisfying $T_k T_m T_n \leq DSP_{\text{avail}}$ and thus maximizes DSP utilization. In SNNs, neuron dynamics $V(t) = V(t-1) + w_{ij}s_j(t) - \theta$ are optimized by reducing spike events with maintaining a balanced temporal workloads. RNNs apply structured pruning $W_{\text{pruned}} = W \odot M$, sparsity scheduling, similarity-based skipping, and quantization $Q(X) = \frac{2^k - 1}{\max X - \min X}X$ to reduce computation and memory precision. 
For GNNs, layer propagation $H^{(l+1)} = \sigma(D^{-1/2} A D^{-1/2} H W)$ involves sparse-dense multiplications with complexity $\mathcal{O}(|E|F)$. In addition to this, there are block-circulant decomposition which reduce weight multiplication to $\mathcal{O}(N \log N)$. Symmetric adjacency $A_{\text{sym}} = \frac{1}{2}(A + A^T)$ avoids redundant edge operations, and topology-aware normalization improves stability while exploiting sparsity, collectively reducing MAC operations, memory bandwidth, and hardware resource usage.

 \section{Performance Analysis of the Accelerators}
 \label{sec5}
For FPGA based accelerators, performance is determined in terms of  throughput, precision achieved, efficiency and resource utilization utilization. Additionally, methodology and operating frequency gives a better perspective of the performance. Tables \ref{comparison_1}, \ref{comparison_2}, \ref{comparison_3} and \ref{comparison_4} provide a comparative analysis of the models used and its implementation details.

\begin{table*}
    \centering
    \caption{Performance Summary of FPGA-based CNN Accelerators}
    \small
    \label{comparison_1}
\newcolumntype{P}[1]{>{\centering\arraybackslash}p{#1}}

\enlargethispage{32pt}
\begin{tabular}{
P{1.4cm}  % Model
P{1.8cm}  % Methodology
P{1.0cm}  % Platform
P{1.8cm}  % Throughput
P{1.0cm}  % Precision
P{1.3cm}  % Frequency
P{0.5cm}  % LUT
P{0.5cm}  % DSP
P{0.5cm}  % BRAM
P{0.5cm}  % FF
P{0.8cm}  % Speedup
P{1.1cm}    % Baseline (CPU)
}
\toprule
\textbf{Model} &  
\textbf{Methodology} & 
\textbf{Platform} & 
\textbf{Throughput} & 
\textbf{Precision} & 
\textbf{Frequency} & 
\multicolumn{4}{c}{\textbf{Resources (Utilization \%)}} & 
\textbf{Speedup} & 
\textbf{Baseline} \\
\midrule
CNP (2009) \cite{5272559} & LeNex-5 & Virtex-4 SX35 & 5.25 GOPS  &  fixed 16 bit & 200 MHz & 90.00\% &  28.00\% & --  & 90.00\% & -- & -- \\
MAPLE (2010) \cite{7851527} & 4 CONV layers & Virtex-5 SX240T & 7 GOPS  & -- & 125 MHz & -- &  -- & 92.00\%  & -- & 0.5 & C870 GPU \\
DLA (2017) \cite{aydonat2017opencltmdeeplearningaccelerator} & AlexNet & Arria 10 GX 1150 & 1,382 GOPS  & fixed 16 bit & 214 MHz & 258.00\% &  97.00\% & 92.00\%  & 40.00\% & -- & -- \\
OpenCL CNN Accelerator (2017) \cite{10.1145/3020078.3021698}   & VGG-16 & Arria 10 GX 1150 & 1790 GOPS  & floating 16 bit & 385 MHz & -- &  87.00\% & 53.40\%  & -- & -- & -- \\
OpenCL CNN Accelerator (2017) \cite{10.1145/3020078.3021698}  & VGG-16 & Arria 10 GX 1150 & 866 GOPS  & floating 32 bit & 370 MHz & -- &  90.80\% & 46.10\%  & -- & -- & -- \\
Automatic RTL Generator (2017) \cite{8056824}  & NiN & Stratix-V GXA7 & 282.67 GOPS  & fixed 16 bit & 150 MHz & 96.00\% &  100.00\% & 59.00\% & -- & -- & -- \\
Automatic RTL Generator (2017) \cite{8056824} & VGG-16 & Stratix-V GXA7 & 352.24 GOPS  & fixed 16 bit & 150 MHz & 90.00\% & 100.00\% & 86.00\%  & -- & -- & -- \\
Automatic RTL Generator (2017) \cite{8056824} & ResNet-50 & Arria 10 GX 1150 & 587.63 GOPS  & fixed 16 bit & 200 MHz & 37.00\% &  100.00\% & 56.00\%  & -- & -- & -- \\
Angel-Eye (2018) \cite{7966660}  & VGG-16 & Zynq XC7Z045 & 84.3 GOPS  & fixed 16 bit & 150 MHz & 184.00\% &  87.000\% & 89.00\%  & 29.00\% & -- & -- \\
Optimized CONV Accelerator (2018) \cite{8330049}  & VGG-16 & Stratix-V GXA7 & 278.2 GOPS  & fixed 16 bit & 150 MHz & 97.00\% &  100.00\% & 59\%  & -- & -- & -- \\
Winograd GEMM Accelerator (2019) \cite{8856248}  & AlexNet and VGG-16  & Virtex-7 VX690T & 433.63 GOPS  & floating 16 bit & 200 MHz & 58.00\% &  39.80\% & 92\%  & 99.0\% & -- & -- \\
Winograd Accelerator (2020) \cite{9421655}  & YOLOv2 & Arria 10 GX 1150 & 278.2 GOPS  & -- & 240 MHz & 35.00\% & 100.00\% & --  & -- & 5.8 & Xeon CPU Gold 5115 \\
Object Recognition Accelerator (2021)  \cite{9476039} & VGG-16 & Xilinx VC709 & 230.1 GOPs  & -- & 200 MHz & 28.00\% &  18.40\% & 31.00\% & 18.50\% & -- & -- \\
Scalable accelerator (2023) \cite{10241078} & MNIST & Xilinx XC7z020 & 63.67 GOP/s  & -- & 100 MHz & 56.00\% &  17.00\% & 11.00\% & 10.00\% & -- & -- \\
1D-CNN accelerator (2024) \cite{10867080}   & 1D-CNN & Xilinx XCZU2CG & 63.67 (GOP/s)  & fixed 16 bit & 200 MHz & -- &  75.00\% & 77.33\% & -- & 21.13 & Intel Core i5-8300\\
 \bottomrule 
\end{tabular}
\end{table*}

\begin{table*}
    \centering
    \caption{Performance Summary of FPGA-based SNN Accelerators}
    \small
    \label{comparison_2}
\newcolumntype{P}[1]{>{\centering\arraybackslash}p{#1}}

\begin{tabular}{
P{1.4cm}  % Model
P{1.8cm}  % Methodology
P{1.0cm}  % Platform
P{1.8cm}  % Throughput
P{1.0cm}  % Precision
P{1.3cm}  % Frequency
P{0.5cm}  % LUT
P{0.5cm}  % DSP
P{0.5cm}  % BRAM
P{0.5cm}  % FF
P{0.8cm}  % Speedup
P{1.1cm}    % Baseline (CPU)
}
\toprule
\textbf{Model} &  
\textbf{Methodology} & 
\textbf{Platform} & 
\textbf{Throughput} & 
\textbf{Precision} & 
\textbf{Frequency} & 
\multicolumn{4}{c}{\textbf{Resources (Utilization \%)}} & 
\textbf{Speedup} & 
\textbf{Baseline} \\
\midrule
& & & & & & \textbf{LUT} & \textbf{DSP} & \textbf{BRAM} & \textbf{FF} & & \\
\midrule

SyncNN (2021) \cite{9556367}  & SCNN & Xilinx ZCU102 & N/A & fixed 4 bit & 200 MHz & 37.45\% &  N/A  & N/A   & N/A  & N/A & N/A\\ 
BP-STDP SNN Accelerator (2022) \cite{9662046}   & SFNN & Xilinx XCVU9P & 5980 & floating 16 bit & 100 MHz & 15.64\% & 0.00\% & N/A  & 12.77\% & 49.42 &  CPU with 3.2GHz CPU and 16GB memory \\
Cerebron (2022) \cite{9855834} & LIF & Xilinx XC7Z100 & 43.07 GSOP/s & N/A & 200 MHz & 30.97\% &  0.00\% & 37.48\% & 12.71\% & N/A  & N/A  \\
Spiker (2022)\cite{9911998}   & SFNN & Artix 7 & N/A & fixed 16 bit & 100 MHz & 55.00\% &  N/A & 32.00\%  & 25.00\% & N/A  & N/A  \\
Skydiver (2022)\cite{9733037}  & SCNN & Xilinx XA7Z020 & 22,600 FPS & N/A & 200 MHz & 21.04\% &  0.00\% & 48.07\%  & 4.70\% & N/A  & N/A  \\
CONVP (2023)\cite{10191153}  & N/A & Xilinx XA7Z020 & 5980 & fixed 8 bit & 100 MHz & 30.47\% &  N/A & N/A  & 10.65\% & 58 &   \\
FireFly (2023) \cite{10143752}  & SCNN(5,7,9,11) & Xilinx XCZU3EG & 1,382.4 GOPS/s & integer 8 bit & 
300 MHz & 21.43\% &  80\% & 75\%  & N/A &  N/A &  N/A    \\
FrameFire (2023)\cite{10168660}  & N/A & Xilinx XC7Z035 & 23.2 GOPS/s & integer 8 bit & 
200 MHz & 24.39\% &  0.00\% & 25.60\% & N/A &  N/A &  N/A \\
DeepFire2 (2023) \cite{10113796} & N/A & Xilinx VCU118 & 17550 &  & 
450 MHz & 12.03\% &  68.73\% & 20.45\%  & N/A &  N/A &  N/A \\
Diff. Time Encoding Accelerator (2023) \cite{windhager2023snnarchitecturedifferentialtime} & N/A & Xilinx ZCU102 & N/A & fixed 6 bit & 
365 MHz & 10.70\% &  0.00\% & 60.67\%  & N/A &  N/A &  N/A \\
ELIF-NHAS (2023) \cite{9785601} & SFNN and SCNN & Xilinx XC7K355T & 3.2 GSOP/s & fixed 6 bit &   200 MHz & 19.24\% &  0.00\% & 11.27\%  & N/A &  N/A &  N/A \\
STBP SNN Accelerator (2024 )\cite{YIN2024106377}  & STBP & Xilinx VC707 & N/A &  13 bit & 200 MHz & 28.15\% &  30.18\% & N/A  & 45.06\% & N/A &  N/A \\
Sparse Weight SNN Acclerator (2024 )\cite{10388553}  & N/A & Xilinx XCKU115 & 5.102 &  8 bit & 120 MHz & 13.90\% & N/A & 1.62\% & 1.86\% & N/A &  N/A \\
Hardware Software Optimised Accelerator (2024) \cite{10737858}   & 38.4 & Xilinx XC7Z020 & 5.102 &  fixed 8 bit & 100 MHz & 22.43\% & 7.67\% & 67.86\%  & 7.67\% & N/A &  N/A \\
Synapse Accelerator (2024) \cite{10558156}  & IF neuron model & Xilinx ZCU102 & 47.3 GSOPS/s &  fixed 11 bit & 30 MHz & 29.06\% & N/A & N/A  & 17.33\% & N/A &  N/A \\
SINK (2024)\cite{10586065}  & N/A & Xilinx ZCU104 & 1575 GSOPS/s & N/A & 350 MHz & 29.06\% & N/A & N/A  & 2.07\% & N/A &  N/A \\
FireFly S (2025)\cite{10754657}  & SCNN(5,7,9) & Xilinx XCZU5EV & 5,352 FPS &  4 bit & 333 MHz & 48.33\% &  N/A & 21.33\%  & 42\% & N/A &  N/A \\ 
        \bottomrule 
\end{tabular}
\end{table*}

\enlargethispage{32pt}

% \begin{table*}
%     \centering
%     \caption{Performance Summary of FPGA-based RNN Accelerators}
%     \small
%     \label{comparison_2.2}
% \newcolumntype{P}[1]{>
% {\centering\arraybackslash}p{#1}}
% \begin{tabular}{
% P{1.4cm}  % Model
% P{1.8cm}  % Methodology
% P{1.0cm}  % Platform
% P{1.8cm}  % Throughput
% P{1.0cm}  % Precision
% P{1.1cm}  % Frequency
% P{0.5cm}  % LUT
% P{0.5cm}  % DSP
% P{0.5cm}  % BRAM
% P{0.5cm}  % FF
% P{0.8cm}  % Speedup
% P{0.8cm}    % Baseline (CPU)
% }
% \toprule
% \textbf{Model} &  
% \textbf{Methodology} & 
% \textbf{Platform} & 
% \textbf{Throughput} & 
% \textbf{Precision} & 
% \textbf{Frequency} & 
% \multicolumn{4}{c}{\textbf{Resources (Utilization \%)}} & 
% \textbf{Speedup} & 
% \textbf{Baseline} \\
% \midrule
% & & & & & & \textbf{LUT} & \textbf{DSP} & \textbf{BRAM} & \textbf{FF} & & \\
% \midrule
% Language Model Accelerator (2015)  \cite{7160054} & Std. RNNLM with hidden layer of 1K nodes & Xilinx
% Virtex6 LX760 & 9.6 GOPS & -- & 150 MHz & 37.00\% & 48.00\% & 39.00\%  & 30.00\% & 14.10 & Intel Xeon CPU E5-2630 \\
% ESE (2017) \cite{han2017eseefficientspeechrecognition} & Std. LSTM & Xilinx XCKU060 & 282 GOPS & integer 16 bit & 200 MHz &  88.60\% & 54.50\% & 87.70\% & 68.30\% & 43.00 & Core i75930k CPU
% \\
% DeepStream (2017) \cite{8050816} & 2 LSTM Layers & Xilinx ZC706 & 145.9 Mops/s/W & integer 16 bit & 142 MHz &  -- & -- & --  & -- & 16.62 & Exynos 54224 Cortex-A15
% \\

% \bottomrule 
%         \end{tabular}
% \end{table*}

\begingroup
\sloppy
\begin{table*}
    \centering
    \caption{Performance Summary of FPGA-based RNN Accelerators}
    \small
    \label{comparison_3}
\newcolumntype{P}[1]{>
{\centering\arraybackslash}p{#1}}
\begin{tabular}{
P{1.4cm}  % Model
P{1.8cm}  % Methodology
P{1.0cm}  % Platform
P{1.8cm}  % Throughput
P{1.0cm}  % Precision
P{1.1cm}  % Frequency
P{0.5cm}  % LUT
P{0.5cm}  % DSP
P{0.5cm}  % BRAM
P{0.5cm}  % FF
P{0.8cm}  % Speedup
P{1.0cm}    % Baseline (CPU)
}
\toprule
\textbf{Model} &  
\textbf{Methodology} & 
\textbf{Platform} & 
\textbf{Throughput} & 
\textbf{Precision} & 
\textbf{Frequency} & 
\multicolumn{4}{c}{\textbf{Resources (Utilization \%)}} & 
\textbf{Speedup} & 
\textbf{Baseline} \\
\midrule
& & & & & & \textbf{LUT} & \textbf{DSP} & \textbf{BRAM} & \textbf{FF} & & \\
\midrule
ESE\cite{han2017eseefficientspeechrecognition} (2017) & Std. LSTM & Xilinx XCKU060 & 282 GOPS & integer 16 bit & 200 MHz &  88.60\% & 54.50\% & 87.70\%  & 68.30\%& 43.00 & Corei 75930k CPU\\
DeepRNN (2017) \cite{8050816} & 2 LSTM Layers & Xilinx ZC706 & 400.98 Mops/s/W & integer 16 bit & 142 MHz & --  & -- & --  & -- & 45.67 & Exynos 54224 Cortex-A15\\
LSTM-RNN (2017) \cite{7858394} & Std. LSTM & Xilinx VC707 & 7.26 GFLOP/s & floating 32 bit & 150 MHz & 65.31\% & 42.00\% & 52.04\%  & 30.08\% & 20.18 & Intel Xeon CPU E5-2430\\
FINN-L (2018) \cite{rybalkin2018finnllibraryextensionsdesign}  & 2 layer Std. LSTM & Xilinx ZCU104 & 746@8bit  & 8 bit & 266 MHz & 26.08\% & -- & 28.84\% & -- & -- & --\\
IndRNN Accelerator (2018) \cite{8780644}  &  2 layers IndRNN &  Xilinx ZC706 & 3.15 GOPS/s/W  & fixed 16 bit & 125 MHz & 2.54\% & 3.44\% & 10.27\% & 2.89\% & Intel CORE i5 CPU & --\\
Transprecision Accelerator (2018)  \cite{8742271}  & Std. LSTM & Xilinx XCKU060 & 515 GOPS & floating 32 bit & 240 MHz & -- & -- & -- & -- & -- & --\\
DeltaRNN (2018) \cite{10.1145/3174243.3174261} & RNN & Xilinx XC7Z100 & 192 GOP/s & integer 16 bits & 125 MHz & 94.22\% & 38.02\% & 60.60\% & 21.50\% & 63.8 & Intel i7-8700K CPU\\
Open-CL based Accelerator (2019) \cite{8977882}   & Std. LSTM & Intel Arria10 GX1150 & 515 GOPS & 16 bit & 242 MHz & -- & -- & -- & -- & -- & -- \\
EdgeDRNN (2020) \cite{9268992} & 2 layer RNN & Xilinx XC7Z007S & 20.2 GOP/s & integer 16 bits & 125 MHz & 30.80\% & 13.6\% & 32.00\%  & 34.10\% & -- & --\\
IMDB LSTM Accelerator (2020) \cite{9295665} & 1 layer LSTM & Xilinx XC7VX485 & 10.9 GOPS & integer 16 bits & 200 MHz & 7.73\% & 5.07\% & 16.07\%  & 3.52\% & 43.13 &  Intel Core i5-6500 \\
Bidirectional accelerator (2022) \cite{10072756}  & 1024 sized LSTM & Virtex-7VC707 & 91.47 GOPS/s & -- & 200 MHz & 21.8\% & 18.29\% & 23.98\% & 6.95\% & 13.4 & Intel(R) Core i7-10700F\\
SIBBS Accelerator (2022) \cite{JIANG2022104417} & Std. LSTM & Xilinx XCKU115 & 712.6 GOPS & fixed 8 bit & 200 MHz & 81.26\% & 75.36\% & 28.33\% & 43.82\% & -- & -- \\
\( S^2\textsc{RNN} \) (2025) \cite{10742088}  & 8 layer LSTM & Altera 10 GX & -- & -- & -- & 9\% & 7\% & -- & -- & 1.79 & -- \\
FSS Accelerator (2025)  \cite{10818746}  & 8 layer LSTM & Xilinx Alveo U280 & 600 TFLOPS  & int 8 bit & 225 MHz & 54.00\% & 81.40\% & 81.90\% & 35.00\% & 2.80 & NVIDIA RTX A6000\\
FSS Accelerator (2025) \cite{10818746}  & 8 layer LSTM & Xilinx Alveo U250 & 600 TFLOPS  & int 8 bit & 225 MHz & 46.00\% & 80.00\% & 84.00\% & 33.00\% & 2.42 & NVIDIA RTX A6000\\
LSTM Accelerator (2025) \cite{ABDELBAKY2025155845} & 8 layer LSTM & Xilinx  KCU105 & 32.5 GOPS  & fixed 20 bit & 166.66 MHz & 6.14\% & 4.90\% & 21.33\% & 4.02\%
& -- & --\\
        \bottomrule 
        \end{tabular}
\end{table*}

\begin{table*}
    \centering
    \caption{Performance Summary of FPGA-based GNN Accelerators}
    \small
    \label{comparison_4}
\newcolumntype{P}[1]{>
{\centering\arraybackslash}p{#1}}
\begin{tabular}{
P{1.4cm}  % Model
P{1.8cm}  % Methodology
P{1.0cm}  % Platform
P{1.8cm}  % Throughput
P{1.0cm}  % Precision
P{1.1cm}  % Frequency
P{0.5cm}  % LUT
P{0.5cm}  % DSP
P{0.5cm}  % BRAM
P{0.5cm}  % FF
P{0.8cm}  % Speedup
P{1.0cm}    % Baseline (CPU)
}
\toprule
\textbf{Model} &  
\textbf{Methodology} & 
\textbf{Platform} & 
\textbf{Throughput} & 
\textbf{Precision} & 
\textbf{Frequency} & 
\multicolumn{4}{c}{\textbf{Resources (Utilization \%)}} & 
\textbf{Speedup} & 
\textbf{Baseline} \\
\midrule
& & & & & & \textbf{LUT} & \textbf{DSP} & \textbf{BRAM} & \textbf{FF} & & \\
\midrule
AWB-GCN (2020) \cite{9252000}  & GCN & Stratix 10SX & 873.1 & floating 32 bit & 330 MHz & 25\% &  71.11\% & N/A  & 13.7\% & 3255 &  Intel Xeon E5-2680 \\
DeepBurning-GL (2020) \cite{9256539}  & GCN, GS-Pool, R-GCN, EdgeConv & Xilinx ZC706 & 8.6 GOPS/W &  fixed 16 bit & 100 MHz & 52.72\% &  85.57\% & 76.21\% & 18.36\%  & 7.43 &  Intel Xeon E5-2680 \\
% DeepBurning-GL (2020) \cite{9256539}  & GCN, GS-Pool, R-GCN, EdgeConv & Xilinx KCU1500 & 47.1 GOPS/W & fixed 16 bit & 200 MHz & 7,27,497 (83.43\%) & 4211 (76.29\%) & 61\% & 33.64\% & 199.98 &  Intel Xeon (Skylake) 6151 \\
DeepBurning-GL \cite{9256539} (2020)  & GCN, GS-Pool, R-GCN, EdgeConv & Xilinx Alveo U50 & 53.5 GOPS/W & fixed 16 bit & 200 MHz & N/A  &  93.57\% & 45.14\% & 32.93\% & 346.98 &  Intel Xeon (Skylake) 6151 \\
BoostGCN (2021) \cite{9444065}  & GCN  & Stratix 10SX & 512 or 1280 &  floating 32 bit & 250 MHz & 10.5\% & 33.33\% & N/A  & N/A  & 45 & Intel Xeon Gold 5120 CPU \\
FP-GNN (2022) \cite{TIAN2022294}  & GCN, Graphsage, GAT & Xilinx VCU128 & 922 & floating 32 bit & 225 MHz & 25.16\% &  90.78\% & 88.88\% & 19.83\% & 665.33 &  Intel Xeon Platinum 8260 \\
GCoD (2022) \cite{9773223} & GCN, Graphsage, GAT, GIN & Xilinx VCU128 & N/A  & fixed 32 bit & 330 MHz & N/A  &  57.62\% & N/A  & N/A  & 15286 &  Intel Xeon E5-2680 \\
I-GCN (2022) \cite{huang2021awb} & GCN, Graphsage, GAT, GIN & Stratix 10SX & 2700 & floating 32 bit & 330 MHz & N/A  &  45.39\% & N/A  & N/A  & 5549 & Intel Xeon E5-2680 \\
FlowGNN (2022) \cite{10071015} & GCN, GIN, GAT, PNA, DGN, VN & Xilinx VCU128 & N/A  & int 8,16,32 bit & 225 MHz & 7.57\% &  17.69\% & 13.76\% & 9.78\% & 3645 & Intel Xeon Gold 6226R \\
QEGCN (2022) \cite{YUAN2022102596} & GCN & Xilinx Alveo U50 & N/A  & N/A  & 300 MHz & 8.05\% & 31.21\% & 2.18\% & 0.48\% & 1009 & Intel Xeon Gold 5120 CPU\\
Dynasparse (2023) \cite{10177394} & GCN, LightGCN, Graphsage, GIN, GAT & Xilinx Alveo U250 & 512 &  floating 32 bit & 225 MHz & 58.51\% & 58.49\% & 42.60\% & N/A  & 306 & Ryzen 3990x \\
Graph-OPU (2023) \cite{10296283}  & GCN, Graphsage, GAT, GIN & Xilinx Alveo U50 & 459.6 &  fixed 32 bit  & 225 MHz & 54.48\% & 46.33\% & 68.97\% & 16.38\% & 1654 & Intel I7-12700KF \\
FTW-GAT (2024) \cite{10136834} & GCN, Graphsage, GAT & Xilinx Alveo U250 & 1080 &  floating 16 bit  & 600 MHz & 65.4\% & 59.3\% & 57.3\% & N/A & 500.13 & AMD Ryzen 3990x \\
        \bottomrule 
        \end{tabular}
\end{table*}

\section{Challenges encountered by FPGA based Accelerators}
\label{sec6}
Over the previous sections, a thorough examination of the state-of-the-art FPGA accelerators for different deep learning models has been presented. It is observed that different types of neural networks do not encounter the same problems and thus has to be tackled according to the network requirements. After an intensive study of all the acceleration techniques and optimizations, FPGA accelerators still face certain challenges. For better understanding, the problems can be categorized as follows:

\begin{enumerate}[label=(\roman*)]
\item Quantization Issues- Fixed bit representation and custom bit not only reduces utilization of hardware resources but also provides for efficient usage. However, using this type of representation for sensitive layers may result in saturation or underflow errors. Quantized models further need additional finetuning for recovery of accuracy. Also, low bit arithmetic may require specialized hardware units. Custom bit representation alleviates some of the above problems but adds lookup overhead which further leads to burdening of the hardware resources.

\item Trade-off between Power Consumption and Efficiency- Some computation techniques like loop unrolling not only increase computational throughput but also lead to a significant rise in power consumption. This happens because when loops are unrolled, multiple PEs perform simultaneous iterations. Also when pipelining is used there is simultaneous switching and routing complexity. This increases dynamic power. While these techniques often promote efficiency but suffer from increased power consumption. This issue has also been raised by Liu et al. \cite{LIU2024128511}.\\

\item Lack of Efficient Co-Design with CPUs - There are a large variety of optimization techniques available for acceleration of the inference and training phases. However, there is absence of sufficient co-design techniques of FPGA with CPU. These face issues such
as stalls in communication and scheduling mismatch of operations. When integrated as a SoC, issues are on a lower scale as the available datapaths can be optimized. But these issues remain unresolved during integration of heterogeneous devices.
\item Communication Inefficiency and Memory Bottlenecks- As Wu et al. \cite{10082175} states that there is still lack in efficient data access techniques between memory and other configurable resources. Though these problems are mitigated to a certain extent with the help of memory partitioning
and NMC, still some bottlenecks exist. These techniques often suffer from memory stalls, computational, area and power overheads which limits the acceleration capability.
\item Scalability Issues and Lack of an Ecosystem - There is an absence of standardized libraries as that of CUDA model. Although high level synthesis tools are present but they are often vendor specific thus leading researchers to search for generalized or customized techniques. On the other hand, in spite of proposed optimizations there is
always a burden on resource usage. These issues are addressed by Fat et al. \cite{11105416}.
\item Security Issues - Although FPGA reconfigurability makes it attractive for hardware acceleration, it also exposes the platform to security threats such as adversarial attacks \cite{rahaman2024secure,nafi2025dash,rahaman2025runtime,rahaman2024samurai}, configuration tampering, and bit-flip injections \cite{rakin2019bit,rahaman2025secure}. Since the configuration bitstream contains detailed structural and functional information about the implemented accelerator, it becomes a critical attack surface. Such attacks can significantly degrade accuracy, induce misclassification, or alter the intended functionality. The functionality and the analysis of the FPGA configuration bitstream consists of plethora of information about the acceleration techniques. The attacks can be launched in such a way that it causes significant reduction in accuracy or leads to misclassification. 

\section{Future Directions for Optimizations}
\label{sec7}
From early FPGA models to the latest System-on-Chip (SoC) platforms, several advanced features and optimizations have been incorporated to enable high performance and efficient computational capability. These developments position FPGAs as strong contenders for optimal hardware acceleration. Section~\ref{sec6} presents an overview of the key challenges faced by FPGA-based accelerators despite extensive optimizations. Addressing these challenges requires accelerator designs that are adaptable to evolving computational demands. In this context, several potential methodologies and future research directions for optimization are discussed in this section.\

Architectural designs for both generalized and customized FPGA-based accelerators have been explored extensively. General-purpose FPGA accelerators support a wide range of workload templates but often lack fine-grained optimizations, resulting in lower throughput compared to customized solutions. In contrast, custom FPGA-based hardware accelerators enable fine-grained optimizations that yield low latency and high peak performance, although they limit architectural reusability. Achieving an effective balance between generality and customization therefore remains an open challenge.

Moreover, compression techniques such as pruning frequently trade model accuracy for improved hardware efficiency. To address this, a risk-analysis framework can be developed to evaluate and balance accuracy and efficiency according to the specific requirements of the neural network and its application. Recent advances in automatic compiler generation, which analyze network characteristics and derive suitable optimization strategies, can inspire the design of such risk-analysis schemes.

As discussed previously, the lack of comprehensive hardware–software co-design strategies often results in mismatches between algorithmic computation and hardware mapping. While modern in-house processing systems employ optimized dataflow mechanisms, similar architectural principles must be incorporated into FPGA accelerator design. Additionally, improved memory-access mechanisms and optimized caching strategies can further enhance throughput while reducing latency.\

Analog and digital in-memory computing have recently emerged as promising approaches for neural-network deployment in analog and ASIC implementations. Analog in-memory computing (AIMC) can be emulated in FPGA-based systems, as demonstrated by Petropoulos et al.~\cite{11021575}, who introduced a Noise Injection Unit (NIU) within conventional processing elements to inject device-level noise into fetched weights instead of performing multiply–accumulate operations. The noisy-weight implementation occurs during inference, thereby eliminating substantial computational overhead encountered in many FPGA-based accelerators.

Similarly, digital in-memory computing (DIMC) solutions for SoC-based architectures have been investigated in~\cite{10596361}, presenting a memory architecture that integrates DMA-enabled system memory, an interconnection fabric, and an embedded accelerator unit. Since hardware-aware models have been proposed to mitigate the limitations of AIMC and DIMC, analogous architectural strategies can be designed to address these challenges while emulating AIMC within FPGA platforms.

Future research should focus on APC-based \cite{rahaman2025runtime,rahaman2024samurai} runtime monitoring to detect anomalies caused by adversarial manipulation, configuration tampering, or malicious bit-flips in FPGA accelerators. Learning-based profiling of trusted configurations combined with lightweight on-chip sensors can enable real-time anomaly detection with minimal overhead.

Overall, future methodologies should aim to achieve an optimal balance among performance, efficiency, and latency. However, performance enhancements may also increase vulnerability to security threats. Consequently, the development of attack resilient FPGA based hardware accelerators remains an important complementary research direction.
\end{enumerate}

\section{Conclusion}
\label{sec8}
This article presents a comprehensive review of recent advances in state-of-the-art hardware accelerators, emphasizing the fundamental trade-off between high performance and power efficiency and the consequent need for optimization across multiple architectural levels, including computation, memory, and system integration. A wide range of acceleration techniques is analyzed, spanning systolic array architectures, customized parallel processing elements, and innovative memory organizations such as weight-transposable buffers, along with key design strategies including tiling, blocking, partitioning, and hardware–software co-design for both on-chip and off-chip resources.

Although several optimization principles are broadly applicable across neural network models, network-specific adaptations remain essential for maximizing throughput and efficiency. The study further examines critical challenges in accelerator design, distinguishing between limitations imposed by inherent architectural constraints and those arising from current optimization methodologies.

To overcome these barriers, future research must develop robust and scalable solutions that fully exploit accelerator capabilities while enhancing performance, scalability, and security. In this context, FPGAs emerge as highly flexible and effective platforms for modern AI acceleration, and continued refinement of their design and optimization frameworks is key to realizing their full potential as efficient hardware accelerators.

\bibliographystyle{ACM-Reference-Format}
\bibliography{sample-base}

@String{BIT = "{BIT}" }

@String{Computing = "Computing" }

@String{Computer = "{IEEE} Computer" }

@String{Springer = "Springer-Verlag" }

@ARTICLE{8632885,
  author={Nassif, Ali Bou and Shahin, Ismail and Attili, Imtinan and Azzeh, Mohammad and Shaalan, Khaled},
  journal={IEEE Access}, 
  title={Speech Recognition Using Deep Neural Networks: A Systematic Review}, 
  year={2019},
  volume={7},
  pages={19143-19165},
  doi={10.1109/ACCESS.2019.2896880}}

@ARTICLE{8594633,
  author={Shawahna, Ahmad and Sait, Sadiq M. and El-Maleh, Aiman},
  journal={IEEE Access}, 
  title={{FPGA}-Based Accelerators of Deep Learning Networks for Learning and Classification: A Review}, 
  year={2019},
  volume={7},
  pages={7823-7859},
  doi={10.1109/ACCESS.2018.2890150}}

@ARTICLE{10924161,
  author={Sharrab, Yousef O. and Attar, Hani and Eljinini, Mohammad Ali H. and Al-Omary, Yasmin and Al-Momani, Wala’A E.},
  journal={IEEE Access}, 
  title={Advancements in Speech Recognition: A Systematic Review of Deep Learning Transformer Models, Trends, Innovations, and Future Directions}, 
  year={2025},
  volume={13},
  number={},
  pages={46925-46940},
  doi={10.1109/ACCESS.2025.3550855}}

@InProceedings{10.1007/978-981-97-1335-6_31,
author="Wu, Meng
and Zhou, Jin
and Peng, Yibin
and Wang, Shuihua
and Zhang, Yudong",
editor="Su, Ruidan
and Zhang, Yu-Dong
and Frangi, Alejandro F.",
title="Deep Learning for Image Classification: A Review",
booktitle="Proceedings of 2023 International Conference on Medical Imaging and Computer-Aided Diagnosis (MICAD 2023)",
year="2024",
publisher="Springer Nature Singapore",
address="Singapore",
pages="352--362",
doi="10.1007/978-981-97-1335-6\_31",
}

@misc{Doshi2025,
  author    = {Riddhi Virendra Doshi and Sagarkumar S. Badhiye and Latika Pinjarkar},
  title     = {Deep Learning Approach for Biomedical Image Classification},
  journal   = {Journal of Imaging Informatics in Medicine},
  year      = {2025},
  doi       = {10.1007/s10278-025-01590-8},
  pages     = {2948-2933},
url={https://doi.org/10.1007/s10278-025-01590-8},
note={{Early access}}
 }

@ARTICLE{9303459,
  author={Dou, Haoran and Karimi, Davood and Rollins, Caitlin K. and Ortinau, Cynthia M. and Vasung, Lana and Velasco-Annis, Clemente and Ouaalam, Abdelhakim and Yang, Xin and Ni, Dong and Gholipour, Ali},
  journal={IEEE Transactions on Medical Imaging}, 
  title={A Deep Attentive Convolutional Neural Network for Automatic Cortical Plate Segmentation in Fetal {MRI}}, 
  year={2021},
  volume={40},
  number={4},
  pages={1123-1133},
  doi={10.1109/TMI.2020.3046579}}

@article{ZHANG2025129395,
title = {Deep convolutional neural networks meet variational shape compactness priors for image segmentation},
journal = {Neurocomputing},
volume = {623},
pages = {129395},
year = {2025},
doi = {https://doi.org/10.1016/j.neucom.2025.129395},
author = {Kehui Zhang and Lingfeng Li and Hao Liu and Jing Yuan and Xue-Cheng Tai},
}

@article{FAN2024e38104,
title = {Deep neural networks for automated damage classification in image-based visual data of reinforced concrete structures},
journal = {Heliyon},
volume = {10},
number = {19},
pages = {e38104},
year = {2024},
doi = {https://doi.org/10.1016/j.heliyon.2024.e38104},
author = {Ching-Lung Fan},
}

@INPROCEEDINGS{8978121,
  author={Weinman, Jerod and Chen, Ziwen and Gafford, Ben and Gifford, Nathan and Lamsal, Abyaya and Niehus-Staab, Liam},
  booktitle={2019 International Conference on Document Analysis and Recognition (ICDAR)}, 
  title={Deep Neural Networks for Text Detection and Recognition in Historical Maps}, 
  year={2019},
  pages={902-909},
  doi={10.1109/ICDAR.2019.00149},
publisher={IEEE},
address = {Piscataway, NJ, USA}}

@article{rahaman2025secure,
  title={Secure and Storage-Efficient Deep Learning Models for Edge AI Using Automatic Weight Generation},
  author={Rahaman, Habibur and Chatterjee, Atri and Bhunia, Swarup},
  journal={arXiv preprint arXiv:2507.06380},
  year={2025}
}

@article{SINTHUJA2024789,
title = {Extraction of Text from Images Using Deep Learning},
journal = {Procedia Computer Science},
volume = {235},
pages = {789-798},
year = {2024},
note = {International Conference on Machine Learning and Data Engineering (ICMLDE 2023)},
doi = {https://doi.org/10.1016/j.procs.2024.04.075},
author = {M Sinthuja and Chirag Ganesh Padubidri and Gaddam Sai Jayachandra and Mudduluru Charan Teja and Golthi Sai Pavan Kumar},
}

@ARTICLE{10098596,
  author={Amjoud, Ayoub Benali and Amrouch, Mustapha},
  journal={IEEE Access}, 
  title={Object Detection Using Deep Learning, {CNNs} and Vision Transformers: A Review}, 
  year={2023},
  volume={11},
  number={},
  pages={35479-35516},
  doi={10.1109/ACCESS.2023.3266093}}

@ARTICLE{5948412,
  author={Wang, Cong and Chen, Tianrui},
  journal={IEEE Transactions on Neural Networks}, 
  title={Rapid Detection of Small Oscillation Faults via Deterministic Learning}, 
  year={2011},
  volume={22},
  number={8},
  pages={1284-1296},
  doi={10.1109/TNN.2011.2159622}}

@article{BATOOL2022107886,
title = {Software fault prediction using data mining, machine learning and deep learning techniques: A systematic literature review},
journal = {Computers and Electrical Engineering},
volume = {100},
pages = {107886},
year = {2022},
issn = {0045-7906},
doi = {https://doi.org/10.1016/j.compeleceng.2022.107886},
author = {Iqra Batool and Tamim Ahmed Khan},
}

@article{YANG2024120797,
title = {Interpretable machine learning for weather and climate prediction: A review},
journal = {Atmospheric Environment},
volume = {338},
pages = {120797},
year = {2024},
issn = {1352-2310},
doi = {https://doi.org/10.1016/j.atmosenv.2024.120797},
author = {Ruyi Yang and Jingyu Hu and Zihao Li and Jianli Mu and Tingzhao Yu and Jiangjiang Xia and Xuhong Li and Aritra Dasgupta and Haoyi Xiong},
}

@ARTICLE{9755930,
  author={Alarfaj, Fawaz Khaled and Malik, Iqra and Khan, Hikmat Ullah and Almusallam, Naif and Ramzan, Muhammad and Ahmed, Muzamil},
  journal={IEEE Access}, 
  title={Credit Card Fraud Detection Using State-of-the-Art Machine Learning and Deep Learning Algorithms}, 
  year={2022},
  volume={10},
  number={},
  pages={39700-39715},
  doi={10.1109/ACCESS.2022.3166891}}

@misc{nvidia1999,
  author       = {{CNNfn}},
  title        = {{nVidia unveils new computer graphics accelerator}},
  year         = {1999},
  howpublished = {\url{http://money.cnn.com/1999/08/31/technology/nvidia/}},
  note         = {Accessed: 2025-07-23}
}

@INPROCEEDINGS{9556367,
  author={Panchapakesan, Sathish and Fang, Zhenman and Li, Jian},
  booktitle={2021 31st International Conference on Field-Programmable Logic and Applications (FPL)}, 
  title={{SyncNN}: Evaluating and Accelerating Spiking Neural Networks on FPGAs}, 
  year={2021},
  pages={286-293},
  doi={10.1109/FPL53798.2021.00058},
publisher={IEEE},
address = {Piscataway, NJ, USA}}

@INPROCEEDINGS{5272559,
  author={Farabet, Clement and Poulet, Cyril and Han, Jefferson Y. and LeCun, Yann},
  booktitle={2009 International Conference on Field Programmable Logic and Applications}, 
  title={{CNP}: An {FPGA}-based processor for Convolutional Networks}, 
  year={2009},
  pages={32-37},
  doi={10.1109/FPL.2009.5272559},
publisher={IEEE},
address = {Piscataway, NJ, USA}}

@INPROCEEDINGS{9252000,
  author={Geng, Tong and Li, Ang and Shi, Runbin and Wu, Chunshu and Wang, Tianqi and Li, Yanfei and Haghi, Pouya and Tumeo, Antonino and Che, Shuai and Reinhardt, Steve and Herbordt, Martin C.},
  booktitle={2020 53rd Annual IEEE/ACM International Symposium on Microarchitecture (MICRO)}, 
  title={{AWB-GCN}: A Graph Convolutional Network Accelerator with Runtime Workload Rebalancing}, 
  year={2020},
  pages={922-936},
  doi={10.1109/MICRO50266.2020.00079},
publisher={IEEE},
address = {Piscataway, NJ, USA}}

@article{TIAN2022294,
title = {{FP-GNN}: Adaptive {FPGA} accelerator for Graph Neural Networks},
journal = {Future Generation Computer Systems},
volume = {136},
pages = {294-310},
year = {2022},
issn = {0167-739X},
doi = {https://doi.org/10.1016/j.future.2022.06.010},
author = {Teng Tian and Letian Zhao and Xiaotian Wang and Qizhe Wu and Wei Yuan and Xi Jin},
}

@INPROCEEDINGS{10177394,
  author={Zhang, Bingyi and Prasanna, Viktor},
  booktitle={2023 IEEE International Parallel and Distributed Processing Symposium (IPDPS)}, 
  title={Dynasparse: Accelerating {GNN} Inference through Dynamic Sparsity Exploitation}, 
  year={2023},
  pages={233-244},
  doi={10.1109/IPDPS54959.2023.00032},
publisher={IEEE},
address = {Piscataway, NJ, USA}}

@INPROCEEDINGS{9444065,
  author={Zhang, Bingyi and Kannan, Rajgopal and Prasanna, Viktor},
  booktitle={2021 IEEE 29th Annual International Symposium on Field-Programmable Custom Computing Machines (FCCM)}, 
  title={{BoostGCN}: A Framework for Optimizing {GCN} Inference on {FPGA}}, 
  year={2021},
  pages={29-39},
  doi={10.1109/FCCM51124.2021.00012},
publisher={IEEE},
address = {Piscataway, NJ, USA}}

@INPROCEEDINGS{9773223,
  author={You, Haoran and Geng, Tong and Zhang, Yongan and Li, Ang and Lin, Yingyan},
  booktitle={2022 IEEE International Symposium on High-Performance Computer Architecture (HPCA)}, 
  title={{GCoD}: Graph Convolutional Network Acceleration via Dedicated Algorithm and Accelerator Co-Design}, 
  year={2022},
  pages={460-474},
  doi={10.1109/HPCA53966.2022.00041},
publisher={IEEE},
address = {Piscataway, NJ, USA}}

@inproceedings{huang2021awb,
  author    = {Shaoyi Huang and Yun Liang and Jinyang Li and Yuze Chi and Jianlei Yang and Yuchen Ma and Weiwen Jiang and Yiyu Shi},
  title     = {{AWB-GCN}: A Graph Convolutional Network Accelerator with Runtime Workload Rebalancing},
  booktitle = {Proceedings of the 2021 ACM/IEEE 48th Annual International Symposium on Computer Architecture (ISCA)},
  year      = {2021},
  pages     = {832--845},
  doi       = {10.1145/3466752.3480113},
  publisher = {ACM},
  address   = {New York, NY, USA}
}

@INPROCEEDINGS{10071015,
  author={Sarkar, Rishov and Abi-Karam, Stefan and He, Yuqi and Sathidevi, Lakshmi and Hao, Cong},
  booktitle={2023 IEEE International Symposium on High-Performance Computer Architecture (HPCA)}, 
  title={{FlowGNN}: A Dataflow Architecture for Real-Time Workload-Agnostic Graph Neural Network Inference}, 
  year={2023},
  pages={1099-1112},
  doi={10.1109/HPCA56546.2023.10071015},
publisher={IEEE},
address = {Piscataway, NJ, USA}}

@ARTICLE{9662046,
  author={Zhang, Jian and Wang, Ran and Pei, Xudong and Luo, Dan and Hussain, Sajjad and Zhang, Guohe},
  journal={IEEE Transactions on Circuits and Systems II: Express Briefs}, 
  title={A Fast Spiking Neural Network Accelerator based on {BP-STDP} Algorithm and Weighted Neuron Model}, 
  year={2022},
  volume={69},
  number={4},
  pages={2271-2275},
  doi={10.1109/TCSII.2021.3137987}}

@article{YIN2024106377,
title = {A reconfigurable {FPGA}-based spiking neural network accelerator},
journal = {Microelectronics Journal},
volume = {152},
pages = {106377},
year = {2024},
issn = {1879-2391},
doi = {https://doi.org/10.1016/j.mejo.2024.106377},
author = {Mingqi Yin and Xiaole Cui and Feng Wei and Hanqing Liu and Yuanyuan Jiang and Xiaoxin Cui},
}

@INPROCEEDINGS{10191153,
  author={Liu, Hanwen and Chen, Yi and Zeng, Zihang and Zhang, Malu and Qu, Hong},
  booktitle={2023 International Joint Conference on Neural Networks (IJCNN)}, 
  title={A Low Power and Low Latency {FPGA}-Based Spiking Neural Network Accelerator}, 
  year={2023},
  pages={1-8},
  doi={10.1109/IJCNN54540.2023.10191153},
publisher={IEEE},
address = {Piscataway, NJ, USA}}

@INPROCEEDINGS{10388553,
  author={Wang, Zilin and Zhong, Yi and Yang, Youming and Cui, Xiaoxin and Wang, Yuan},
  booktitle={2023 IEEE Biomedical Circuits and Systems Conference (BioCAS)}, 
  title={An Efficient Spiking Neural Network Accelerator with Sparse Weight}, 
  year={2023},
  pages={1-5},
  doi={10.1109/BioCAS58349.2023.10388553},
publisher={IEEE},
address = {Piscataway, NJ, USA}}

@article{YUAN2022102596,
title = {{QEGCN}: An {FPGA}-based accelerator for quantized {GCNs} with edge-level parallelism},
journal = {Journal of Systems Architecture},
volume = {129},
pages = {102596},
year = {2022},
issn = {1383-7621},
doi = {https://doi.org/10.1016/j.sysarc.2022.102596},
author = {Wei Yuan and Teng Tian and Qizhe Wu and Xi Jin},
}

@INPROCEEDINGS{10296283,
  author={Chen, Ruiqi and Zhang, Haoyang and Li, Shun and Tang, Enhao and Yu, Jun and Wang, Kun},
  booktitle={2023 33rd International Conference on Field-Programmable Logic and Applications (FPL)}, 
  title={{Graph-OPU}: A Highly Integrated {FPGA}-Based Overlay Processor for Graph Neural Networks}, 
  year={2023},
  pages={228-234},
  doi={10.1109/FPL60245.2023.00039},
publisher={IEEE},
address = {Piscataway, NJ, USA}}

@INPROCEEDINGS{9256539,
  author={Liang, Shengwen and Liu, Cheng and Wang, Ying and Li, Huawei and Li, Xiaowei},
  booktitle={2020 IEEE/ACM International Conference On Computer Aided Design (ICCAD)}, 
  title={{DeepBurning-GL}: an Automated Framework for Generating Graph Neural Network Accelerators}, 
  year={2020},
  pages={1-9},  doi={10.1109/VLSID64188.2025.00052},
publisher={IEEE},
address = {Piscataway, NJ, USA}}

@ARTICLE{10136834,
  author={He, Zerong and Tian, Teng and Wu, Qizhe and Jin, Xi},
  journal={IEEE Transactions on Circuits and Systems II: Express Briefs}, 
  title={{FTW-GAT}: An {FPGA}-Based Accelerator for Graph Attention Networks With Ternary Weights}, 
  year={2023},
  volume={70},
  number={11},
  pages={4211-4215},
  doi={10.1109/TCSII.2023.3280180}}

@INPROCEEDINGS{10737858,
  author={Nimbekar, Anagha and Katti, Prabodh and Li, Chen and Al-Hashimi, Bashir M. and Acharyya, Amit and Rajendran, Bipin},
  booktitle={2024 IEEE 37th International System-on-Chip Conference (SOCC)}, 
  title={Hardware-Software Co-optimised Fast and Accurate Deep Reconfigurable Spiking Inference Accelerator Architecture Design Methodology}, 
  year={2024},
  pages={1-6},
  doi={10.1109/VLSID64188.2025.00052},
publisher={IEEE},
address = {Piscataway, NJ, USA}}

@ARTICLE{10754657,
  author={Li, Tenglong and Li, Jindong and Shen, Guobin and Zhao, Dongcheng and Zhang, Qian and Zeng, Yi},
  journal={IEEE Transactions on Circuits and Systems I: Regular Papers}, 
  title={{FireFly-S}: Exploiting Dual-Side Sparsity for Spiking Neural Networks Acceleration With Reconfigurable Spatial Architecture}, 
  year={2025},
  volume={72},
  number={8},
  pages={4007-4020},
  doi={10.1109/TCSI.2024.3496554}}

@ARTICLE{10143752,
  author={Li, Jindong and Shen, Guobin and Zhao, Dongcheng and Zhang, Qian and Zeng, Yi},
  journal={IEEE Transactions on Very Large Scale Integration (VLSI) Systems}, 
  title={{FireFly}: A High-Throughput Hardware Accelerator for Spiking Neural Networks With Efficient {DSP} and Memory Optimization}, 
  year={2023},
  volume={31},
  number={8},
  pages={1178-1191},
  doi={10.1109/TVLSI.2023.3279349}}

@INPROCEEDINGS{10558156,
  author={Li, Mingyang and Kan, Yirong and Zhang, Renyuan and Nakashima, Yasuhiko},
  booktitle={2024 IEEE International Symposium on Circuits and Systems (ISCAS)}, 
  title={A Fully-Parallel Reconfigurable Spiking Neural Network Accelerator with Structured Sparse Connections}, 
  year={2024},
  pages={1-5},
  doi=
{10.1109/ISCAS58744.2024.10558156},
publisher={IEEE},
address = {Piscataway, NJ, USA}}

@INPROCEEDINGS{10586065,
  author={Karakchi, Rasha},
  booktitle={2024 IEEE 3rd International Conference on Computing and Machine Intelligence (ICMI)}, 
  title={A Scratchpad Spiking Neural Network Accelerator}, 
  year={2024},
  pages={1-5},
  doi={10.1109/ICMI60790.2024.10586065},
publisher={IEEE},
address = {Piscataway, NJ, USA}}

@ARTICLE{9855834,
  author={Chen, Qinyu and Gao, Chang and Fu, Yuxiang},
  journal={IEEE Transactions on Very Large Scale Integration (VLSI) Systems}, 
  title={Cerebron: A Reconfigurable Architecture for Spatiotemporal Sparse Spiking Neural Networks}, 
  year={2022},
  volume={30},
  number={10},
  pages={1425-1437},
  doi={10.1109/TVLSI.2022.3196839}}

@INPROCEEDINGS{10168660,
  author={Chen, Qinyu and Sun, Congyi and Gao, Chang and Fang, Xinyuan and Luan, Haitao},
  booktitle={2023 IEEE 5th International Conference on Artificial Intelligence Circuits and Systems (AICAS)}, 
  title={FrameFire: Enabling Efficient Spiking Neural Network Inference for Video Segmentation}, 
  year={2023},
  pages={1-5},
  doi={10.1109/AICAS57966.2023.10168660},
publisher={IEEE},
address = {Piscataway, NJ, USA}}

@article{ZHANG2025106616,
title = {A power-efficient spiking convolutional neural network accelerator based on temporal parallelism and streaming dataflow},
journal = {Microelectronics Journal},
volume = {158},
pages = {106616},
year = {2025},
issn = {1879-2391},
doi = {https://doi.org/10.1016/j.mejo.2025.106616},
author = {Jian Zhang and Yong Wang and Yanlong Zhang and Bo Bi and Qiliang Chen and Yimao Cai},
}

@ARTICLE{6701396,
  author={Neil, Daniel and Liu, Shih-Chii},
  journal={IEEE Transactions on Very Large Scale Integration (VLSI) Systems}, 
  title={Minitaur, an Event-Driven {FPGA}-Based Spiking Network Accelerator}, 
  year={2014},
  volume={22},
  number={12},
  pages={2621-2628},
  doi={10.1109/TVLSI.2013.2294916}}

@misc{windhager2023snnarchitecturedifferentialtime,
      title={{SNN }Architecture for Differential Time Encoding Using Decoupled Processing Time}, 
      author={Daniel Windhager and Bernhard A. Moser and Michael Lunglmayr},
      year={2023},
doi={10.48550/arXiv.2311.14447}, 
}

@ARTICLE{10113796,
  author={Aung, Myat Thu Linn and Gerlinghoff, Daniel and Qu, Chuping and Yang, Liwei and Huang, Tian and Goh, Rick Siow Mong and Luo, Tao and Wong, Weng-Fai},
  journal={IEEE Transactions on Computers}, 
  title={{DeepFire2}: A Convolutional Spiking Neural Network Accelerator on {FPGAs}}, 
  year={2023},
  volume={72},
  number={10},
  pages={2847-2857},
  doi={10.1109/TC.2023.3272284}}

@ARTICLE{9785601,
  author={Ye, Wujian and Chen, Yuehai and Liu, Yijun},
  journal={IEEE Transactions on Computer-Aided Design of Integrated Circuits and Systems}, 
  title={The Implementation and Optimization of Neuromorphic Hardware for Supporting Spiking Neural Networks With {MLP} and {CNN} Topologies}, 
  year={2023},
  volume={42},
  number={2},
  pages={448-461},
  doi={10.1109/TCAD.2022.3179246}}

@INPROCEEDINGS{9911998,
  author={Carpegna, Alessio and Savino, Alessandro and Di Carlo, Stefano},
  booktitle={2022 IEEE Computer Society Annual Symposium on VLSI (ISVLSI)}, 
  title={Spiker: an {FPGA}-optimized Hardware accelerator for Spiking Neural Networks}, 
  year={2022},
  pages={14-19},
  doi={10.1109/ISVLSI54635.2022.00016},
publisher={IEEE},
address = {Piscataway, NJ, USA}}

@ARTICLE{9733037,
  author={Chen, Qinyu and Gao, Chang and Fang, Xinyuan and Luan, Haitao},
  journal={IEEE Transactions on Computer-Aided Design of Integrated Circuits and Systems}, 
  title={Skydiver: A Spiking Neural Network Accelerator Exploiting Spatio-Temporal Workload Balance}, 
  year={2022},
  volume={41},
  number={12},
  pages={5732-5736},
  doi={10.1109/TCAD.2022.3158834}}

@inproceedings{10.1145/2541940.2541967,
author = {Chen, Tianshi and Du, Zidong and Sun, Ninghui and Wang, Jia and Wu, Chengyong and Chen, Yunji and Temam, Olivier},
title = {{DianNao}: a small-footprint high-throughput accelerator for ubiquitous machine-learning},
year = {2014},
isbn = {9781450323055},
publisher = {Association for Computing Machinery},
address = {New York, NY, USA},
doi = {10.1145/2541940.2541967},
booktitle = {Proceedings of the 19th International Conference on Architectural Support for Programming Languages and Operating Systems},
pages = {269–284},
numpages = {16},
location = {Salt Lake City, Utah, USA},
series = {ASPLOS '14}
}

@INPROCEEDINGS{7011421,
  author={Chen, Yunji and Luo, Tao and Liu, Shaoli and Zhang, Shijin and He, Liqiang and Wang, Jia and Li, Ling and Chen, Tianshi and Xu, Zhiwei and Sun, Ninghui and Temam, Olivier},
  booktitle={2014 47th Annual IEEE/ACM International Symposium on Microarchitecture}, 
  title={{DaDianNao}: A Machine-Learning Supercomputer}, 
  year={2014},
  pages={609-622},
  doi={10.1109/MICRO.2014.58},
publisher={IEEE},
address = {Piscataway, NJ, USA}}

@inproceedings{10.1145/2694344.2694358,
author = {Liu, Daofu and Chen, Tianshi and Liu, Shaoli and Zhou, Jinhong and Zhou, Shengyuan and Teman, Olivier and Feng, Xiaobing and Zhou, Xuehai and Chen, Yunji},
title = {{PuDianNao}: A Polyvalent Machine Learning Accelerator},
year = {2015},
isbn = {9781450328357},
publisher = {Association for Computing Machinery},
address = {New York, NY, USA},
doi = {10.1145/2694344.2694358},
booktitle = {Proceedings of the Twentieth International Conference on Architectural Support for Programming Languages and Operating Systems},
pages = {369–381},
numpages = {13},
location = {Istanbul, Turkey},
series = {ASPLOS '15}
}

@INPROCEEDINGS{7284058,
  author={Du, Zidong and Fasthuber, Robert and Chen, Tianshi and Ienne, Paolo and Li, Ling and Luo, Tao and Feng, Xiaobing and Chen, Yunji and Temam, Olivier},
  booktitle={2015 ACM/IEEE 42nd Annual International Symposium on Computer Architecture (ISCA)}, 
  title={{ShiDianNao}: Shifting vision processing closer to the sensor}, 
  year={2015},
  pages={92-104},
  doi={10.1145/2749469.2750389},
publisher={IEEE},
address = {Piscataway, NJ, USA}}

@ARTICLE{11048618,
  author={Zhang, Tao and Zhong, Yi and Yang, Youming and Wang, Zilin and Zhang, Zhaotong and Wang, Yuan},
  journal={IEEE Transactions on Circuits and Systems II: Express Briefs}, 
  title={{UniPRE}: An {SNN-ANN} Accelerator With Unified Max-Pooling Prediction and Redundancy Elimination}, 
  year={2025},
  volume={72},
  number={8},
  pages={1088-1092},
  doi={10.1109/TCSII.2025.3582265}}

@INPROCEEDINGS{7920855,
  author={Lu, Wenyan and Yan, Guihai and Li, Jiajun and Gong, Shijun and Han, Yinhe and Li, Xiaowei},
  booktitle={2017 IEEE International Symposium on High Performance Computer Architecture (HPCA)}, 
  title={{FlexFlow}: A Flexible Dataflow Accelerator Architecture for Convolutional Neural Networks}, 
  year={2017},
  pages={553-564},
  doi={10.1109/HPCA.2017.29},
publisher={IEEE},
address = {Piscataway, NJ, USA}}

@INPROCEEDINGS{11043330,
  author={Chen, Bo-Yu and Chang, Tian-Sheuan},
  booktitle={2025 IEEE International Symposium on Circuits and Systems (ISCAS)}, 
  title={Hardware Efficient Accelerator for Spiking Transformer With Reconfigurable Parallel Time Step Computing}, 
  year={2025},
  pages={1-5},
  doi={10.1109/ISCAS56072.2025.11043330},
publisher={IEEE},
address = {Piscataway, NJ, USA}}

@INPROCEEDINGS{8192463,
  author={Jouppi, Norman P. and Young, Cliff and Patil, Nishant and Patterson, David and Agrawal, Gaurav and Bajwa, Raminder and Bates, Sarah and Bhatia, Suresh and Boden, Nan and Borchers, Al and Boyle, Rick and Cantin, Pierre-luc and Chao, Clifford and Clark, Chris and Coriell, Jeremy and Daley, Mike and Dau, Matt and Dean, Jeffrey and Gelb, Ben and Ghaemmaghami, Tara Vazir and Gottipati, Rajendra and Gulland, William and Hagmann, Robert and Ho, C. Richard and Hogberg, Doug and Hu, John and Hundt, Robert and Hurt, Dan and Ibarz, Julian and Jaffey, Aaron and Jaworski, Alek and Kaplan, Alexander and Khaitan, Harshit and Killebrew, Daniel and Koch, Andy and Kumar, Naveen and Lacy, Steve and Laudon, James and Law, James and Le, Diemthu and Leary, Chris and Liu, Zhuyuan and Lucke, Kyle and Lundin, Alan and MacKean, Gordon and Maggiore, Adriana and Mahony, Maire and Miller, Kieran and Nagarajan, Rahul and Narayanaswami, Ravi and Ni, Ray and Nix, Kathy and Norrie, Thomas and Omernick, Mark and Penukonda, Narayana and Phelps, Andy and Ross, Jonathan and Ross, Matt and Salek, Amir and Samadiani, Emad and Severn, Chris and Sizikov, Gregory and Snelham, Matthew and Souter, Jed and Steinberg, Dan and Swing, Andy and Tan, Mercedes and Thorson, Gregory and Tian, Bo and Toma, Horia and Tuttle, Erick and Vasudevan, Vijay and Walter, Richard and Wang, Walter and Wilcox, Eric and Yoon, Doe Hyun},
  booktitle={2017 ACM/IEEE 44th Annual International Symposium on Computer Architecture (ISCA)}, 
  title={In-datacenter performance analysis of a tensor processing unit}, 
  year={2017},
  pages={1-12},
  doi={10.1145/3079856.3080246},
publisher={IEEE},
address = {Piscataway, NJ, USA}}

@ARTICLE{9469808,
  author={Jeon, Won and Lee, Jiwon and Kang, Dongseok and Kal, Hongju and Ro, Won Woo},
  journal={IEEE Access}, 
  title={{PIMCaffe}: Functional Evaluation of a Machine Learning Framework for In-Memory Neural Processing Unit}, 
  year={2021},
  volume={9},
  pages={96629-96640},
  doi={10.1109/ACCESS.2021.3094043}}

@ARTICLE{8358031,
  author={Jouppi, Norman and Young, Cliff and Patil, Nishant and Patterson, David},
  journal={IEEE Micro}, 
  title={Motivation for and Evaluation of the First Tensor Processing Unit}, 
  year={2018},
  volume={38},
  number={3},
  pages={10-19},
  doi={10.1109/MM.2018.032271057}}

@INPROCEEDINGS{6100451,
  author={Xie, Ermai and McGinnity, Martin and Wu, QingXiang and Cai, Jianyong and Cai, Rontai},
  booktitle={2011 4th International Congress on Image and Signal Processing}, 
  title={{GPU} implementation of spiking neural networks for color image segmentation}, 
  year={2011},
  volume={3},
  pages={1246-1250},
  doi={10.1109/CISP.2011.6100451},
publisher={IEEE},
address = {Piscataway, NJ, USA}}

@ARTICLE{6887358,
  author={Juang, Chia-Feng and Chen, Wei-Yuan and Liang, Chung-Wei},
  journal={IEEE Transactions on Fuzzy Systems}, 
  title={Speedup of Learning in Interval Type-2 Neural Fuzzy Systems Through Graphic Processing Units}, 
  year={2015},
  volume={23},
  number={4},
  pages={1286-1298},
  doi={10.1109/TFUZZ.2014.2353136}}

@INPROCEEDINGS{9826082,
  author={Fei, Xiang and Han, Jianhui and Huang, Jianqiang and Zheng, Weimin and Zhang, Youhui},
  booktitle={2022 22nd IEEE International Symposium on Cluster, Cloud and Internet Computing (CCGrid)}, 
  title={Accelerating Neural Network Training with Processing-in-Memory {GPU}}, 
  year={2022},
  pages={414-421},
  doi={10.1109/CCGrid54584.2022.00051},
publisher={IEEE},
address = {Piscataway, NJ, USA}}

@INPROCEEDINGS{7828382,
  author={Li, Shijie and Dou, Yong and Lv, Qi and Wang, Qiang and Niu, Xin and Yang, Ke},
  booktitle={2016 IEEE 18th International Conference on High Performance Computing and Communications; IEEE 14th International Conference on Smart City; IEEE 2nd International Conference on Data Science and Systems (HPCC/SmartCity/DSS)}, 
  title={Optimized {GPU} Acceleration Algorithm of Convolutional Neural Networks for Target Detection}, 
  year={2016},
  pages={224-230},
  doi={10.1109/HPCC-SmartCity-DSS.2016.0041},
publisher={IEEE},
address = {Piscataway, NJ, USA}}

@misc{ibm2024npu,
  author       = {{Josh Schneider and Ian Smalley}},
  title        = {{What is a neural processing unit {(NPU)}?}},
  year         = {2024},
  howpublished = {\url{https://www.ibm.com/think/topics/neural-processing-unit}},
  note         = {Published: 27 September 2024; Accessed: 2025-08-08}
}

@article{OUYANG2025106684,
title = {{PRLM}: A parallel loading mechanism for a deep neural network accelerator based on {NoC}},
journal = {Microelectronics Journal},
volume = {160},
pages = {106684},
year = {2025},
issn = {1879-2391},
doi = {https://doi.org/10.1016/j.mejo.2025.106684},
author = {Yiming Ouyang and Chengming An and Jianhua Li and HuaGuo Liang},
}

@article{WANG2025109887,
title = {An Efficient Large Kernel Convolution Network Designed for Neural Processing Unit},
journal = {Engineering Applications of Artificial Intelligence},
volume = {142},
pages = {109887},
year = {2025},
issn = {0952-1976},
doi = {https://doi.org/10.1016/j.engappai.2024.109887},
author = {Jiawen Wang and Chenfei Liao and Dewei Li and Zhongqi Zhao and Jingchuan Chen and Kehu Yang},
}

@INPROCEEDINGS{10326073,
  author={Soto-Chirinos, Jean Pierre and Condori-Alejo, Henry Ivan and Alzamora, Guina Sotomayor},
  booktitle={2023 IEEE XXX International Conference on Electronics, Electrical Engineering and Computing (INTERCON)}, 
  title={Neural Rendering in the Cloud with Tensor Processing Unit}, 
  year={2023},
  pages={1-7},
  doi={10.1109/INTERCON59652.2023.10326073},
publisher={IEEE},
address = {Piscataway, NJ, USA}}

@INPROCEEDINGS{10212679,
  author={Kim, Seongwook and Kim, Yongjun and Byeon, Gwangeun and Hong, Seokin},
  booktitle={2023 International Technical Conference on Circuits/Systems, Computers, and Communications (ITC-CSCC)}, 
  title={{CAESAR}: A {CNN} Accelerator Exploiting Sparsity and Redundancy Pattern}, 
  year={2023},
  pages={1-5},
  doi={10.1109/ITC-CSCC58803.2023.10212679},
publisher={IEEE},
address = {Piscataway, NJ, USA}}

@INPROCEEDINGS{9556422,
  author={Meng, Jian and Venkataramanaiah, Shreyas Kolala and Zhou, Chuteng and Hansen, Patrick and Whatmough, Paul and Seo, Jae-sun},
  booktitle={2021 31st International Conference on Field-Programmable Logic and Applications (FPL)}, 
  title={{FixyFPGA}: Efficient FPGA Accelerator for Deep Neural Networks with High Element-Wise Sparsity and without External Memory Access}, 
  year={2021},
  pages={9-16},
  doi={10.1109/FPL53798.2021.00010},
publisher={IEEE},
address = {Piscataway, NJ, USA}}

@ARTICLE{9120209,
  author={Yu, Ye and Jha, Niraj K.},
  journal={IEEE Transactions on Emerging Topics in Computing}, 
  title={{SPRING}: A Sparsity-Aware Reduced-Precision Monolithic {3D} {CNN} Accelerator Architecture for Training and Inference}, 
  year={2022},
  volume={10},
  number={1},
  pages={237-249},
  doi={10.1109/TETC.2020.3003328}}

@ARTICLE{10255124,
  author={Tang, Kaifei and Wang, Jiantao and Xu, Wenqu and Ji, Xiang and Liu, Jiahui and Huang, Xiaobin and Xin, Yu and Dai, Pan and Sun, Guozhu and Zeng, Zhaobang and Xiao, Rulei and Chen, Xiangfei and Jiang, Wei},
  journal={Journal of Lightwave Technology}, 
  title={Photonic Tensor Processing Unit With Single Dataflow and Programmable High-Precision Weighting Control}, 
  year={2024},
  volume={42},
  number={2},
  pages={659-669},
  doi={10.1109/JLT.2023.3317090}}

@INPROCEEDINGS{8682197,
  author={Yuan, Longhao and Li, Chao and Cao, Jianting and Zhao, Qibin},
  booktitle={ICASSP 2019 - 2019 IEEE International Conference on Acoustics, Speech and Signal Processing (ICASSP)}, 
  title={Randomized Tensor Ring Decomposition and Its Application to Large-scale Data Reconstruction}, 
  year={2019},
  pages={2127-2131},
  doi={10.1109/ICASSP.2019.8682197},
publisher={IEEE},
address = {Piscataway, NJ, USA}}

@INPROCEEDINGS{10020427,
  author={Mummoju, Pranava and Wolff, Anna and Perdacher, Martin and Plant, Claudia and Böhm, Christian},
  booktitle={2022 IEEE International Conference on Big Data (Big Data)}, 
  title={Enhancing k-Means Algorithm with Tensor Processing Unit}, 
  year={2022},
  pages={194-200},
  doi={10.1109/BigData55660.2022.10020427},
publisher={IEEE},
address = {Piscataway, NJ, USA}}

@INPROCEEDINGS{10369087,
  author={Tang, Kaifei and Wang, Jiantao and Ji, Xiang and Liu, Jiahui and Xin, Yu and Cao, Haijiang and Zeng, Zhaobang and Xiao, Rulei and Jiang, Wei},
  booktitle={2023 Asia Communications and Photonics Conference/2023 International Photonics and Optoelectronics Meetings (ACP/POEM)}, 
  title={Complete photonic tensor convolution driven by single dataflow}, 
  year={2023},
  pages={1-3},
  doi={10.1109/ACP/POEM59049.2023.10369087},
publisher={IEEE},
address = {Piscataway, NJ, USA}}

@INPROCEEDINGS{10809912,
  author={Li, Yixuan and Wang, Benshan and Xu, Tengji and Liu, Shaojie and Xiao, Qiarong and Huang, Chaoran},
  booktitle={2024 Asia Communications and Photonics Conference (ACP) and International Conference on Information Photonics and Optical Communications (IPOC)}, 
  title={Scalable and Energy-Efficient Photonic Neural Networks Through Convolution Compression}, 
  year={2024},
  pages={1-6},
  doi={10.1109/ACP/IPOC63121.2024.10809912},
publisher={IEEE},
address = {Piscataway, NJ, USA}}

@misc{lebedev2015speedingupconvolutionalneuralnetworks,
      title={Speeding-up Convolutional Neural Networks Using Fine-tuned CP-Decomposition}, 
      author={Vadim Lebedev and Yaroslav Ganin and Maksim Rakhuba and Ivan Oseledets and Victor Lempitsky},
      year={2015},
      eprint={1412.6553},
      archivePrefix={arXiv},
      primaryClass={cs.CV},
      doi={
https://doi.org/10.48550/arXiv.1412.6553}, 
}

@INPROCEEDINGS{10614519,
  author={Schwartz, Russell L. T. and Yang, Hangbo and Peserico, Nicola and Sorger, Volker J.},
  booktitle={2024 IEEE Photonics Society Summer Topicals Meeting Series (SUM)}, 
  title={The Von Neumann Bottleneck in Photonic Tensor Core Systems}, 
  year={2024},
  pages={1-2},
  doi={10.1109/SUM60964.2024.10614519},
publisher={IEEE},
address = {Piscataway, NJ, USA}}

@INPROCEEDINGS{10543636,
  author={Schwartz, Russell L. T. and Jahannia, Belal and Peserico, Nicola and Dalir, Hamed and Sorger, Volker J.},
  booktitle={2024 IEEE Silicon Photonics Conference (SiPhotonics)}, 
  title={Data Throughput for Efficient Photonic Neural Network Accelerators}, 
  year={2024},
  pages={1-2},
  doi={10.1109/SiPhotonics60897.2024.10543636},
publisher={IEEE},
address = {Piscataway, NJ, USA}}

@INPROCEEDINGS{9975540,
  author={Fardoost, Alireza and Vanani, Fatemeh Ghaedi and Zhu, Zheyuan and Doerr, Christopher and Pang, Shuo and Li, Guifang},
  booktitle={2022 IEEE Photonics Conference (IPC)}, 
  title={A High-Speed Photonic Tensor Accelerator}, 
  year={2022},
  pages={1-2},
  doi={10.1109/IPC53466.2022.9975540},
publisher={IEEE},
address = {Piscataway, NJ, USA}}

@INPROCEEDINGS{10456805,
  author={Ezekiel, Aniebiet Micheal and Onwuchekwa, Daniel and Obermaisser, Roman},
  booktitle={2023 26th Euromicro Conference on Digital System Design (DSD)}, 
  title={Optimization of the Versatile Tensor Accelerator ({VTA)} Load Module in a Time-Triggered Memory Access}, 
  year={2023},
  pages={146-152},
  doi={10.1109/DSD60849.2023.00030},
publisher={IEEE},
address = {Piscataway, NJ, USA}}

@INPROCEEDINGS{10946782,
  author={Lin, Yujun and Zhang, Zhekai and Han, Song},
  booktitle={2025 IEEE International Symposium on High Performance Computer Architecture (HPCA)}, 
  title={{LEGO}: Spatial Accelerator Generation and Optimization for Tensor Applications}, 
  year={2025},
  pages={1335-1347},
  doi={10.1109/HPCA61900.2025.00101},
publisher={IEEE},
address = {Piscataway, NJ, USA}}

@INPROCEEDINGS{10323722,
  author={Xie, Xi and Peng, Hongwu and Hasan, Amit and Huang, Shaoyi and Zhao, Jiahui and Fang, Haowen and Zhang, Wei and Geng, Tong and Khan, Omer and Ding, Caiwen},
  booktitle={2023 IEEE/ACM International Conference on Computer Aided Design (ICCAD)}, 
  title={{Accel-GCN}: High-Performance {GPU} Accelerator Design for Graph Convolution Networks}, 
  year={2023},
  pages={01-09},
  doi={10.1109/ICCAD57390.2023.10323722},
publisher={IEEE},
address = {Piscataway, NJ, USA}}

@ARTICLE{10620224,
  author={Kahng, Andrew B. and Wang, Zhiang},
  journal={IEEE Transactions on Computer-Aided Design of Integrated Circuits and Systems}, 
  title={{DG-RePlAce}: A Dataflow-Driven {GPU}-Accelerated Analytical Global Placement Framework for Machine Learning Accelerators}, 
  year={2025},
  volume={44},
  number={2},
  pages={696-708},
  doi={10.1109/TCAD.2024.3436521}}

@INPROCEEDINGS{8988598,
  author={Guo, Jinrong and Liu, Wantao and Wang, Wang and Yao, Chunrong and Han, Jizhong and Li, Ruixuan and Lu, Yijun and Hu, Songlin},
  booktitle={2019 IEEE 37th International Conference on Computer Design (ICCD)}, 
  title={{AccUDNN}: A {GPU} Memory Efficient Accelerator for Training Ultra-Deep Neural Networks}, 
  year={2019},
  pages={65-72},
  doi={10.1109/ICCD46524.2019.00017},
publisher={IEEE},
address = {Piscataway, NJ, USA}}

@INPROCEEDINGS{9586181,
  author={Zhou, Zhe and Shi, Bizhao and Zhang, Zhe and Guan, Yijin and Sun, Guangyu and Luo, Guojie},
  booktitle={2021 58th ACM/IEEE Design Automation Conference (DAC)}, 
  title={{BlockGNN}: Towards Efficient {GNN} Acceleration Using Block-Circulant Weight Matrices}, 
  year={2021},
  pages={1009-1014},
  doi={10.1109/DAC18074.2021.9586181},
publisher={IEEE},
address = {Piscataway, NJ, USA}}

@INPROCEEDINGS{8742271,
  author={Diamantopoulos, Dionysios and Hagleitner, Christoph},
  booktitle={2018 International Conference on Field-Programmable Technology (FPT)}, 
  title={A System-Level Transprecision {FPGA} Accelerator for {BLSTM} Using On-chip Memory Reshaping}, 
  year={2018},
  pages={338-341},
  doi={10.1109/FPT.2018.00068},
publisher={IEEE},
address = {Piscataway, NJ, USA}}

@INPROCEEDINGS{10595963,
  author={Khan, Fatima Hameed and Adeel Pasha, Muhammad and Masud, Shahid},
  booktitle={2024 IEEE 6th International Conference on AI Circuits and Systems (AICAS)}, 
  title={Exploring Memory Access Techniques for Efficient {FPGA} based {3D CNN} Accelerator Design}, 
  year={2024},
  pages={218-222},
  doi={10.1109/AICAS59952.2024.10595963},
publisher={IEEE},
address = {Piscataway, NJ, USA}}

@INPROCEEDINGS{6239808,
  author={Cheng, Shaoyi and Lin, Mingjie and Liu, Hao Jun and Scott, Simon and Wawrzynek, John},
  booktitle={2012 IEEE 20th International Symposium on Field-Programmable Custom Computing Machines}, 
  title={Exploiting Memory-Level Parallelism in Reconfigurable Accelerators}, 
  year={2012},
  pages={157-160},
  doi={10.1109/FCCM.2012.35},
publisher={IEEE},
address = {Piscataway, NJ, USA}}

@ARTICLE{9826871,
  author={Kokkinis, Argyris and Diamantopoulos, Dionysios and Siozios, Kostas},
  journal={IEEE Computer Architecture Letters}, 
  title={Dynamic Optimization of On-Chip Memories for {HLS} Targeting Many-Accelerator Platforms}, 
  year={2022},
  volume={21},
  number={2},
  pages={41-44},
  doi={10.1109/LCA.2022.3190048}}

@INPROCEEDINGS{9951617,
  author={Tao, Guanchen and Li, Yonggen and Xu, Yanfeng and Fan, Jicong and Shen, Haibin and Huang, Kejie},
  booktitle={2022 2nd International Conference on Frontiers of Electronics, Information and Computation Technologies (ICFEICT)}, 
  title={A Near Memory Computing {FPGA} Architecture for Neural Network Acceleration}, 
  year={2022},
  pages={543-548},
  doi={10.1109/ICFEICT57213.2022.00100},
publisher={IEEE},
address = {Piscataway, NJ, USA}}

@INPROCEEDINGS{8839401,
  author={Dhar, Ashutosh and Huang, Sitao and Xiong, Jinjun and Jamsek, Damir and Mesnet, Bruno and Huang, Jian and Kim, Nam Sung and Hwu, Wen-mei and Chen, Deming},
  booktitle={2019 IEEE Computer Society Annual Symposium on VLSI (ISVLSI)}, 
  title={Near-Memory and In-Storage {FPGA} Acceleration for Emerging Cognitive Computing Workloads}, 
  year={2019},
  pages={68-75},
  doi={10.1109/ISVLSI.2019.00021},
publisher={IEEE},
address = {Piscataway, NJ, USA}}

@article{LIU2024106197,
title = {Improving the computational efficiency and flexibility of {FPGA}-based {CNN} accelerator through loop optimization},
journal = {Microelectronics Journal},
volume = {147},
pages = {106197},
year = {2024},
issn = {1879-2391},
doi = {https://doi.org/10.1016/j.mejo.2024.106197},
author = {Yuhao Liu and Yanhua Ma and Bowei Zhang and Lu Liu and Jie Wang and Shibo Tang},
}

@INPROCEEDINGS{10702206,
  author={Zhang, Qianqian and Zhang, Xin},
  booktitle={2024 20th International Conference on Natural Computation, Fuzzy Systems and Knowledge Discovery (ICNC-FSKD)}, 
  title={Design of A Low-Latency General-Purpose CNN Hardware Accelerator Based on Pulsed Arrays on {FPGAs}}, 
  year={2024},
  pages={1-8},
  doi={10.1109/ICNC-FSKD64080.2024.10702206},
publisher={IEEE},
address = {Piscataway, NJ, USA}}

@INPROCEEDINGS{10900711,
  author={Rai, Himanshu and Sridhar, Aishwarya and Ecker, Wolfgang and Rao, Nanditha},
  booktitle={2025 38th International Conference on VLSI Design and 2024 23rd International Conference on Embedded Systems (VLSID)}, 
  title={{FPUGen}: A FrameWork to Generate Custom Floating Point {FMA} Accelerators on {FPGAs}}, 
  year={2025},
  pages={225-230},
  doi={10.1109/VLSID64188.2025.00052},
publisher={IEEE},
address = {Piscataway, NJ, USA}}

@INPROCEEDINGS{10822997,
  author={Guo, Shaoshan and Huang, Yunqian and Mai, Kailan and Mo, Guopeng},
  booktitle={2024 6th International Conference on Artificial Intelligence and Computer Applications (ICAICA)}, 
  title={A Superscalar Six-Stage Pipeline Neural Network Accelerator: Design and Implementation Based on {FPGA}}, 
  year={2024},
  pages={126-133},
  doi={10.1109/ICAICA63239.2024.10822997},
publisher={IEEE},
address = {Piscataway, NJ, USA}}

@INPROCEEDINGS{9360939,
  author={Kang, Lei and Li, Hui and Li, Xin and Zheng, Haowei},
  booktitle={2020 2nd International Conference on Machine Learning, Big Data and Business Intelligence (MLBDBI)}, 
  title={Design of Convolution Operation Accelerator based on {FPGA}}, 
  year={2020},
  pages={80-84},
  doi={10.1109/MLBDBI51377.2020.00021},
publisher={IEEE},
address = {Piscataway, NJ, USA}}

@ARTICLE{9211770,
  author={Baranwal, Akhil Raj and Ullah, Salim and Sahoo, Siva Satyendra and Kumar, Akash},
  journal={IEEE Transactions on Computer-Aided Design of Integrated Circuits and Systems}, 
  title={ReLAccS: A Multilevel Approach to Accelerator Design for Reinforcement Learning on FPGA-Based Systems}, 
  year={2021},
  volume={40},
  number={9},
  pages={1754-1767},
  doi={10.1109/TCAD.2020.3028350}}

@INPROCEEDINGS{10871700,
  author={Wang, Bo and Han, Qing and Shi, Xiupeng},
  booktitle={2024 International Conference on Intelligent Robotics and Automatic Control (IRAC)}, 
  title={Multi-Objective Optimization of {ML} Algorithms and {FPGA} Accelerators for Robotic Intelligence}, 
  year={2024},
  pages={130-137},
  doi={10.1109/IRAC63143.2024.10871700},
publisher={IEEE},
address = {Piscataway, NJ, USA}}

@INPROCEEDINGS{7927161,
  author={Zhong, Guanwen and Prakash, Alok and Wang, Siqi and Liang, Yun and Mitra, Tulika and Niar, Smail},
  booktitle={Design, Automation \& Test in Europe Conference \& Exhibition (DATE), 2017}, 
  title={Design Space exploration of {FPGA}-based accelerators with multi-level parallelism}, 
  year={2017},
  pages={1141-1146},
  doi={10.23919/DATE.2017.7927161},
publisher={IEEE},
address = {Piscataway, NJ, USA}}

@INPROCEEDINGS{10082175,
  author={Wu, Yuhao},
  booktitle={2023 IEEE 6th Information Technology,Networking,Electronic and Automation Control Conference (ITNEC)}, 
  title={Review on {FPGA}-Based Accelerators in Deep learning}, 
  year={2023},
  volume={6},
  pages={452-456},
  doi={10.1109/ITNEC56291.2023.10082175},
publisher={IEEE},
address = {Piscataway, NJ, USA}}

@article{LIU2024128511,
title = {Review of neural network model acceleration techniques based on {FPGA} platforms},
journal = {Neurocomputing},
volume = {610},
pages = {128511},
year = {2024},
issn = {0925-2312},
doi = {https://doi.org/10.1016/j.neucom.2024.128511},
author = {Fang Liu and Heyuan Li and Wei Hu and Yanxiang He},
}

@ARTICLE{11105416,
  author={Fata, John and Elmannai, Wafa and Elleithy, Khaled},
  journal={IEEE Access}, 
  title={Balancing Performance and Cost—{FPGA}-Based {CNN} Accelerators for Edge Computing: Status Quo, Key Challenges, and Prospective Innovations}, 
  volume={13},
  number={},
  year={2025},
  pages={1-1},
}

@INPROCEEDINGS{8977882,
  author={Sun, Yunfei and Liu, Brian and Xu, Xianchao},
  booktitle={2019 International Conference on Field-Programmable Technology (ICFPT)}, 
  title={An {OpenCL}-Based Hybrid {CNN-RNN} Inference Accelerator On {FPGA}}, 
  year={2019},
  pages={283-286},
  doi={10.1109/ICFPT47387.2019.00048},
publisher={IEEE},
address = {Piscataway, NJ, USA}}

@ARTICLE{11027895,
  author={Pacini, Tommaso and Nannipieri, Pietro and Moranti, Silvia and Fanucci, Luca},
  journal={IEEE Access}, 
  title={{FPG-AI RNN}: A Technology-Agnostic Framework for the Automatic Acceleration of {LSTM/GRU}-Based Models on {FPGAs}}, 
  year={2025},
  volume={13},
  pages={100353-100369},
  doi={10.1109/ACCESS.2025.3577908}}

@ARTICLE{10041123,
  author={Kim, Taesu and Ahn, Daehyun and Lee, Dongsoo and Kim, Jae-Joon},
  journal={IEEE Transactions on Computer-Aided Design of Integrated Circuits and Systems}, 
  title={{V-LSTM}: An Efficient {LSTM} Accelerator Using Fixed Nonzero-Ratio Viterbi-Based Pruning}, 
  year={2023},
  volume={42},
  number={10},
  pages={3327-3337},
  doi={10.1109/TCAD.2023.3243879}}

@INPROCEEDINGS{8780644,
  author={Gao, Chen and Zhang, Fan},
  booktitle={2018 IEEE 4th International Conference on Computer and Communications (ICCC)}, 
  title={{FPGA}-based Accelerator for Independently Recurrent Neural Network}, 
  year={2018},
  pages={2075-2080},
  doi={10.1109/CompComm.2018.8780644},
publisher={IEEE},
address = {Piscataway, NJ, USA}}

@INPROCEEDINGS{8050816,
  author={Chang, Andre Xian Ming and Culurciello, Eugenio},
  booktitle={2017 IEEE International Symposium on Circuits and Systems (ISCAS)}, 
  title={Hardware accelerators for recurrent neural networks on {FPGA}}, 
  year={2017},
  pages={1-4},
  doi={10.1109/ISCAS.2017.8050816},
publisher={IEEE},
address = {Piscataway, NJ, USA}}

@article{JIANG2022104417,
title = {A low-latency {LSTM} accelerator using balanced sparsity based on {FPGA}},
journal = {Microprocessors and Microsystems},
volume = {89},
pages = {104417},
year = {2022},
issn = {0141-9331},
doi = {https://doi.org/10.1016/j.micpro.2021.104417},
author = {Jingfei Jiang and Tao Xiao and Jinwei Xu and Dong Wen and Lei Gao and Yong Dou},
}

@article{GUO2025128871,
title = {{FPGA}-based component-wise {LSTM} training accelerator for neural granger causality analysis},
journal = {Neurocomputing},
volume = {615},
pages = {128871},
year = {2025},
issn = {0925-2312},
doi = {https://doi.org/10.1016/j.neucom.2024.128871},
author = {Chuliang Guo and Yufei Chen and Yu Fu},
}

@article{10.1063/5.0168089,
    author = {Le Gallo, Manuel and Lammie, Corey and Büchel, Julian and Carta, Fabio and Fagbohungbe, Omobayode and Mackin, Charles and Tsai, Hsinyu and Narayanan, Vijay and Sebastian, Abu and El Maghraoui, Kaoutar and Rasch, Malte J.},
    title = {Using the {IBM} analog in-memory hardware acceleration kit for neural network training and inference},
    journal = {APL Machine Learning},
    volume = {1},
    number = {4},
    pages = {041102},
    year = {2023},
    month = {11},
    issn = {2770-9019},
    doi = {10.1063/5.0168089},
}

@INPROCEEDINGS{8742284,
  author={Xie, Liang and Fan, Xitian and Cao, Wei and Wang, Lingli},
  booktitle={2018 International Conference on Field-Programmable Technology (FPT)}, 
  title={High Throughput {CNN} Accelerator Design Based on {FPGA}}, 
  year={2018},
  pages={274-277},
  doi={10.1109/FPT.2018.00052},
publisher={IEEE},
address = {Piscataway, NJ, USA}}

@inproceedings{10.1145/2684746.2689060,
author = {Zhang, Chen and Li, Peng and Sun, Guangyu and Guan, Yijin and Xiao, Bingjun and Cong, Jason},
title = {Optimizing {FPGA}-based Accelerator Design for Deep Convolutional Neural Networks},
year = {2015},
isbn = {9781450333153},
publisher = {Association for Computing Machinery},
address = {New York, NY, USA},
doi = {10.1145/2684746.2689060},
booktitle = {Proceedings of the 2015 ACM/SIGDA International Symposium on Field-Programmable Gate Arrays},
pages = {161–170},
numpages = {10},
location = {Monterey, California, USA},
series = {FPGA '15}
}

@ARTICLE{10148988,
  author={Kim, Victoria Heekyung and Choi, Kyuwon Ken},
  journal={IEEE Access}, 
  title={A Reconfigurable {CNN}-Based Accelerator Design for Fast and Energy-Efficient Object Detection System on Mobile {FPGA}}, 
  year={2023},
  volume={11},
  pages={59438-59445},
  doi={10.1109/ACCESS.2023.3285279}}

@INPROCEEDINGS{8892195,
  author={Kolala Venkataramanaiah, Shreyas and Ma, Yufei and Yin, Shihui and Nurvithadhi, Eriko and Dasu, Aravind and Cao, Yu and Seo, Jae-Sun},
  booktitle={2019 29th International Conference on Field Programmable Logic and Applications (FPL)}, 
  title={Automatic Compiler Based {FPGA} Accelerator for {CNN} Training}, 
  year={2019},
  pages={166-172},
  doi={10.1109/FPL.2019.00034},
publisher={IEEE},
address = {Piscataway, NJ, USA}}

@ARTICLE{8558097,
  author={Ma, Yufei and Cao, Yu and Vrudhula, Sarma and Seo, Jae-Sun},
  journal={IEEE Transactions on Computer-Aided Design of Integrated Circuits and Systems}, 
  title={Automatic Compilation of Diverse {CNNs} Onto High-Performance {FPGA} Accelerators}, 
  year={2020},
  volume={39},
  number={2},
  pages={424-437},
  doi={10.1109/TCAD.2018.2884972}}

@INPROCEEDINGS{9465504,
  author={Varadarajulu, Swetha and Mariammal, K.},
  booktitle={2021 5th International Conference on Computer, Communication and Signal Processing (ICCCSP)}, 
  title={Design of {SentiNet RTL} Library for {CNN} based Hardware Accelerator}, 
  year={2021},
  volume={},
  number={},
  pages={100-108},
  doi={10.1109/ICCCSP52374.2021.9465504},
publisher={IEEE},
address = {Piscataway, NJ, USA}}

@ARTICLE{9610053,
  author={El-Maksoud, Ahmed J. Abd and Ebbed, Mohamed and Khalil, Ahmed H. and Mostafa, Hassan},
  journal={IEEE Access}, 
  title={Power Efficient Design of High-Performance Convolutional Neural Networks Hardware Accelerator on {FPGA}: A Case Study With {GoogLeNet}}, 
  year={2021},
  volume={9},
  number={},
  pages={151897-151911},
  doi={10.1109/ACCESS.2021.3126838}}

@INPROCEEDINGS{10927809,
  author={G, Sakthi and Tripathi, Abhishek N},
  booktitle={2025 International Conference on Computational, Communication and Information Technology (ICCCIT)}, 
  title={Systolic Array Design for Efficient {FPGA} Implementation of {CNN} Accelerators: Power and Area Optimizations}, 
  year={2025},
  pages={483-485},
  doi={10.1109/ICCCIT62592.2025.10927809},
publisher={IEEE},
address = {Piscataway, NJ, USA}}

@INPROCEEDINGS{5537908,
  author={Farabet, Clément and Martini, Berin and Akselrod, Polina and Talay, Selçuk and LeCun, Yann and Culurciello, Eugenio},
  booktitle={Proceedings of 2010 IEEE International Symposium on Circuits and Systems}, 
  title={Hardware accelerated convolutional neural networks for synthetic vision systems}, 
  year={2010},
  pages={257-260},
  doi={10.1109/ISCAS.2010.5537908},
publisher={IEEE},
address = {Piscataway, NJ, USA}}

@INPROCEEDINGS{5200010,
  author={Sankaradas, Murugan and Jakkula, Venkata and Cadambi, Srihari and Chakradhar, Srimat and Durdanovic, Igor and Cosatto, Eric and Graf, Hans Peter},
  booktitle={2009 20th IEEE International Conference on Application-specific Systems, Architectures and Processors}, 
  title={A Massively Parallel Coprocessor for Convolutional Neural Networks}, 
  year={2009},
  pages={53-60},
  doi={10.1109/ASAP.2009.25},
publisher={IEEE},
address = {Piscataway, NJ, USA}}

@inproceedings{NIPS2008_31fefc0e,
 author = {Graf, Hans and Cadambi, Srihari and Jakkula, Venkata and Sankaradass, Murugan and Cosatto, Eric and Chakradhar, Srimat and Dourdanovic, Igor},
 booktitle = {Advances in Neural Information Processing Systems},
 editor = {D. Koller and D. Schuurmans and Y. Bengio and L. Bottou},
 publisher = {Curran Associates Inc.},
address = {Red Hook, NY, USA},
 title = {A Massively Parallel Digital Learning Processor},
 volume = {21},
 year = {2008},
pages = {529–536},
numpages = {8},
}

@ARTICLE{11021575,
  author={Petropoulos, Anastasios and Antonakopoulos, Theodore},
  journal={IEEE Transactions on Very Large Scale Integration (VLSI) Systems}, 
  title={A Scalable {FPGA} Architecture With Adaptive Memory Utilization for {GEMM}-Based Operations}, 
  year={2025},
  volume={33},
  number={8},
  pages={2334-2338},
  doi={10.1109/TVLSI.2025.3571677}}

@misc{rybalkin2018finnllibraryextensionsdesign,
      title={{FINN-L}: Library Extensions and Design Trade-off Analysis for Variable Precision {LSTM} Networks on {FPGAs}}, 
      author={Vladimir Rybalkin and Alessandro Pappalardo and Muhammad Mohsin Ghaffar and Giulio Gambardella and Norbert Wehn and Michaela Blott},
      year={2018},
      eprint={1807.04093},
      archivePrefix={arXiv},
      primaryClass={cs.CV},
}

@ARTICLE{10742088,
  author={Khalil, Kasem and Dey, Bappaditya and Bayoumi, Magdy},
  journal={IEEE Internet of Things Journal}, 
  title={{S²RNN}: Self-Supervised Reconfigurable Neural Network Hardware Accelerator for Machine Learning Applications}, 
  year={2025},
  volume={12},
  number={6},
  pages={6708-6720},
  doi={10.1109/JIOT.2024.3490893}}

@INPROCEEDINGS{10072756,
  author={Wang, Hao and Qiu, Danfeng and Ge, Fen and Yang, Ying},
  booktitle={2022 IEEE 22nd International Conference on Communication Technology (ICCT)}, 
  title={Implementation of Bidirectional {LSTM} Accelerator Based on {FPGA}}, 
  year={2022},
  pages={1512-1516},
  doi={10.1109/ICCT56141.2022.10072756},
publisher={IEEE},
address = {Piscataway, NJ, USA}}

@INPROCEEDINGS{7858394,
  author={Guan, Yijin and Yuan, Zhihang and Sun, Guangyu and Cong, Jason},
  booktitle={2017 22nd Asia and South Pacific Design Automation Conference (ASP-DAC)}, 
  title={{FPGA}-based accelerator for long short-term memory recurrent neural networks}, 
  year={2017},
  pages={629-634},
  doi={10.1109/ASPDAC.2017.7858394},
publisher={IEEE},
address = {Piscataway, NJ, USA}}

@ARTICLE{10818746,
  author={Zhang, Chen and Cao, Shijie and Dai, Guohao and Geng, Chenbo and Yao, Zhuliang and Xiao, Wencong and Liu, Yunxin and Wu, Ming and Zhang, Lintao and Sun, Guangyu and Ji, Zhigang and Wang, Runsheng and Huang, Ru},
  journal={IEEE Transactions on Computer-Aided Design of Integrated Circuits and Systems}, 
  title={Fine-Grained Structured Sparse Computing for {FPGA}-Based {AI} Inference}, 
  year={2025},
  volume={44},
  number={7},
  pages={2544-2557},
  doi={10.1109/TCAD.2024.3524356},
publisher={IEEE},
address = {Piscataway, NJ, USA}}

@ARTICLE{9268992,
  author={Gao, Chang and Rios-Navarro, Antonio and Chen, Xi and Liu, Shih-Chii and Delbruck, Tobi},
  journal={IEEE Journal on Emerging and Selected Topics in Circuits and Systems}, 
  title={{EdgeDRNN}: Recurrent Neural Network Accelerator for Edge Inference}, 
  year={2020},
  volume={10},
  number={4},
  pages={419-432},
  doi={10.1109/JETCAS.2020.3040300}}

@article{ABDELBAKY2025155845,
title = {High-performance {FPGA}-accelerated {LSTM} neural network for chaotic time series prediction},
journal = {AEU - International Journal of Electronics and Communications},
volume = {199},
pages = {155845},
year = {2025},
issn = {1434-8411},
doi = {https://doi.org/10.1016/j.aeue.2025.155845},
author = {Mahmoud H. AbdElbaky and Mohammed H. Yacoub and Wafaa S. Sayed and Lobna A. Said},
}

@INPROCEEDINGS{9295665,
  author={Zhang, Weifeng and Ge, Fen and Cui, Chenchen and Yang, Ying and Zhou, Fang and Wu, Ning},
  booktitle={2020 IEEE 20th International Conference on Communication Technology (ICCT)}, 
  title={Design and Implementation of {LSTM} Accelerator Based on {FPGA}}, 
  year={2020},
  pages={1675-1679},
  doi={10.1109/ICCT50939.2020.9295665},
publisher={IEEE},
address = {Piscataway, NJ, USA}}

@inproceedings{10.1145/3174243.3174261,
author = {Gao, Chang and Neil, Daniel and Ceolini, Enea and Liu, Shih-Chii and Delbruck, Tobi},
title = {{DeltaRNN}: A Power-efficient Recurrent Neural Network Accelerator},
year = {2018},
isbn = {9781450356145},
publisher = {Association for Computing Machinery},
address = {New York, NY, USA},
doi = {10.1145/3174243.3174261},
booktitle = {Proceedings of the 2018 ACM/SIGDA International Symposium on Field-Programmable Gate Arrays},
pages = {21–30},
numpages = {10},
location = {Monterey, CALIFORNIA, USA},
series = {FPGA '18}
}

@misc{han2017eseefficientspeechrecognition,
      title={{ESE}: Efficient Speech Recognition Engine with Sparse {LSTM} on {FPGA}}, 
      author={Song Han and Junlong Kang and Huizi Mao and Yiming Hu and Xin Li and Yubin Li and Dongliang Xie and Hong Luo and Song Yao and Yu Wang and Huazhong Yang and William J. Dally},
      year={2017},
      eprint={1612.00694},
      archivePrefix={arXiv},
      primaryClass={cs.CL},
}

@ARTICLE{8330049,
  author={Ma, Yufei and Cao, Yu and Vrudhula, Sarma and Seo, Jae-sun},
  journal={IEEE Transactions on Very Large Scale Integration (VLSI) Systems}, 
  title={Optimizing the Convolution Operation to Accelerate Deep Neural Networks on {FPGA}}, 
  year={2018},
  volume={26},
  number={7},
  pages={1354-1367},
  doi={10.1109/TVLSI.2018.2815603}}

@INPROCEEDINGS{7966660,
  author={Lu, Liqiang and Liang, Yun and Xiao, Qingcheng and Yan, Shengen},
  booktitle={2017 IEEE 25th Annual International Symposium on Field-Programmable Custom Computing Machines (FCCM)}, 
  title={Evaluating Fast Algorithms for Convolutional Neural Networks on {FPGAs}}, 
  year={2017},
  pages={101-108},
  doi={10.1109/FCCM.2017.64},
publisher={IEEE},
address = {Piscataway, NJ, USA}}

@INPROCEEDINGS{10867080,
  author={Chen, Defu and Liu, Xiaohu and Sang, Yijian and Zhou, Hao and Lu, Xiaoai and Song, Jingyang},
  booktitle={2024 4th International Conference on Electronic Information Engineering and Computer (EIECT)}, 
  title={A {FPGA}-Based Accelerator Design for Speaker Verification using {1D-CNN}}, 
  year={2024},
  pages={80-84},
  doi={10.1109/EIECT64462.2024.10867080},
publisher={IEEE},
address = {Piscataway, NJ, USA}}

@inproceedings{10.1145/3020078.3021698,
author = {Zhang, Jialiang and Li, Jing},
title = {Improving the Performance of {OpenCL}-based {FPGA} Accelerator for Convolutional Neural Network},
year = {2017},
isbn = {9781450343541},
publisher = {Association for Computing Machinery},
address = {New York, NY, USA},
doi = {10.1145/3020078.3021698},
booktitle = {Proceedings of the 2017 ACM/SIGDA International Symposium on Field-Programmable Gate Arrays},
pages = {25–34},
numpages = {10},
location = {Monterey, California, USA},
series = {FPGA '17}
}

@misc{aydonat2017opencltmdeeplearningaccelerator,
      title={An {OpenCL(TM) Deep Learning Accelerator on Arria 10}}, 
      author={Utku Aydonat and Shane O'Connell and Davor Capalija and Andrew C. Ling and Gordon R. Chiu},
      year={2017},
      eprint={1701.03534},
      archivePrefix={arXiv},
      primaryClass={cs.DC},
}

@INPROCEEDINGS{8056824,
  author={Ma, Yufei and Cao, Yu and Vrudhula, Sarma and Seo, Jae-sun},
  booktitle={2017 27th International Conference on Field Programmable Logic and Applications (FPL)}, 
  title={An automatic {RTL compiler for high-throughput FPGA} implementation of diverse deep convolutional neural networks}, 
  year={2017},
  pages={1-8},
  doi={10.23919/FPL.2017.8056824},
publisher={IEEE},
address = {Piscataway, NJ, USA}}

@INPROCEEDINGS{7851527,
  author={Cadambi, Srihari and Majumdar, Abhinandan and Becchi, Michela and Chakradhar, Srimat and Graf, Hans Peter},
  booktitle={2010 19th International Conference on Parallel Architectures and Compilation Techniques (PACT)}, 
  title={A programmable parallel accelerator for learning and classification}, 
  year={2010},
  pages={273-283},
  doi={},
publisher={IEEE},
address = {Piscataway, NJ, USA}}

@INPROCEEDINGS{10241078,
  author={Ye, Jinlin and Zhang, Wei},
  booktitle={2023 42nd Chinese Control Conference (CCC)}, 
  title={A Scalable {ARM+FPGA-Based CNN} Accelerator with Limited Hardware Resources}, 
  year={2023},
  pages={2498-2503},
  doi={10.23919/CCC58697.2023.10241078},
publisher={IEEE},
address = {Piscataway, NJ, USA}}

@INPROCEEDINGS{9421655,
  author={Lv, Peng and Liu, Wei and Li, Jinghui},
  booktitle={2020 5th International Conference on Mechanical, Control and Computer Engineering (ICMCCE)}, 
  title={A {FPGA-based accelerator implementaion for YOLOv2} object detection using Winograd algorithm}, 
  year={2020},
  pages={1894-1898},
  doi={10.1109/ICMCCE51767.2020.00415},
publisher={IEEE},
address = {Piscataway, NJ, USA}}

@ARTICLE{9476039,
  author={Li, Jixuan and Un, Ka-Fai and Yu, Wei-Han and Mak, Pui-In and Martins, Rui P.},
  journal={IEEE Transactions on Circuits and Systems II: Express Briefs}, 
  title={An {FPGA}-Based Energy-Efficient Reconfigurable Convolutional Neural Network Accelerator for Object Recognition Applications}, 
  year={2021},
  volume={68},
  number={9},
  pages={3143-3147},
  doi={10.1109/TCSII.2021.3095283}}

@ARTICLE{8856248,
  author={Kala, S. and Jose, Babita R. and Mathew, Jimson and Nalesh, S.},
  journal={IEEE Transactions on Very Large Scale Integration (VLSI) Systems}, 
  title={High-Performance {CNN} Accelerator on {FPGA} Using Unified Winograd-{GEMM} Architecture}, 
  year={2019},
  volume={27},
  number={12},
  pages={2816-2828},
  doi={10.1109/TVLSI.2019.2941250}}

@inproceedings{10.1145/3316781.3326334,
author = {Ajayi, Tutu and Chhabria, Vidya A. and Foga\c{c}a, Mateus and Hashemi, Soheil and Hosny, Abdelrahman and Kahng, Andrew B. and Kim, Minsoo and Lee, Jeongsup and Mallappa, Uday and Neseem, Marina and Pradipta, Geraldo and Reda, Sherief and Saligane, Mehdi and Sapatnekar, Sachin S. and Sechen, Carl and Shalan, Mohamed and Swartz, William and Wang, Lutong and Wang, Zhehong and Woo, Mingyu and Xu, Bangqi},
title = {Toward an Open-Source Digital Flow: First Learnings from the {OpenROAD} Project},
year = {2019},
isbn = {9781450367257},
publisher = {Association for Computing Machinery},
address = {New York, NY, USA},
doi = {10.1145/3316781.3326334},
booktitle = {Proceedings of the 56th Annual Design Automation Conference 2019},
articleno = {76},
numpages = {4},
location = {Las Vegas, NV, USA},
series = {DAC '19}
}

@ARTICLE{10372220,
  author={Kahng, Andrew B. and Varadarajan, Ravi and Wang, Zhiang},
  journal={IEEE Transactions on Computer-Aided Design of Integrated Circuits and Systems}, 
  title={{Hier-RTLMP}: A Hierarchical Automatic Macro Placer for Large-Scale Complex {IP} Blocks}, 
  year={2024},
  volume={43},
  number={5},
  pages={1552-1565},

  doi={10.1109/TCAD.2023.3346284}}

@INPROCEEDINGS{7753296,
  author={Han, Xushen and Zhou, Dajiang and Wang, Shihao and Kimura, Shinji},
  booktitle={2016 IEEE 34th International Conference on Computer Design (ICCD)}, 
  title={{CNN-MERP}: An {FPGA}-based memory-efficient reconfigurable processor for forward and backward propagation of convolutional neural networks}, 
  year={2016},
  volume={},
  number={},
  pages={320-327},
  doi={10.1109/ICCD.2016.7753296},
publisher={IEEE},
address = {Piscataway, NJ, USA}}

@inproceedings{markidis2018nvidia,
  title={Nvidia tensor core programmability, performance \& precision},
  author={Markidis, Stefano and Der Chien, Steven Wei and Laure, Erwin and Peng, Ivy Bo and Vetter, Jeffrey S},
  booktitle={2018 IEEE international parallel and distributed processing symposium workshops (IPDPSW)},
  pages={522--531},
  year={2018},
  publisher={IEEE},
address      = {Vancouver, BC, Canada}
}

@article{shi2018graph,
  title={Graph processing on GPUs: A survey},
  author={Shi, Xuanhua and Zheng, Zhigao and Zhou, Yongluan and Jin, Hai and He, Ligang and Liu, Bo and Hua, Qiang-Sheng},
  journal={ACM Computing Surveys (CSUR)},
  volume={50},
  number={6},
  pages={1--35},
  year={2018},
  publisher={ACM New York, NY, USA}
}

@inproceedings{jouppi2017datacenter,
  title={In-datacenter performance analysis of a tensor processing unit},
  author={Jouppi, Norman P and Young, Cliff and Patil, Nishant and Patterson, David and Agrawal, Gaurav and Bajwa, Raminder and Bates, Sarah and Bhatia, Suresh and Boden, Nan and Borchers, Al and others},
  booktitle={Proceedings of the 44th annual international symposium on computer architecture},
  pages={1--12},
  year={2017},
publisher={IEEE},
address = {Piscataway, NJ, USA}
}

@article{chen2016eyeriss,
  title={Eyeriss: A spatial architecture for energy-efficient dataflow for convolutional neural networks},
  author={Chen, Yu-Hsin and Emer, Joel and Sze, Vivienne},
  journal={ACM SIGARCH computer architecture news},
  volume={44},
  number={3},
  pages={367--379},
  year={2016},
  publisher = {Curran Associates Inc.},
address = {Red Hook, NY, USA}
}

@inproceedings{zhang2015optimizing,
  title={Optimizing FPGA-based accelerator design for deep convolutional neural networks},
  author={Zhang, Chen and Li, Peng and Sun, Guangyu and Guan, Yijin and Xiao, Bingjun and Cong, Jason},
  booktitle={Proceedings of the 2015 ACM/SIGDA international symposium on field-programmable gate arrays},
  pages={161--170},
  year={2015},
publisher = {Curran Associates Inc.},
address = {Red Hook, NY, USA}
}

@article{venieris2018toolflows,
  title={Toolflows for mapping convolutional neural networks on FPGAs: A survey and future directions},
  author={Venieris, Stylianos I and Kouris, Alexandros and Bouganis, Christos-Savvas},
  journal={ACM Computing Surveys (CSUR)},
  volume={51},
  number={3},
  pages={1--39},
  year={2018},
  publisher={ACM New York, NY, USA}
}

@INPROCEEDINGS{10596361,
  author={Perri, Stefania and Zambelli, Cristian and Ielmini, Daniele and Silvano, Cristina},
  booktitle={2024 IEEE International Parallel and Distributed Processing Symposium Workshops (IPDPSW)}, 
  title={Digital In-Memory Computing to Accelerate Deep Learning Inference on the Edge}, 
  year={2024},
  pages={130-133},
  doi={10.1109/IPDPSW63119.2024.00037},
publisher={IEEE},
address = {Piscataway, NJ, USA}}

@ARTICLE{7738524,
  author={Chen, Yu-Hsin and Krishna, Tushar and Emer, Joel S. and Sze, Vivienne},
  journal={IEEE Journal of Solid-State Circuits}, 
  title={Eyeriss: An Energy-Efficient Reconfigurable Accelerator for Deep Convolutional Neural Networks}, 
  year={2017},
  volume={52},
  number={1},
  pages={127-138},
  doi={10.1109/JSSC.2016.2616357}}

@inproceedings{rahaman2024secure,
  title={Secure ai systems: Emerging threats and defense mechanisms},
  author={Rahaman, Habibur and Chatterjee, Atri and Bhunia, Swarup},
  booktitle={2024 IEEE 33rd Asian Test Symposium (ATS)},
  pages={1--6},
  year={2024},
  organization={IEEE}
}

@article{nafi2025dash,
  title={DASH: A Meta-Attack Framework for Synthesizing Effective and Stealthy Adversarial Examples},
  author={Nafi, Abdullah Al Nomaan and Rahaman, Habibur and Haider, Zafaryab and Mahfuz, Tanzim and Suya, Fnu and Bhunia, Swarup and Chakraborty, Prabuddha},
  journal={arXiv preprint arXiv:2508.13309},
  year={2025}
}

@article{rahaman2025runtime,
  title={Runtime Detection of Adversarial Attacks in AI Accelerators Using Performance Counters},
  author={Rahaman, Habibur and Chatterjee, Atri and Bhunia, Swarup},
  journal={arXiv preprint arXiv:2503.07568},
  year={2025}
}

@inproceedings{rahaman2024samurai,
  title={Samurai: A framework for safeguarding against malicious usage and resilience of ai},
  author={Rahaman, Habibur and Chatterjee, Atri and Bhunia, Swarup},
  booktitle={2024 IEEE 33rd Asian Test Symposium (ATS)},
  pages={1--6},
  year={2024},
  organization={IEEE}
}

@inproceedings{rakin2019bit,
  title={Bit-flip attack: Crushing neural network with progressive bit search},
  author={Rakin, Adnan Siraj and He, Zhezhi and Fan, Deliang},
  booktitle={Proceedings of the IEEE/CVF International Conference on Computer Vision},
  pages={1211--1220},
  year={2019}
}

\end{document}